\documentclass[apj, numberedappendix]{emulateapj}

\usepackage{apjfonts}
\usepackage{wrapfig}
\usepackage{amsmath}

\newcommand{\agnjnk}{AGNs}

\newcommand{\tnm}{\tablenotemark}
\newcommand{\tnt}{\tablenotetext}

\newcommand{\fnm}{\footnotemark}
\newcommand{\fnt}{\footnotetext}

\newcommand{\ergs}{erg s$^{-1}$}
\newcommand{\flux}{erg cm$^{-2}$ s$^{-1}$}

\newcommand{\pers}{s$^{-1}$}
\newcommand{\cdens}{cm$^{-2}$}

\newcommand{\xmmnewton}{{\it XMM-Newton}}
\newcommand{\chandra}{{\it Chandra}}

\newcommand{\im}{\item}

\newcommand{\lumfour}{$L_{\rm 4.5\mu m}$}

\newcommand{\average}[1]{\ensuremath{\langle#1\rangle} }

\newcommand{\msun}{M_{\sun}}

\newcommand{\W}{\hphantom{0}}

\newcommand{\lbol}{$L_{\rm bol}$}
\newcommand{\lx}{$L_{\rm X}$}

\newcommand{\bootes}{Bo\"{o}tes} 
\newcommand{\xbootes}{XBo\"{o}tes} 
\newcommand{\spitzer}{{\it Spitzer}}

\newcommand{\hminus}{$h^{-1}$}
\newcommand{\hmpc}{$h^{-1}$ Mpc}
\newcommand{\hmsun}{$h^{-1}$ $M_{\sun}$}

\newcommand{\itot}{$I_{\rm tot}$}
\newcommand{\urs}{$^{0.1}(u-r)$}
\newcommand{\mrone}{$M_{^{0.1}r}$}
\newcommand{\dccorr}{$(\Delta C)_{\rm corr}$}

\begin{document}
\slugcomment{Accepted for publication in The Astrophysical Journal}

\title{Host galaxies, clustering,  Eddington ratios, and evolution \\ of  radio, X-ray, and infrared-selected \agnjnk}

\shorttitle{AGN HOST GALAXIES AND CLUSTERING}
\shortauthors{HICKOX ET AL.}
\author{Ryan C. Hickox\altaffilmark{1}}
\author{Christine Jones\altaffilmark{1}}
\author{William R. Forman\altaffilmark{1}}
\author{Stephen S. Murray\altaffilmark{1}}
\author{Christopher S. Kochanek\altaffilmark{2}}
\author{Daniel Eisenstein\altaffilmark{3}}
\author{Buell T. Jannuzi\altaffilmark{4}}
\author{Arjun Dey\altaffilmark{4}}
\author{Michael J.~I. Brown\altaffilmark{5}}
\author{Daniel Stern\altaffilmark{6}}
\author{Peter R. Eisenhardt\altaffilmark{6}}
\author{Varoujan Gorjian\altaffilmark{6}}
\author{Mark Brodwin\altaffilmark{1,4}}
\author{Ramesh Narayan\altaffilmark{1}}
\author{Richard J. Cool\altaffilmark{7}}
\author{Almus Kenter\altaffilmark{1}}
\author{Nelson Caldwell\altaffilmark{1}}
\author{Michael E. Anderson\altaffilmark{8,1}}

\altaffiltext{1}{Harvard-Smithsonian Center for Astrophysics, 60 Garden Street,
 Cambridge, MA 02138; rhickox@cfa.harvard.edu.}
\altaffiltext{2}{Department of Astronomy and Center for Cosmology and Astroparticle
Physics, The Ohio State University, 140 West 18th Avenue, Columbus, OH 43210-1173.}
\altaffiltext{3}{Steward Observatory, 933 North Cherry Avenue, Tucson, AZ 85721.}
\altaffiltext{4}{National Optical Astronomy Observatory, Tucson, AZ
 85726-6732.}
\altaffiltext{5}{School of Physics, Monash
University, Clayton 3800, Victoria, Australia.}
\altaffiltext{6}{Jet Propulsion Laboratory, California Institute of Technology, Pasadena, CA 91109.}
\altaffiltext{7}{Princeton University Observatory, Peyton Hall,
 Princeton, NJ  08544-1001.}
\altaffiltext{8}{Department of Astronomy, California Institute of Technology, Pasadena, CA 91109.}

\begin{abstract}
We explore the connection between different classes of active galactic
nuclei (AGNs) and the evolution of their host galaxies, by deriving
host galaxy properties, clustering, and Eddington ratios of AGNs
selected in the radio, X-ray, and infrared (IR) wavebands.  We study a
sample of 585 AGNs at $0.25<z<0.8$ using redshifts from the AGN and
Galaxy Evolution Survey (AGES).  We select AGNs with observations in
the radio at 1.4 GHz from the Westerbork Synthesis Radio Telescope,
X-rays from the \chandra\ X\bootes\ Survey, and mid-IR from the {\em
Spitzer} IRAC Shallow Survey.  The radio, X-ray, and IR AGN samples
show modest overlap, indicating that to the flux limits of the
survey, they represent largely distinct classes of AGNs.  We derive
host galaxy colors and luminosities, as well as Eddington ratios, for
obscured or optically faint AGNs.  We also measure the two-point
cross-correlation between AGNs and galaxies on scales of 0.3--10
\hmpc, and derive typical dark matter halo masses.  We find that: (1)
radio AGNs are mainly found in luminous red sequence galaxies, are
strongly clustered (with $M_{\rm halo}\sim3\times10^{13}$ \hmsun), and
have very low Eddington ratios ($\lambda\lesssim10^{-3}$); (2)
X-ray-selected AGNs are preferentially found in galaxies that lie in
the "green valley" of color-magnitude space and are clustered
similar to typical AGES galaxies ($M_{\rm halo}\sim10^{13}$ \hmsun),
with $10^{-3}\lesssim \lambda \lesssim 1$; (3) IR AGNs reside in
slightly bluer, slightly less luminous galaxies than X-ray AGNs, are
weakly clustered ($M_{\rm halo}\lesssim 10^{12}$ \hmsun), and have
$\lambda>10^{-2}$.  We interpret these results in terms of a simple
model of AGN and galaxy evolution, whereby a ``quasar'' phase and the
growth of the stellar bulge occurs when a galaxy's dark matter halo
reaches a critical mass between $\sim$$10^{12}$ and $10^{13}$ $\msun$.
After this event, star formation ceases and AGN accretion shifts from
radiatively efficient (optical- and IR- bright) to radiatively
inefficient (optically faint, radio-bright) modes.
\end{abstract}


\section{Introduction}
\label{intro}

There is increasing evidence that the evolution of galaxies is related
to the evolution of their supermassive black holes (SMBHs), which
primarily grow through accretion of material as active galactic nuclei
(AGNs).  This connection between galaxies and SMBHs is suggested by
the observed tight correlation between SMBH and galaxy bulge masses
\citep[e.g.,][]{mago98, ferr00, gebh00}, and the similar redshift
evolution of star formation and AGN activity
\citep[e.g.,][]{mada96,ueda03}.  Galaxies show a well-established
bimodality in color that separates blue, generally disk-dominated,
star-forming galaxies (the ``blue cloud'') from red, generally
bulge-dominated galaxies (the ``red sequence'') \citep[see e.g.,][and
references therein]{stra01galcol,blan06blue, fabe07lfunc}.  Recent work has
suggested that AGNs may play a crucial role in the origin of this
bimodality, and particularly the quenching of star formation in blue
galaxies and their transition to the red sequence
\citep[e.g.,][]{hopk06apjs, crot06,khal08feedback}.

Valuable clues to the associated evolution of galaxies and AGNs come
from the properties of AGN hosts.  It is well-established that radio
AGNs are most commonly found in massive early-type galaxies
\citep[e.g.,][]{yee87radio,best05}, and that a majority of bright
local ellipticals show low-power nuclear radio sources
\citep{sadl89radio}.  Among optically selected AGNs,
\citet{kauf03host} found that low-redshift, low-luminosity AGNs reside
in massive galaxies with generally old stellar populations, and that
the stellar population becomes younger for galaxies with increasing
[\ion{O}{3}] luminosity.  At higher redshifts ($0.5\lesssim z \lesssim
1.5$), X-ray-selected AGNs are also found in massive galaxies with
generally red colors \citep{alon08agn}, although some studies have
also found an excess of AGNs in the ``green valley'' between the blue
cloud and red sequence, compared to the distribution of quiescent
galaxies \citep{sanc04agnhost, nand07host, geor08agn, silv08host,
koce09agnclust, scha09agn, trei09ecdfs}.  The differences in host
galaxies between AGNs of different types suggest that the process of
nuclear accretion is related to the mass and age of the host galaxy.

Further constraints on the links between AGNs and galaxies come from
the clustering of AGNs and their local galaxy environments.  Among
quiescent galaxies, red galaxies are more strongly clustered, and
therefore are found in denser environments, than blue galaxies
\citep[for some recent results see e.g.,][]{zeha05a, coil08galclust}.
Clustering measurements on scales $\gtrsim 1$ Mpc allow us to estimate
the masses of the dark matter halos in which galaxies reside
\citep[e.g.,][]{shet99b}.  There is evidence that the evolution of
galaxies is strongly linked to host halo mass
\citep[e.g.,][]{fabe07lfunc}, and similar relationships may also exist
for AGN activity.  Previous works have found somewhat differing
results for AGN clustering and environments, largely owing to
different samples of sources \citep[see][ for a summary of previous
results]{brow01clust}.  However, a general picture has emerged in
which radio-loud AGNs are found in dense environments, and massive
halos, similar to galaxy clusters, while radio-quiet AGNs are found in
poorer environments similar to field galaxies.

Recently, with large surveys such as the Sloan Digital Sky Survey
\citep[SDSS,][]{york00sdss}, the 2dF QSO Redshift Survey
\citep{croo04twodfqz}, and the DEEP2 survey \citep{davi03}, AGN-galaxy
clustering results have been placed on firmer statistical
footing. Studies of the autocorrelation of optically selected quasars
from 2dF and SDSS showed that their clustering amplitude increases
with redshift out to $z\sim3$, indicating that at all redshifts they
inhabit dark matter halos of similar characteristic mass
$\sim$$3\times10^{12}$ $M_{\sun}$ \citep{porc04clust,croo05,coil07a,
myer07clust1,shen07clust, daan08clust,padm08qsored}.  This suggests
that luminous quasar activity is related to the mass of the
surrounding dark matter halo.  Among lower luminosity
optically selected AGNs from SDSS, the Seyfert galaxies are generally
less clustered than the full galaxy sample \citep{kauf04}, but show
comparable large-scale clustering to control samples of quiescent
galaxies with similar properties to the AGN hosts \citep{li06agnclust,
mand09agnclust}.  Very low luminosity AGNs (low ionization nuclear
emission regions (LINERs)) show no significant bias relative to
normal galaxies\fnm\ \citep{mill03, cons06clust}.  At higher
redshifts, \citet{coil07a} found that quasars in the DEEP2 fields at
$z\sim 1$ show a slight antibias (although at the 1--$2\sigma$ level)
relative to all galaxies, similar to the antibias observed for blue
galaxies.  

In contrast, using X-ray data in the Extended Groth Strip region and
DEEP2 redshifts, \citet{geor07} found that X-ray selected AGNs
generally lie in overdense regions at $z\sim 1$, while in the Extended
Chandra Deep Field South, \citet{silv08host} found that AGNs in
``green'' hosts at $0.4<z<1.1$ preferentially reside in large-scale
structures such as walls or filaments.  

X-ray AGNs are generally strongly clustered and reside in relatively
dense environments.  Studies of the spatial auto-correlation of X-ray
AGNs detected with \chandra\ and \xmmnewton\
\citep[e.g.,][]{basi04xmmclust, gill05, yang06, pucc06xmmclust,
miya07xmmclust} generally obtain relatively large clustering
amplitudes \citep[although with differing results that may reflect a
variation in clustering with X-ray flux;][]{plio08xclust}.  Recently,
\citet{coil09xclust} found that X-ray AGNs in the Extended Groth Strip
are significantly biased relative to galaxies in the DEEP2 survey, in
contrast to the results for optical quasars.  Correspondingly,
\citet{geor07} showed that X-ray AGNs in DEEP2 at $z\sim 1$ generally
lie in overdense regions, while in the Extended {\em Chandra} Deep
Field South, \citet{silv08host} found that AGNs in "green" hosts at
$0.4 < z < 1.1$ preferentially reside in large-scale structures such
as walls or filaments.  Other recent works
\citep{lehm09proto, koce09xsuperclust, gala09agnclust} have found an
enhancement of the density of X-ray selected AGNs in large-scale
structures at relatively high redshifts ($z\gtrsim 0.9$).  However,
the significance of the results for X-ray AGNs has been limited by
relatively small AGN samples.

Compared to X-ray selected sources, radio-selected AGN are found in
even denser environments including X-ray groups and clusters
\citep[e.g.,][]{cros08radiox}.  Radio AGNs show strong spatial
clustering similar to local elliptical galaxies
\citep[e.g.][]{mand09agnclust, wake08radio}, with evidence for higher
amplitude for more powerful sources \citep[e.g.,][]{over03radio}.

\footnotetext{Throughout this paper, the terms ``quiescent'' or
``normal'' refer to galaxies with no detected AGN activity, but do
include star-forming galaxies.}

These results indicate that AGNs selected using different techniques
represent separate populations, with differences in host galaxies,
environments, and accretion modes. Large redshift surveys with
extensive multiwavelength coverage now make it possible to explore,
{\em within the same uniform data set}, (1) host galaxies, (2)
clustering, and thus dark matter halo masses, and (3) Eddington ratios
of AGNs selected by these different techniques.

In this paper, we study a sample of AGNs in the redshift interval
$0.25<z<0.8$, using spectroscopic redshifts from the AGN and Galaxy
Evolution Survey (AGES).  We focus on AGNs selected in the radio,
X-ray, and infrared (IR) bands, which provide relatively efficient and
unbiased ways of identifying certain populations of AGNs \citep[for a
review see][]{mush04book}.  We do not, in general, use optical
spectral diagnostics for selecting AGNs, because of selection effects
that can arise from the low-resolution fiber spectroscopy of some of
the sources.  We place particular emphasis on obscured or optically
faint AGNs, for which we can study the optical properties of their
host galaxies.  For these objects we can
also measure host galaxy bulge luminosities and thus estimate black
hole masses (assuming an $L_{\rm bul}-M_{\rm BH}$ relation), in order
to derive Eddington ratios.

This paper is organized as follows.  In Section~\ref{data}, we describe the
\bootes\ multiwavelength data set.  In Section~\ref{galsample} we describe the main
AGES galaxy sample, and in Section~\ref{agnsample} we discuss the samples
of radio, X-ray, and IR selected AGNs.  In Section~\ref{agnhost} we present
results on the colors and luminosities of AGN host galaxies.  We
describe our correlation analysis in Section~\ref{corr}, and present the
results, along with estimates of dark matter halo mass, in
Section~\ref{results}.  In Section~\ref{xedd} we calculate average X-ray spectra
and Eddington ratios for the different samples of AGNs.  In
Section~\ref{discussion} we discuss our results in terms of a simple
picture of AGN and galaxy evolution, in which luminous quasar activity
and the termination of star formation occur at a characteristic dark
matter halo mass.  We summarize our results in Section~\ref{summary}.  We
also include two appendices that describe (Appendix \ref{appendix}) our
correction for AGN contamination of host galaxy colors, and
(Appendix \ref{appendix_ir}) limits on the contamination of the IR AGN sample
by star-forming galaxies.

Throughout this paper we assume a lambda cold dark matter ($\Lambda$CDM) cosmology with
$\Omega_{\rm m}=0.3$ and $\Omega_{\Lambda}=0.7$.  For direct
comparison with other works, we assume $H_0=70$ km s$^{-1}$
Mpc$^{-1}$, except for comoving distances and absolute magnitudes,
which are explicitly given in terms of $h=H_0/(100$ km s$^{-1}$
Mpc$^{-1})$.  The NOAO Deep Wide-Field Survey (NDWFS) photometry in the
$B_W$, $R$, and $I$ bands is presented in Vega magnitudes, while
magnitudes in the SDSS $u$ and $r$ bands are AB normalized.  All
quoted uncertainties are $1\sigma$ (68\% confidence).

\section{Observations}
\label{data}

The 9 deg$^{2}$ survey region in \bootes\ covered by the NDWFS
\citep{jann99} is unique among extragalactic multiwavelength surveys
in its wide field and uniform coverage using space- and ground-based
observatories, including the {\it Chandra X-Ray Observatory} and the
{\em Spitzer Space Telescope}.  Extensive optical spectroscopy makes
this field especially well suited for studying the statistical
properties of a large number of AGNs (C.\ Kochanek et~al. 2009, in
preparation).  The areas covered by the X-ray, optical, infrared, and
radio surveys are shown in Figure 1.

Redshifts for this study come from AGES, which used the Hectospec
multifiber spectrograph on the MMT \citep{fabr05}.  We use AGES Data
Release 2 (DR 2), which consists of all the AGES spectra taken in
2004--2006.  The 2004 observations covered 15 pointings for a total
area of 7.9 deg$^2$, which we define to be the main AGES survey
region.  In the 2004 observations, targets include (1) all extended
sources (i.e. galaxies, see Section~\ref{counterparts}) with $R\le
19.2$ (2) a randomly selected sample of 20\% of all extended sources
with $19<R\le 20$, (3) all extended sources with $R\le 20$ and
Infrared Arary Camera (IRAC) 3.6, 4.5, 5.8, and 8.0 $\mu$m magnitudes
$\le$15.2, 15.2, 14.7, and 13.2, respectively.  In addition, (4)
fainter sources were observed, selected mainly from objects that are
counterparts of \chandra\ X-ray sources \citep{murr05, brand06a,
kent05}, radio sources from the Very Large Array (VLA) FIRST survey \citep{beck95}, and
objects selected from 24 $\mu$m observations with the Multiband
Imaging Photometer for \spitzer\ (MIPS) (E.\ Le Floc'h et al. 2009, in
preparation).  AGES DR 2 contains $I$-selected targets with $I\le
21.5$ for point sources and $I\le 20.5$ for extended sources.

AGES redshifts and spectral classifications follow the procedures used
by SDSS.  Redshifts are measured by comparing the extracted
spectroscopy to the grid of galaxy templates, and choosing the
combination of template and redshift that minimizes the $\chi^2$
between the data and the model.  Good redshifts are defined to be those that
do not have a second minimum in $\chi^2$ that is within 0.01 in
reduced $\chi^2$ to the best-fit value.  All redshifts were examined
by eye and bad fits were flagged manually.  

Optical photometry from NDWFS was used for the selection of AGES
targets and to derive optical colors and fluxes for AGES sources.
NDWFS images were obtained with the Mosaic-1 camera on the 4-m Mayall
Telescope at Kitt Peak National Observatory, with 50\% completeness
limits of 26.7, 25.0, and 24.9 mag, in the $B_{W}$, $R$, and $I$
bands, respectively.  Photometry is derived using SExtractor
\citep{bert96}.

Radio data are taken from the Westerbork Synthesis Radio Telescope
(WSRT) 1.4 GHz radio survey that covers $\approx$7 deg$^2$ in the
center of the NDWFS field \citep{devr02}.  3035 radio sources are
detected in the AGES survey region, to a limiting flux of $\approx$0.1
mJy and beam size $13\arcsec \times 27\arcsec$.  Centroid positions of
the radio sources are measured to 0\farcs4.

X-ray data are taken from the X\bootes\ survey, which covers the full
AGES spectroscopic region.  The X\bootes\ mosaic consists of 126 5 ks
\chandra\ ACIS-I exposures and is the largest contiguous field
observed to date with \chandra\ \citep{murr05}. Owing to the shallow
exposures and low background in the ACIS detector, X-ray sources can
be detected to high significance with as few as four counts. The
X\bootes\ catalog contains 3293 X-ray point sources with four or more
counts \citep{kent05},  and of these, 2724 lie in the AGES survey
region.

Mid-IR observations are taken from the \spitzer\ IRAC Shallow
Survey \citep{eise04}, which covers the full AGES field in all four
IRAC bands (3.6, 4.5, 5.8, and 8 $\mu$m), with $5\sigma$ flux limits
of 6.4, 8.8, 51 and 50 $\mu$Jy respectively.  The IRAC photometry is
described in detail in \citet{brod06}.  The catalog contains 15,488
sources detected in all four IRAC bands, and of these, 14,069 lie
within the AGES survey region.

\begin{figure}
\epsscale{1.1}
\plotone{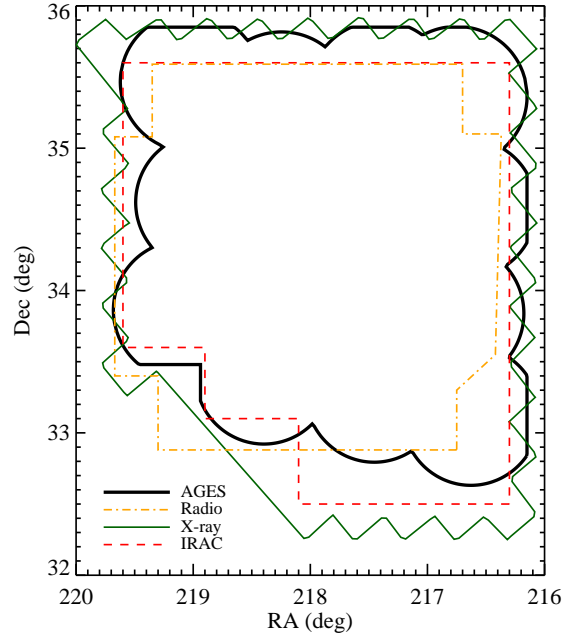}
\caption{Map of the \bootes\ survey region, showing the approximate
areas covered by the AGES, WSRT (radio), \xbootes\ (X-ray), and IRAC
Shallow Survey (mid-IR) observations.
\label{fbound}}
\end{figure}

\section{Galaxy sample}
\label{galsample}

The first step in our analysis is to define a statistical sample of
galaxies from the AGES catalog.  These galaxies will be used to
provide a comparison sample against which we can compare the
properties of AGN host galaxies.  These galaxies will also be used to
determine the spatial cross-correlation between AGNs and normal
galaxies, and so to derive AGN clustering.

The statistical galaxy sample consists of AGES galaxy targets with the
flux limit $I<20$.  We outline the key aspects of the statistical
galaxy sample here, while full details will be presented by
D.\ Eisenstein et~al.\ (2009; in preparation).  AGES galaxy targets
must satisfy quality cuts in the $I$ band and either the $B_W$ and $R$
bands in the NDWFS imaging.  In addition, all objects in the sample
must have good detections (SExtractor ${\rm FLAGS} < 8$) in all three
bands, to allow for $K$ corrections (this eliminates very few
sources).  In addition, objects must be detected as an extended source
(see Section~\ref{counterparts}) in at least one of the three bands.

\begin{figure}
\epsscale{1.1}
\plotone{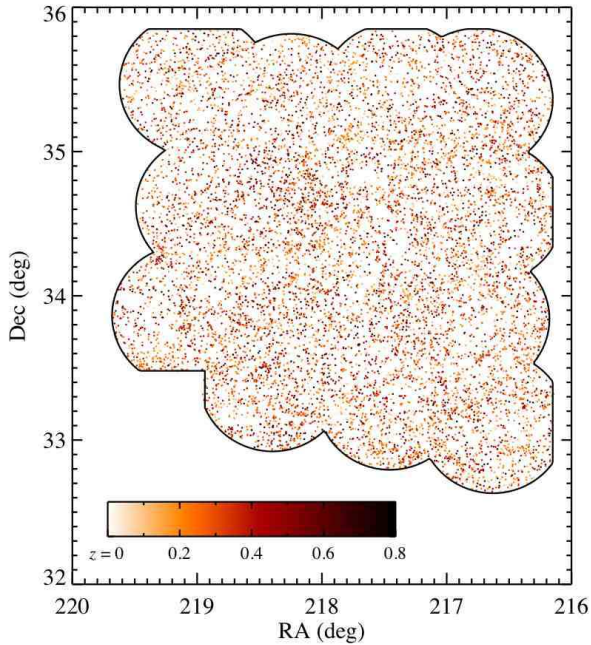}
\caption{ Positions of 6262 AGES galaxies in the main statistical
sample (Section~\ref{galsample}) at redshifts $0.25<z<0.8$.  Points are
color-coded by redshift.  The black line shows the boundary of the 15
main AGES pointings.
\label{fsky}}
\end{figure}

\begin{figure}
\epsscale{1.1}
\plotone{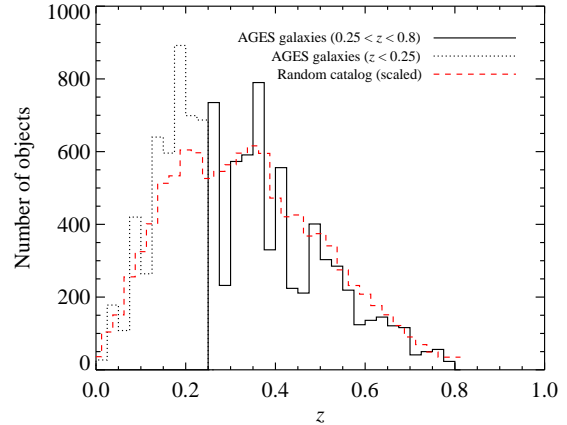}
\caption{Redshift distribution for galaxies in the main AGES sample
and for the AGES random catalog (see Section~\ref{random}).  The solid and
dotted black histograms show the redshift distribution for AGES main
sample galaxies (in our analysis we only include objects at
$0.25<z<0.8$, shown by the solid black histogram; for completeness the
dotted black histogram shows AGES main sample galaxies at $z<0.25$).
The red dashed histogram shows the redshift distribution for the
random galaxy catalog.  For clarity, the histogram values for the
random galaxies are divided by 200, since the random sample contains
200 times as many objects as the (statistically weighted) observed
galaxy sample.  Note that the observed galaxy histogram is not
normalized by statistical weights.
\label{fz}} \vskip0.5cm
\end{figure}

In this analysis we use the auto (Kron) $I$-band magnitude; however in
some regions the $I$-band photometry suffers from
low-surface-brightness halos around bright stars, which causes the
Kron galaxy magnitudes to be overly bright.  To correct for this, we
define the $I_R$ magnitude constructed from the sum of the $R$ band
Kron magnitude and the $R-I$ color derived from 6\arcsec\ aperture
magnitudes.  We then compare $I_R$ and the $I$ (Kron) magnitude and
compute \itot, which consists of the fainter magnitude (if the two
differ significantly) and otherwise is the average of the two
magnitudes.  $I_{\rm tot}$ is typically 0.01--0.09 mag fainter
than the $I$ (Kron) magnitude.  We cut the sample at $I_{\rm
tot}<19.95$, which excludes 2\% of the galaxies with Kron magnitude
$I<20$.  The positions on the sky of the AGES galaxies with
$0.25<z<0.8$ are shown in Figure~\ref{fsky}, and their redshift
distribution is shown in Figure~\ref{fz}.

\subsection{Sampling weights}
\label{weights}
For the targets that pass the above criteria, a complicated set of
random sparse sampling criteria were used to select objects for
observation.  In general, 100\% of bright ($I<18.5$) galaxies were
observed, while the sampling rate for sources with $18.5<I<20$ was
20\%--30\% (depending on whether they pass flux cuts in interesting UV,
optical, or IR bands).  In addition, many galaxies were assigned
fibers (and obtained redshifts) as part of the AGES observations, but
were not systematically targeted.  Some of these redshifts correspond
to X-ray or radio-selected AGNs and so are used in this analysis.
However, these objects are not included in the ``main'' statistical
galaxy sample that we use for AGN-galaxy cross-correlation.  The main
AGES sample includes only those galaxies that were systematically
targeted in the random sampling.

For each target in the main AGES sample that has a well-determined
redshift, we assign a series of sampling weights.  First we assign a
sparse sampling weight, equal to the inverse of the random sampling
rate that is known exactly from the target selection (that is, if 20\%
of objects in a class are targeted, the sampling weight is 5).
Second, we determine a target assignment weight, given by the inverse
of the probability that a target was assigned a fiber in the
2004--2006 observing runs.  Owing to the high completeness of the AGES
observing strategy for the main galaxy sample, this fiber weight
averages only 1.05.  Finally, we assign a redshift weight, given by
the inverse probability that an object that is assigned to a fiber
succeeds in yielding a well-determined redshift.  We determine this
weight for various subsamples of the targets in bins of magnitude,
surface brightness, color, and sparse sampling class, by computing the
ratio of the number of attempts to the number of well-determined
redshifts.  The total statistical weight $W$ associated with each
galaxy is the product of these three weights.

\subsection{Bright star exclusion}
\label{brightstar}
Owing to difficulties in obtaining redshifts for objects in the
vicinity of bright stars, objects around bright sources were not
targeted for AGES spectroscopy.  The exclusion was performed using a
bright source list taken from objects in the USNO-B catalog
\citep{mone03} with $R<17$.  A number of these bright ``stars''
correspond to extended sources, some of which are AGES target
galaxies.  To minimize the exclusion of targets around AGES galaxies,
the 475 objects that match extended sources in the Two Micron All Sky
Survey (2MASS) catalog \citep{skru06} were removed from the bright
star list before the AGES targets were selected.  This leaves 6690
bright sources around which objects were not targeted for AGES
spectroscopy. Possible target objects were excluded around these
bright sources, with an exclusion radius that varied with the flux of
the bright source and the flux of the target object.  Around a bright
source of magnitude $R$, faint target objects were excluded out to a
radius
\begin{equation}
r_{\rm limit}=20\arcsec + 5\arcsec(15-R),
\end{equation}
while brighter target objects were excluded inside smaller radii.  

To minimize biases in our spatial clustering analysis, we  avoid
the regions on the sky from which AGES targets were excluded.  For simplicity, we
remove from our sample all AGES sources (regardless of flux) within
$r_{\rm limit}$ of a bright source, including the bright source itself
if it is an AGES galaxy.  We exclude the same regions in our random
catalog (Section~\ref{random}).  This exclusion removes 776 AGES target
galaxies that correspond to objects in the bright source catalog, but
were not matched to 2MASS extended sources.  However, all but 19 of
these bright AGES targets lie at $z<0.25$. To minimize the
effects of removing the bright AGES galaxies, we restrict our analysis to
objects at $z>0.25$, which leaves us with 6262 galaxies, with total
statistical weight equal to 15,653.  We refer to these galaxies as the
``main'' AGES statistical sample and focus on this sample in our
analysis.

\begin{figure*}
\epsscale{1.1}
\plottwo{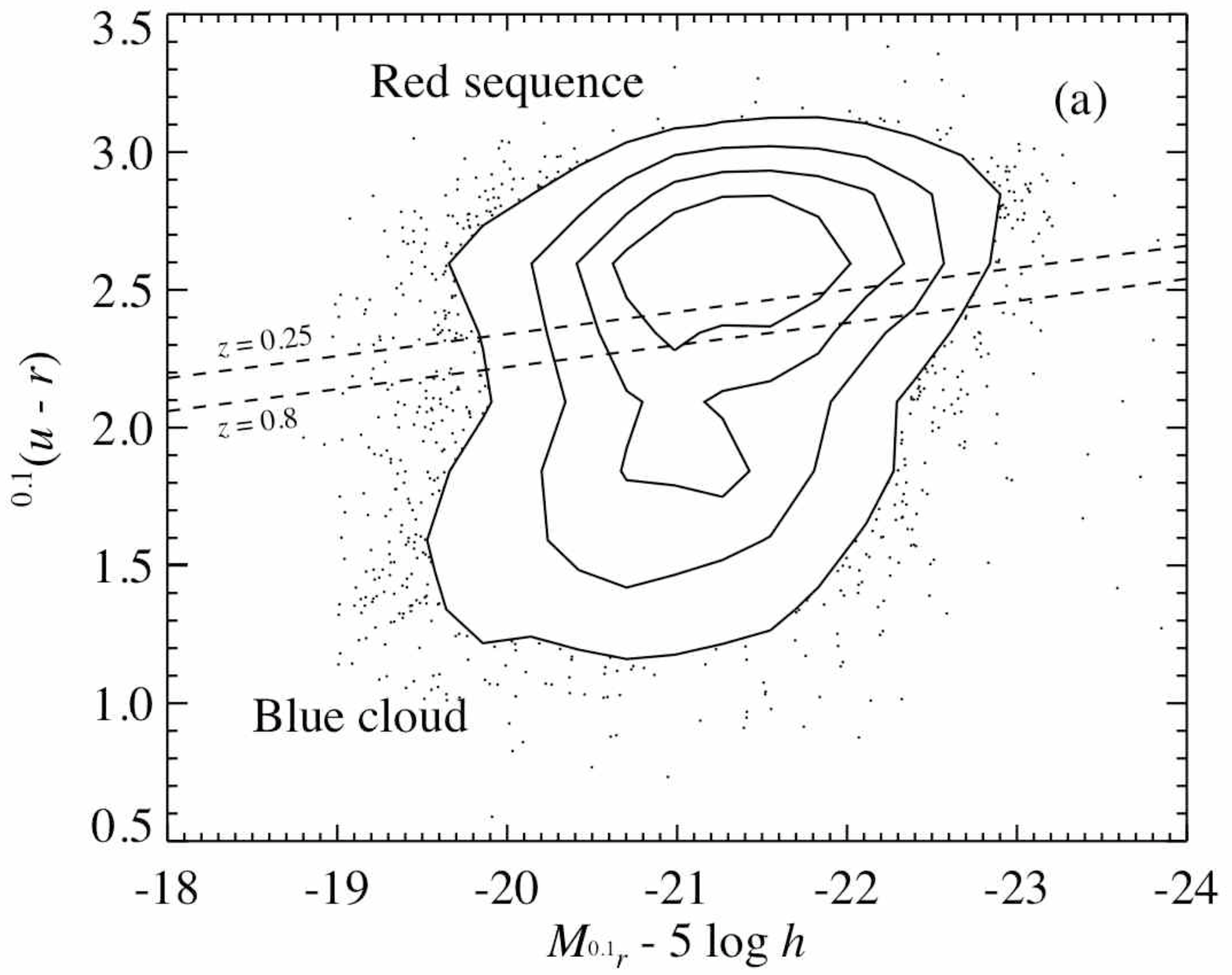}{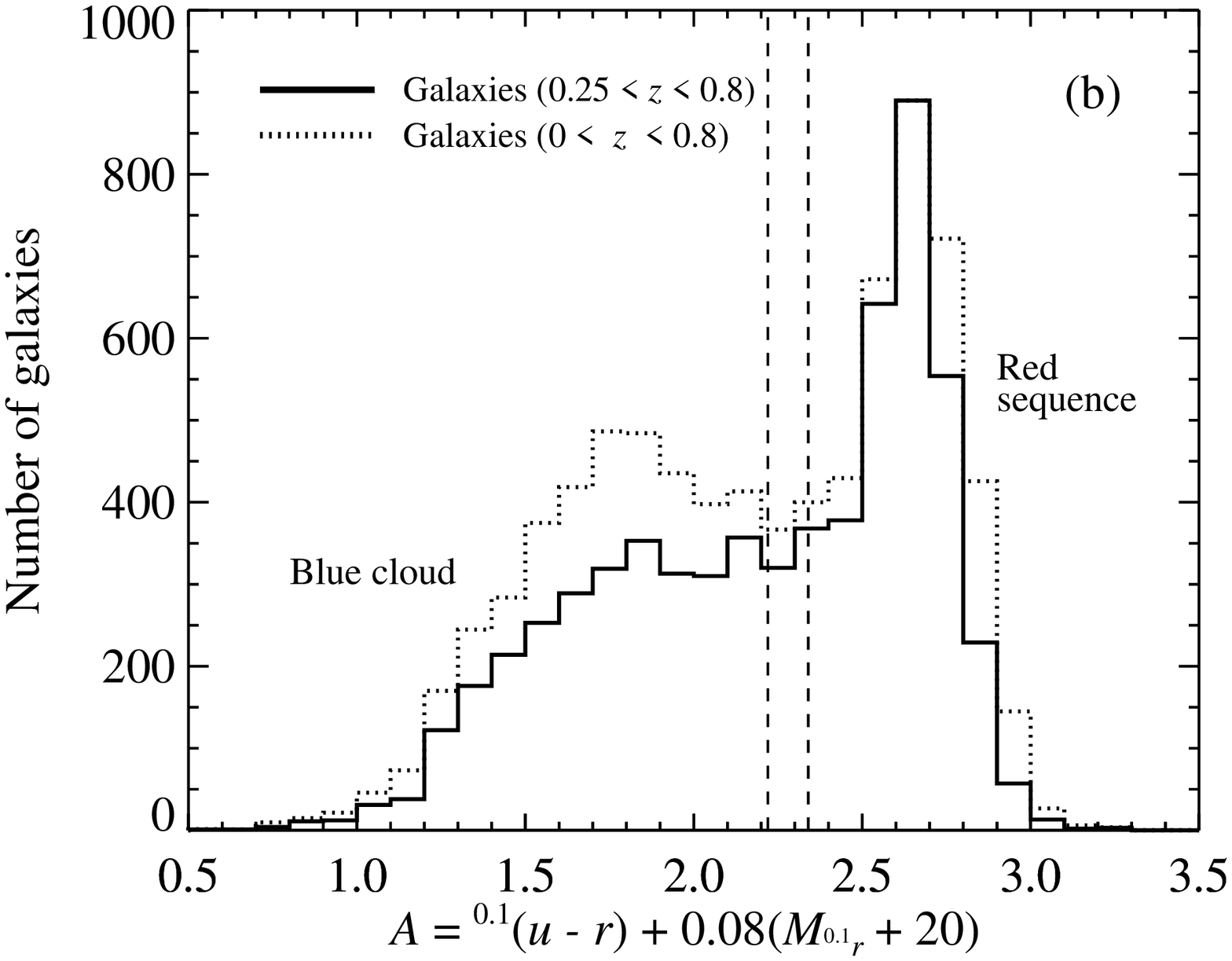}
\caption{({\em a}) Rest-frame optical luminosities and colors for AGES
galaxies.  The contours and points show the distribution of rest-frame
\urs\ color versus \mrone\ absolute magnitude (not corrected for passive
evolution), for AGES galaxies in the statistical sample in the
redshift interval $0.25<z<0.8$ (points are left out in the region
covered by contours).  The diagonal dashed lines show the empirical
division between red and blue galaxy samples for $z=0.25$ (top) and
$z=0.8$ (bottom).  ({\em b}) Distribution in the color parameter $A$ (see Eqn.~\ref{eqna})
for AGES galaxies at $0.25<z<0.8$, showing the ``red sequence'' and
``blue cloud''.  The boundaries between red and blue galaxies at
$z=0.25$ and $z=0.8$ are shown as dashed lines, as in ({\em a}).  The
dotted line shows the distribution including galaxies at $z<0.25$,
which include many lower luminosity blue galaxies, and which more
clearly shows the ``green valley''.
\label{fcol}} \vskip0.5cm
\end{figure*}

\subsection{$K$-corrections and colors}
\label{kcorr}
One powerful diagnostic for understanding the properties of galaxies
is their distribution in color and absolute magnitude.  Many previous
studies have found that the color-magnitude distribution of galaxies
is bimodal, separated into a red sequence of luminous,
bulge-dominated, passively evolving galaxies, and a blue cloud of less
luminous, star-forming, disk-dominated galaxies
\citep[e.g.,][]{stra01galcol,blan03galopt, kauf03mass}.

For the AGES galaxies, we use the 4\arcsec\ aperture magnitudes and
$I_{\rm tot}$ to perform $K$-corrections and thus derive the
rest-frame colors and absolute magnitudes.  As discussed above, only
galaxies with good $B_W$, $R$, and $I$ photometry are included in the
galaxy sample.  We obtain $B_W$ and $R$ magnitudes by adding the
$B_W-I$ and $R-I$ 4\arcsec\ aperture colors, respectively, to $I_{\rm
tot}$.  These magnitudes are in general close to the Kron magnitudes;
the mean and standard deviation in $\Delta m=m_{\rm tot}-m_{\rm Kron}$
are (0.01, 0.10) and (-0.003, 0.038) in the $B_W$ and $R$ bands,
respectively.  We convert Vega magnitudes to AB using corrections of
+0.02 mag, +0.19 mag, and +0.44 mag for the $B_W$, $R$, and $I$ bands,
respectively, and correct absolute magnitudes for interstellar
extinction \citep{schl98}, which for this field is only $A_I=0.02$
mag.

We define rest-frame colors and absolute magnitudes in terms of
photometric bands that are close to the wavelengths probed by the
NDWFS photometry at typical AGES redshifts.  We convert the $I$ and
$B_W$ magnitudes to $^{0.1}r$ and $^{0.1}u$ respectively, which are
defined as the AB magnitudes in the SDSS $r$ and $u$ bands, shifted
blueward by $z=0.1$.  These conversions are insensitive to the galaxy
spectral energy distribution (SED) at $z=0.42$ and 0.27, respectively.
Calculating rest-frame colors in terms of these bands allows for
straightforward comparison to results from SDSS
\citep[e.g.,][]{blan03kcorr,blan03galopt, blan03glf,kauf03stel}.  The
$K$-corrections are performed using kcorrect version 3.2 \citep{blan03kcorr};
using an updated version of kcorrect (version 4.1.4) produces no significant
change in the $K$-corrections.

In addition to calculating rest-frame colors for each galaxy, we
derive the absolute magnitude ($M_{^{0.1}r}$) in the $^{0.1}r$ band at
$z=0.1$.  We also calculate the absolute magnitude corrected for
passive evolution, assuming 1.6 mag of evolution per unit
redshift relative to the observed redshift \citep{eise05small}.  This
correction varies from 0 at $z=0.1$ (by definition) to +1.1 mag at
$z=0.8$, in the sense that the evolution-corrected absolute magnitude
at $z=0.1$ is fainter than the observed \mrone.  The
evolution-corrected absolute magnitudes are used only in the estimates
of galaxy bulge and black hole masses (Section~\ref{edd}); elsewhere we use
absolute magnitudes that are not corrected for evolution.

Figure~\ref{fcol} shows rest-frame $^{0.1}(u-r)$ color versus
$M_{^{0.1}r}$ for galaxies in the AGES main sample in the redshift
interval $0.25<z<0.8$.  The bimodal distribution in galaxy color (the
red sequence and blue cloud) is clearly evident.  We divide the
galaxies into red and blue subsamples; to account for the slope of the
red sequence, we define the quantity
\begin{equation}
A = {^{0.1}(u-r)}+0.08(M_{^{0.1}r}+20),
\label{eqna}
\end{equation}
(where $M_{^{0.1}r}$ is calculated for $h=1$) and divide red and blue
galaxies by a cut in $A$.  The red sequence evolves slightly in color
with redshift, so we empirically determine the median value in $A$ for
the red sequence as a function of $z$; this varies from $A_{\rm
red}=2.64$ at $z=0.25$ to $A_{\rm red}=2.52$ at $z=0.8$.  At each
redshift, we set the boundary between red and blue samples to be $0.3$
mag blueward of $A_{\rm red}$.  The selection boundaries for $z=0.25$
and $z=0.8$ are shown in Figure~\ref{fcol}.  This criterion selects 3119
red galaxies and 3143 blue galaxies in the redshift interval
$0.25<z<0.8$.  After a small correction for nuclear contamination from
AGNs (Appendix \ref{appendix}), we select 3146 red galaxies and 3119
blue galaxies.

\subsection{Random galaxy catalog}
\label{random}
To derive the spatial correlation of galaxies and AGNs (Section~\ref{corr})
requires an estimate of the number of galaxy-galaxy or AGN-galaxy
pairs expected if the galaxies were randomly distributed in comoving
space.  We therefore produce a catalog of galaxies that are spatially
uncorrelated, but reflect the selection function of the AGES sample.
To ensure that the fractional uncertainties of the AGN-random catalog
pairs are negligible, the density of random galaxies must be
significantly larger than in the data catalog.  This is especially
true on small scales (less than 1 \hminus\ Mpc), where real galaxies are most
strongly clustered.  We design the random catalog to contain 200 times
the (statistically weighted) number of galaxies in the data catalog.

For each galaxy in the AGES statistical sample with statistical weight
$W^i$, we generate $N_i$ duplicate galaxies, where $N_i=200W^i$,
rounded to the nearest integer.  We then assign each of these random
galaxies a statistical weight equal to $W_{R}^i=W^i/N_i$, so that the
total statistical weight in the random sample is equal to the total
weight in the data.  

We place each random galaxy in a random sky position inside the main
AGES field (and outside the bright star masks) and offset its redshift
by a random $-0.05<\Delta z<+0.05$.  This effectively applies a boxcar
smoothing to the line of sight distances on scales of 200--300
\hminus\ Mpc, larger than the largest observed structures, and so
eliminates any significant correlation along the line of sight.  The
redshift distribution of the random catalog is shown in Figure~\ref{fz},
along with the distribution for the observed galaxies.

\section{AGN sample}
\label{agnsample}

\begin{figure}
\epsscale{1.1}
\plotone{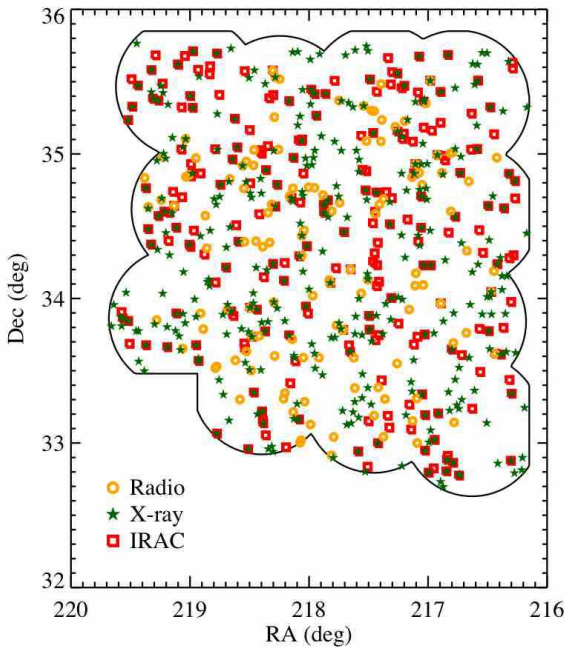}
\caption{Positions of 585 AGNs in the main AGES field.  We show AGNs
selected in the radio (orange circles) X-rays (green stars), and IR
(red squares) that lie in the main AGES field and
away from bright stars, and in the redshift interval $0.25<z<0.8$. The
solid line shows the boundary of the main AGES field.
\label{fagnsky}}
\vskip0.5cm
\end{figure}

We next define the samples of AGNs for which we will determine host
galaxy, clustering, and accretion properties.  For complete as
possible an AGN census, we select AGNs independently in the radio,
X-ray, and IR wavebands.  As discussed above, we do not use
optical spectroscopic selection in order to avoid systematic effects
in the analysis of the fiber spectroscopy.  We restrict ourselves to
sources that have redshifts from AGES in the interval $0.25<z<0.8$,
are located in the main AGES field, and are away from bright stars, as
described in Section~\ref{galsample}.  In this section we describe the
selection criteria and AGN samples.

\subsection{Radio AGNs}
\label{radio}
We first select AGNs in the radio.  The AGES catalog contains 525
matches (within 2\arcsec) to WSRT radio sources, of which 245 are in
the AGES field and in the interval $0.25<z<0.8$ (note that the WSRT
field is $\approx$7 deg$^2$ and does not cover the full AGES
spectroscopic region, as shown in Figure~\ref{fbound}). None of the
radio sources match more than one AGES galaxy.  

For each radio source we calculate the radio power $P_{1.4\;
\rm{GHz}}$ (in W Hz$^{-1}$).  To obtain the rest-frame power we
include a small $K$-correction of 0.1 dex, appropriate for the average
redshift of the radio-detected objects between $0.25<z<0.8$, and the
typical spectrum of faint 1.4 GHz sources \citep[$\alpha\approx0.5$,
where $S_\nu\propto \nu^{-\alpha}$;][]{pran06radio}.  The
distributions in $P_{1.4\; \rm{GHz}}$ and redshift are shown in
Figure~\ref{flumz}{\em a}.  For objects with low radio power, the
observed emission can be powered by either AGN activity or star
formation.  It is possible to separate AGNs from star-forming galaxies
based on a combination of $P_{1.4\; \rm{GHz}}$ and optical spectral
features such as H$\alpha$ luminosity \citep[e.g.,][]{kauf08radio};
however the required fits to the optical spectra for the AGES galaxies
are not currently available.  To minimize the number of star-forming
galaxies, we restrict our sample to the 122 sources with
$\log{P_{1.4\; \rm{GHz}}}>23.8$.  In low-redshift samples, almost all
galaxies with radio power above this cut are AGNs \citep[see Figure~1
of][]{kauf08radio}\footnotemark.  All the radio AGNs in our sample
have $P_{1.4\; \rm{GHz}}<2\times10^{26}$ W Hz$^{-1}$, and so are in
the luminosity range typical of \citet{fana74} type I sources
\citep{ledl96radio}.  The positions on the sky of the radio AGNs are
shown in Figure~\ref{fagnsky}.  To test for spurious matches of the
radio sources to AGES galaxies, we offset the radio source positions
by 1\arcmin\ and reperformed the source matching.  We find only four
matches, for which we calculated $P_{1.4\; \rm{GHz}}$ corresponding to
the redshift of the matched AGES galaxy.  None of the matches have
$\log{P_{1.4\; \rm{GHz}}}>23.8$, indicating that there is minimal
contamination from spurious matches in our radio AGN sample.

\footnotetext{Note that \citet{kauf08radio} assume $H_0=50$ km s$^{-1}$ Mpc$^{-1}$; our cut is scaled to our assumed cosmology.}

\subsection{X-ray AGNs}
\label{xray}
We next define the X-ray AGN sample.  Among the 3293 X-ray sources
with $\geq$4 X-ray counts in the \citet{kent05} catalog, 362 of these
lie within the AGES field, are not close to bright stars (see
Section~\ref{brightstar}), and are matched within 3\farcs5 to objects
with good AGES redshifts at $0.25<z<0.8$. We have excluded four
objects that show evidence for extended X-ray emission that could
arise from galaxy groups or clusters (we verify explicitly that this
does not significantly affect the clustering results).  Nine of the
362 sources in our X-ray AGN sample have two AGES galaxies within
3\farcs5; for these we selected the closer galaxy as the optical
counterpart.  We test for spurious matches by offsetting the
IR-selected AGN positions by 1\arcmin\ and reperforming the source
matching.  We find five matches, indicating contamination from spurious
matches of only $\approx$1\%.

For each X-ray source we calculate an 0.5--7 keV luminosity from the
number of detected source counts in the 0.5--2 keV and 2--7 keV bands
separately, and assuming a power-law X-ray spectrum with photon index
$\Gamma=1.8$, typical of AGNs \citep{tozz06}.  The corresponding
conversions from count rates (in counts s$^{-1}$) to flux (in \flux)
are $5.9\times10^{-12}$ erg cm$^{-2}$ count$^{-1}$ in the 0.5--2 keV
band and $1.9\times10^{-11}$ erg cm$^{-2}$ count$^{-1}$ in the 2--7
keV band \citep{kent05}.  For simplicity we do not include
$K$-corrections, which are generally small ($<$5\%); for most sources
uncertainty in the flux and $\Gamma$ are significantly larger.

\begin{figure}
\epsscale{0.9}
\plotone{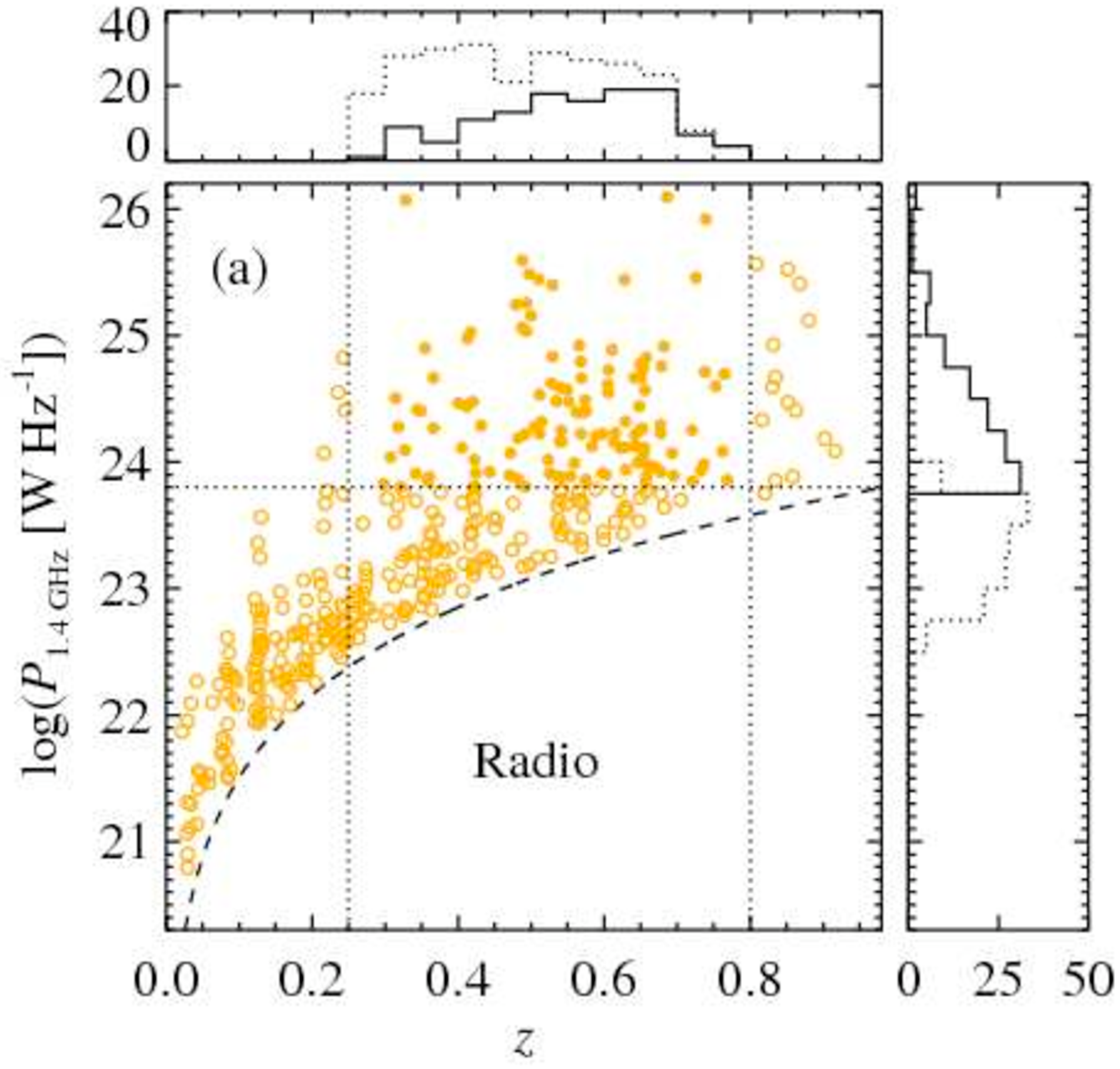}
\plotone{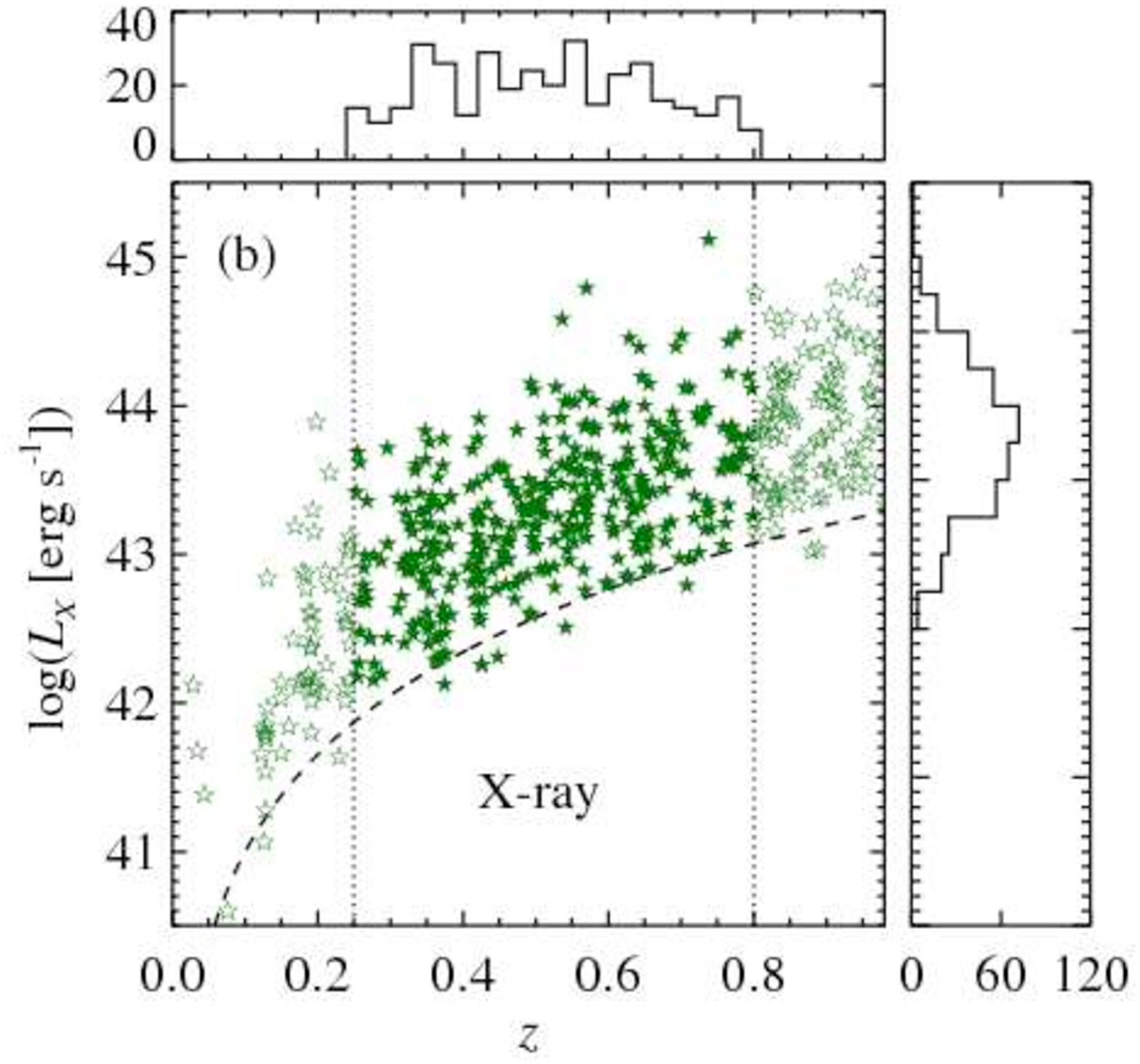}
\plotone{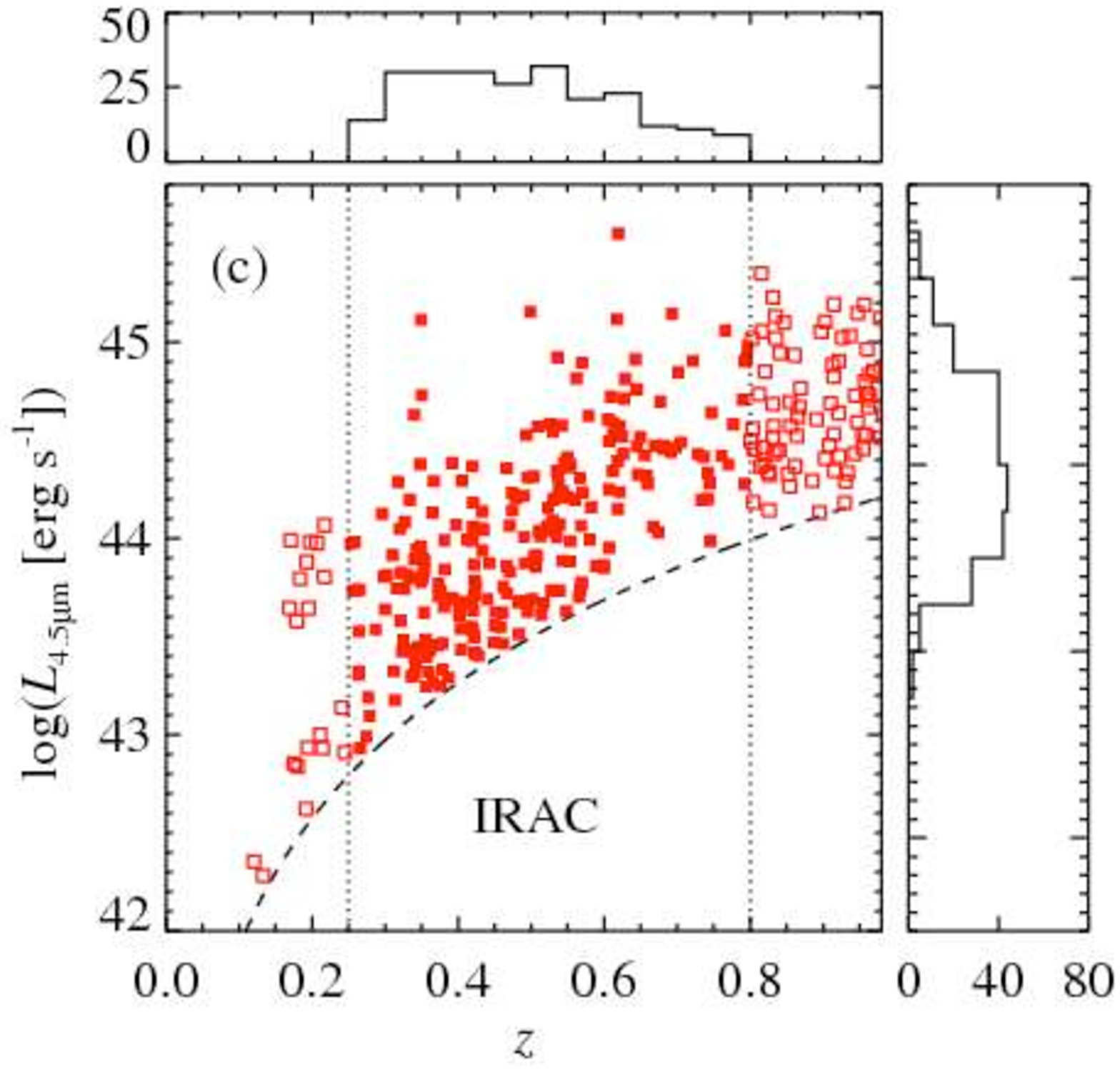}
\caption{Distribution in redshift and luminosity for ({\em a}) radio, ({\em
 b}) X-ray, and ({\em c}) IR AGNs with AGES redshifts, in the main AGES
 field and away from bright stars.  Luminosities are the observed 1.4
 GHz radio power ($P_{1.4\; \rm{GHz}}$), the 0.5--7 keV luminosity ($L_X$), and $\nu
 L_\nu$ at 4.5 \micron\ n the rest frame, ($L_{4.5\micron}$).  We focus on
 the redshift interval $0.25<z<0.8$, shown by the vertical dotted
 lines, and in the radio, sources with $\log{P_{1.4\; \rm{GHz}}}>23.8$
 (see Section~\ref{agnsample}), shown by a horizontal line.  Objects
 included in the analyses are shown by filled symbols.  The dashed lines
 show approximate luminosity limits for the survey flux limits (note
 that the X-ray limit is for an X-ray spectrum with photon index
 $\Gamma=1.8$). The histograms show the distribution in $z$ and luminosity
 for sources in this redshift interval.  For the radio sources, the
 dashed histograms show the distributions in $z$ and $P_{1.4\; \rm{GHz}}$ for the
 complete sample of radio sources in the interval $0.25<z<0.8$, while
 the solid histograms show those sources with $\log{P_{1.4\; \rm{GHz}}}>23.8$.
\label{flumz}}
\end{figure}

The X-ray AGN positions are shown in
Figure~\ref{fagnsky}, and their distribution in luminosity and redshift
is shown in Figure~\ref{flumz}{\em b}.  The luminosities of these
sources are in the range $10^{42}<L_X<10^{45}$ \ergs,
characteristic of moderate- to high luminosity AGNs and significantly
larger than is typically found for star-forming galaxies
\citep[e.g.,][]{rana03}.  We are therefore confident that the X-ray
emission from these objects is powered by nuclear accretion rather than star formation.

We note that since X\bootes\ sources have as few as 4 counts in 5 ks,
the fluxes of the faintest sources are affected by the Eddington bias.
For a differential flux distribution with $dN/dS\propto S^{-\beta}$,
the true mean flux for a source observed with $n$ counts is
approximately $n-\beta$ \citep{eddi13}.  At the X\bootes\ flux limits
($S_{\rm 0.5-2\; keV}\sim5\times10^{-15}$ \ergs), $\beta\simeq2$
\citep{vikh95a, kent05}, so a source with 4 counts has a true mean
flux of $\simeq$2 counts \citep{kent05}.  This correction is
relatively small for most sources, and does not affect our
interpretation of the X-ray sources as accretion-dominated.
Correcting for the Eddington bias, all of the sources have
$L_X>10^{42}$ \ergs, so we conclude that contamination by
starburst-dominated galaxies will be small.  In what follows we use
\lx\ values that are corrected for Eddington bias.

\subsection{IR AGNs}
\label{irac}
We finally select AGNs using IRAC observations.  With the launch of
{\em Spitzer}, the mid-IR has provided a sensitive new window for
identifying AGNs.  The mid-IR SEDs of broad-line quasars and Seyfert
galaxies typically show a featureless, roughly power-law continuum
that rises to longer wavelengths \citep[e.g.,][]{glik06}.  This SED
can be used to separate AGNs from normal and starburst galaxies, whose
SEDs exhibit a stellar bump at 1.6 \micron\ and then fall at longer
wavelengths.  Several criteria have been developed to select AGNs
based on IRAC colors \citep{lacy04, ster05} or SED fitting
\citep[e.g.,][]{alon06, donl07plaw}.  In this paper we use the color
selection of \citet{ster05}, which was derived using the \bootes\
multiwavelength data and so is well matched to the depth of the
survey.  Because mid-IR wavelengths are not as strongly affected by
dust extinction as the optical or UV, IRAC selection is particularly
useful for identifying dust-obscured AGNs \citep[e.g.,][]{ster05,
rowa05, poll06, hick07abs}.

\begin{figure}
\epsscale{1.2}
\plotone{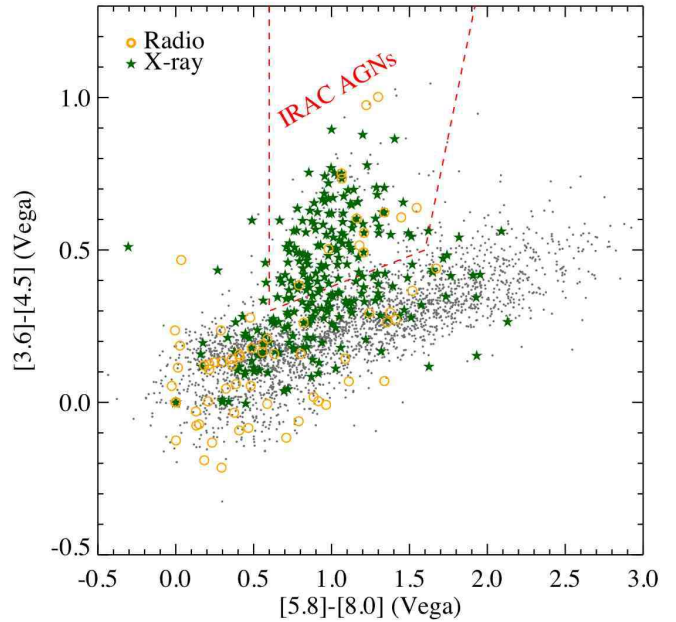}
\caption{IRAC color-color diagram for all AGES sources at
$0.25<z<0.8$ that lie inside the main AGES field and away from bright stars,
and have $5\sigma$ detections in all four IRAC bands.  The dashed line
shows the selection region defined by \citet{ster05}.  Symbols for
radio and X-ray AGNs are as in Figure~\ref{fagnsky}.  Most radio AGNs
are found to the lower left, in the region occupied by stars and
quiescent galaxies \citep{ster05}, while X-ray AGNs are found throughout the distribution, with a significant number in the IRAC AGN selection region.
\label{fstern}}
\end{figure}

The \citet{ster05} AGN color-color criteria requires that a source be
detected in all four IRAC bands.  Of the IRAC sources detected in all
four bands, 2952 are matched (within 3\farcs5) to AGES sources at
$0.25<z<0.8$.  The distribution in IRAC colors for these objects are
shown in Figure~\ref{fstern}.  The objects in the bottom left of this
diagram have IRAC colors similar to stars ($[5.8]-[8.0] < 0.8$ and
$[3.6]-[4.5] < 0.3$); these are generally passively evolving galaxies
that are dominated by starlight in the mid-IR.  The diagram also shows
a continuous sequence of sources with redder $[5.8]-[8.0]$ and
$[3.6]-[4.5]$ colors, which are generally star-forming galaxies with
dust features (mainly due to polycyclic aromatic hydrocarbons (PAHs)) that
move through the IRAC bands with redshift \citep[see Figure~1 of][ for
evolutionary tracks of star-forming galaxies on this diagram]{ster05}.
A separate sequence of sources with very red $[3.6]-[4.5]$ colors is
selected by the \citet{ster05} AGN criteria (shown by the dashed
line).  These SEDs are inconsistent with typical emission from dust heated by
star formation, and are characteristic of broad-line quasars and Seyfert
galaxies.

Of the AGES sources at $0.25<z<0.8$ matched to the IRAC photometry,
these criteria select 238 objects.  Six of these sources have two AGES
galaxies within 3\farcs5; we selected the closer galaxy as the
optical counterpart.  We test for spurious matches by offsetting the
IR-selected AGN positions by 1\arcmin\ and find six matches, indicating
contamination from spurious matches of only $\approx$2\%.  The
positions of the IR-selected AGNs are shown in Figure~\ref{fagnsky}.
For each IR-selected AGN, we calculate the rest-frame luminosity ($\nu
L_\nu$) at 4.5 \micron, \lumfour.  The SED at rest-frame 4.5 \micron\
is probed by the IRAC photometry at all redshifts in our sample, and
since this wavelength is significantly redward of the 1.6 \micron\
peak in the stellar emission from galaxies, the observed flux is
likely dominated by the AGN.  We estimate \lumfour\ by logarithmically
interpolating between the observed $\nu L_\nu$ in the 4.5 \micron, 5.8
\micron, and 8 \micron\ IRAC bands.  The distribution of the
IRAC-selected AGNs in \lumfour\ and redshift is shown in
Figure~\ref{flumz}{\em c}.

One possible limitation of IRAC color selection is in reliability.
For the radio and X-ray samples, the high radio and X-ray luminosities
are robust indicators of the presence of nuclear accretion.  The
reliability of IRAC selection is less well studied [although there are
discussions of this in \citet{barm06} and \citet{gorj08}].  The most
likely contaminants are star-forming galaxies, which in some extreme
cases, can have IR luminosities and colors similar to our IR-selected
AGN sample.

Our sample of AGNs with AGES spectroscopy appears to suffer little
contamination from star-forming galaxies, owing primarily to the
relatively bright optical and IRAC flux limits.  \citet{donl08spitz}
examined \spitzer\ 24 \micron\ sources in the GOODS-S field, and found
that IRAC color-selected samples in deep surveys have significant
contamination from star-forming galaxies, but this contamination
decreases for higher IR and optical fluxes.  \citet{donl08spitz} argue
that the selection of AGNs by fitting a power-law SED to the observed IRAC
emission reduces the contamination in their sample.  To check whether
power-law selection would improve the reliability of our sample, we apply the
power-law criteria of \citet{donl08spitz} to objects with AGES
spectra at $0.25<z<0.8$ and detections in all four IRAC bands.  We
identify 54 power-law AGNs, of which all but one are also selected
using the \citet{ster05} color selection.  Among the power-law AGNs,
53 lie in the X-ray field, and 21 (40\%) are detected in X-rays,
indicating that they are unambiguously AGNs.  In comparison, 184
color-selected AGNs are {\em not} selected by the power-law criteria
and have X-ray coverage, and of these 97 (53\%) are detected in
X-rays.  Therefore in our sample, the IRAC color criteria are equally
or more likely to identify X-ray-detected AGNs as the power-law
criteria.  We conclude that for our relatively shallow IRAC
observations and bright optical spectroscopic flux limit, the
\citet{ster05} color criteria are as reliable as power-law criteria in
selecting AGNs.

As another check, we carefully analyze the average X-ray emission from
the IRAC color-selected sources that are not detected in X-rays.  The
average fluxes, stacked spectrum, and distribution of observed X-ray
counts indicate that star-forming contamination in our IR-selected AGN
sample is at most $20\%$, and most likely $<$10\% (see Appendix
\ref{appendix_ir} for a detailed discussion).  We conclude that
contamination of the IRAC AGN sample from star-forming galaxies is
small.

\subsection{Optical counterparts}
\label{counterparts}

Useful observational clues as to the nature of AGNs come from the
morphologies of their optical counterparts.  Objects that are
unresolved in the NDWFS images are generally dominated in the optical
by the active nucleus, while objects that are optically extended are
dominated by their host galaxies and have optically faint or obscured
nuclei.  For optically extended sources, we can determine the optical
properties of the host galaxies, and indirectly, estimate their
central black hole masses.  We therefore divide the AGN sample into
two categories based on whether their optical counterparts are
extended or unresolved.

The sample of AGNs with extended counterparts is defined as sources
having the SExtractor parameter ${\rm CLASS\_STAR}<0.8$ in all of the
$B_W$, $R$, and $I$ bands.  92\%, 97\%, and 99\% of these objects have
${\rm CLASS\_STAR}<0.3$ in the $B_W$, $R$, and $I$ bands,
respectively.  These will be referred to as AGNs with ``galaxy'' (or
``extended'') counterparts.  Almost all (113) of the radio AGNs, roughly
two-thirds (238) of the X-ray AGNs, and half (133) of the IR AGNs have
extended optical counterparts.  For simplicity, and to obtain the best
possible statistics, we include all the optically extended sources in
our analysis.  We note that, 74, 157, and 111, of the radio, X-ray,
and IR AGNs, respectively were systematically targeted by AGES and are
included in the main galaxy sample.  We verify explicitly that if we
limit the sample only to these sources (which may suffer less
complicated selection effects), there is no significant change to our
results.

Although the optical counterparts of these sources are extended, the
derivation of host galaxy properties is complicated by the fact that
an optically weak AGN can still affect the integrated colors from the
galaxy. By comparing the aperture photometry for these AGNs to that of
normal galaxies, we carefully estimate the AGN color contamination and
derive corrections to the host galaxy colors.  Details of these
corrections are given in Appendix \ref{appendix}.  The color
corrections are small (typically $\simeq$0.1 mag), and do not
significantly affect our conclusions.

The second category of AGN optical counterparts are those with
pointlike optical morphologies.  We define pointlike sources as
those with ${\rm CLASS\_STAR}\geq 0.8$ in any of the $B_W$, $R$, and
$I$ bands.  The flux from most of these sources is dominated by
nuclear emission in the $B_W$; for simplicity, we will refer to them
as ``unresolved'', but we note that many of these sources are extended
in the $R$ and $I$ bands in which there is a larger contribution from
the host galaxy.  Although these objects are dominated by the nucleus
at blue optical wavelengths, they would not historically be defined as
``quasars''.  Their absolute magnitudes in the $B$ band (estimated
roughly from their observed $R$-band flux) are $-18<M_B<-24$.  Above
the usual luminosity threshold defined for quasars ($M_B<-23$),
the unresolved AGN sample only includes 11 sources (or 23 sources if
we include unresolved objects not selected in the radio, X-ray, or
IR).  Therefore, our sample of optically unresolved AGNs would primarily
be classified as luminous Seyfert galaxies.  We note that it is not
surprising that there are few luminous optical quasars in our sample
at $0.25<z<0.8$.  From the optical quasar luminosity function of
\citet{rich05}, we would expect only 18 objects with $M_B<-23$ in this
redshift interval over the area of 7.3 deg$^2$ (after excluding bright stars),
which probes a volume of $5\times10^6$ $h^{-3}$ Mpc$^3$ at
$0.25<z<0.8$.

\subsection{Overlap between samples}
\begin{figure}
\epsscale{0.8}
\plotone{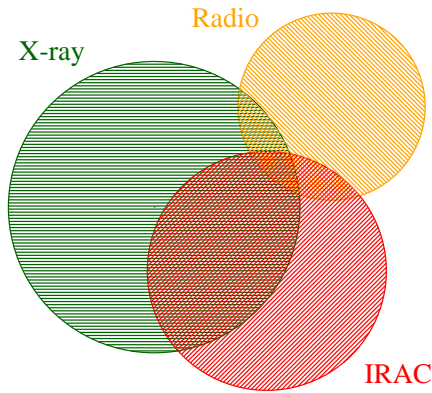}
\caption{Venn diagram showing the relative number of AGNs with AGES
redshifts at $0.25<z<0.8$, selected in the radio, X-ray, and IR (see also Table~\ref{tabsample}).  The
figure shows only those sources that lie within the regions covered by
radio, X-ray, and IRAC observations (Figure~\ref{fbound}).  The overlapping areas between
the samples correspond to the relative numbers selected in multiple
wavebands, except for the region where all three overlap, which is
slightly too small in the diagram.  There is some overlap between the X-ray and IRAC samples, but little overlap between these and the radio AGNs.
\label{fvenn}} \vskip0.5cm
\end{figure}

One complication of selecting AGNs in different wavebands is that the
samples overlap.  Some AGNs are selected in more than one
waveband, so the different AGN samples are not entirely
independent.  The various samples and their overlaps are given in
Table~\ref{tabsample}, and a Venn diagram depicting the relative sizes
and overlaps of the samples is shown in Figure~\ref{fvenn}.  This figure
shows the sample of sources that lie in the area covered by the radio,
X-ray, and IRAC observations, to allow a fair comparison between the
samples.

It is immediately apparent that there is very little overlap between
the radio AGNs and those selected in the X-ray and IR.  This may
provide an important clue as to the nature of the accretion in these
systems (Section~\ref{discussion}).  However, there is significant overlap
between the X-ray and IR samples.  Roughly one-third of X-ray AGNs are in
the IR sample, and half of the IR AGNs are detected in X-rays.  These
fractions are smaller (roughly 20\% and 40\%, respectively) when we
consider only optically extended AGNs.  The sources that are selected
as both X-ray and IRAC AGNs have relatively high luminosities; their
median $L_X$ and \lumfour\ are $\simeq$50\% higher than for AGNs
selected in only the X-ray and IR bands, respectively.  At some stages
in the following analysis we study the properties of this joint X-ray
and IR sample, but in most of the paper we consider the X-ray and IR
AGN samples separately.  This allows our results to be compared
directly to other works that study X-ray or IR AGNs only, but it is
important to keep in mind that the samples are not entirely
independent.

It is also essential to stress that AGN selection is strongly
affected by the flux limits of the observations in each waveband.  In
all three bands, our AGN selection criteria identify objects that are
{\em dominated} by the AGN in that band.  The luminosity cut in the
radio, the X-ray flux limit, and the color criteria in the IR all
select objects where the nucleus dominates over stellar processes at
the observed wavelengths.  However, as discussed in detail by
\citet{ho08llagn}, if observations go deep enough, then the majority of
AGNs show signatures of nuclear activity in multiple bands,
particularly in the X-ray and radio.  Indeed, we show in Appendix
\ref{appendix_ir} that many, if not most, of IR AGNs that are
undetected in the shallow X-ray survey still have low-level X-ray
emission.

We also note that most X-ray AGNs are {\em not} selected by the IRAC
color-color criteria, but in these objects the IRAC emission from the
host galaxy may mask a relatively low-luminosity, red power-law-type
SED from the AGN.  Figure~\ref{fstern} shows that in the IRAC
color-color diagram, many X-ray AGNs lie just below the \citet{ster05}
selection region, which could be evidence for a low-level AGN
contribution \citep{gorj08}. It is possible that these objects contain
relatively IR-faint AGNs that redden the colors of a quiescent galaxy
and move it toward (but not all the way into) the AGN color-color
selection region.  However, IRAC color-color space below the
\citet{ster05} region is also populated by quiescent galaxies with
``green'' colors similar to many X-ray AGN hosts (Section~\ref{agnhost}),
so the colors alone do not provide evidence for a weak AGN.  In the
future, a detailed study of the multiwavelength SEDs of these sources
could put limits on the nuclear contribution to the IRAC flux.

\begin{deluxetable*}{lcccccc}
\tabletypesize{\scriptsize}
\tablecaption{AGN samples \label{tabsample}}
\tablehead{
\colhead{} &
\colhead{} &
\colhead{} &
\multicolumn{4}{c}{Also selected in} \\
\colhead{Sample} &
\colhead{Total\tnm{a}} &
\colhead{Only\tnm{b}} &
\colhead{Radio\tnm{c}} &
\colhead{X-ray\tnm{c}} &
\colhead{IRAC\tnm{c}} &
\colhead{All\tnm{d}}
}
\startdata
\multicolumn{7}{c}{All AGNs} \\
Radio &122 {\em (122)} &103 {\em (103)} &\nodata &\W\W6 {\em \W\W(6)} &\W\W6 {\em \W\W(6)} &\W\W7 \\
X-ray &362 {\em (296)} &238 {\em (192)} &\W\W6 {\em \W\W(6)} &\nodata &111 {\em \W(91)} &\W\W7 \\
IRAC  &238 {\em (199)} &114 {\em \W(95)} &\W\W6 {\em \W\W(6)} &111 {\em \W(91)} &\nodata &\W\W7 \\
\multicolumn{7}{c}{Optically extended AGNs} \\
Radio &113 {\em (113)} &\W98 {\em \W(98)} &\nodata &\W\W4 {\em \W\W(4)} &\W\W5 {\em \W\W(5)} &\W\W6 \\
X-ray &241 {\em (200)} &187 {\em (153)} &\W\W4 {\em \W\W(4)} &\nodata &\W44 {\em \W(37)} &\W\W6 \\
IRAC  &133 {\em (114)} &\W78 {\em \W(66)} &\W\W5 {\em \W\W(5)} &\W44 {\em \W(37)} &\nodata &\W\W6 \\
\multicolumn{7}{c}{Optically unresolved AGNs} \\
Radio &\W\W9 {\em \W\W(9)} &\W\W5 {\em \W\W(5)} &\nodata &\W\W2 {\em \W\W(2)} &\W\W1 {\em \W\W(1)} &\W\W1 \\
X-ray &121 {\em \W(96)} &\W51 {\em \W(39)} &\W\W2 {\em \W\W(2)} &\nodata &\W67 {\em \W(54)} &\W\W1 \\
IRAC  &105 {\em \W(85)} &\W36 {\em \W(29)} &\W\W1 {\em \W\W(1)} &\W67 {\em \W(54)} &\nodata &\W\W1
\enddata

\tnt{a}{Total number of AGNs with AGES redshifts $0.25<z<0.8$,  selected in waveband for that row, inside the main AGES field, away from bright stars.}
\tnt{b}{The number of AGNs selected in the waveband for that row {\em only}.}
\tnt{c}{The number of AGNs selected in the waveband for that row, and also in the band for that column  {\em only}.}
\tnt{d}{The number of AGNs selected in all three wavebands.}

\tablecomments{Values in parentheses are for sources inside the region covered by radio, X-ray, and IRAC observations (Fig.~\ref{fbound}).}

\end{deluxetable*}

\section{AGN host galaxies}
\label{agnhost}

Links between AGN activity and the evolution of galaxies are reflected
in the characteristics of the galaxies that host different classes of
AGNs.  As discussed in Section~\ref{kcorr}, a powerful diagnostic for
understanding galaxy properties is their positions in optical
color-magnitude space.  In this section, we examine the optical colors
and luminosities for AGNs with extended optical counterparts
(corrected for nuclear contamination as described in
Appendix~\ref{appendix}), and compare to the distribution for the full
galaxy sample.

\begin{figure*}
\epsscale{1.15}
\plottwo{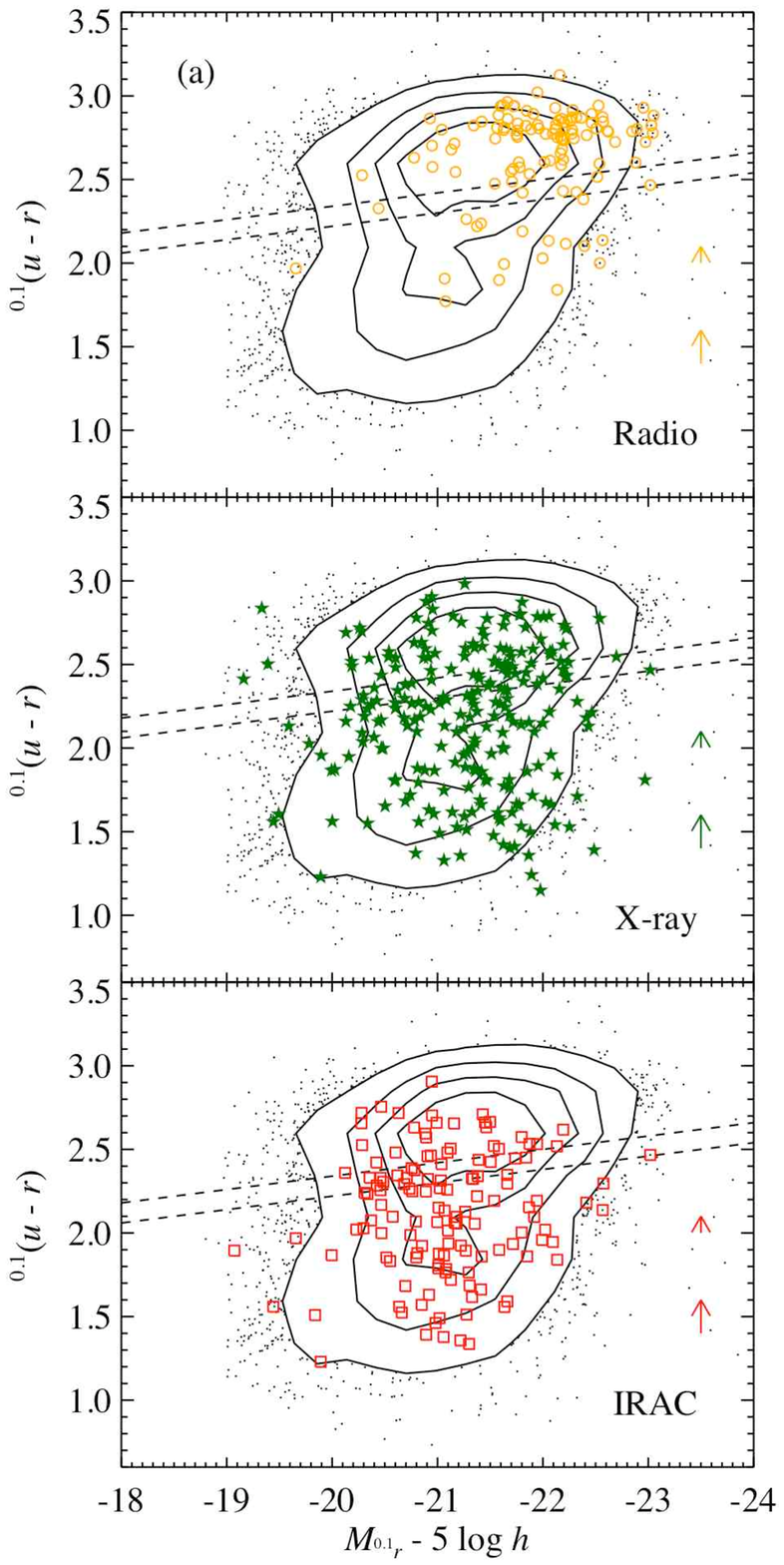}{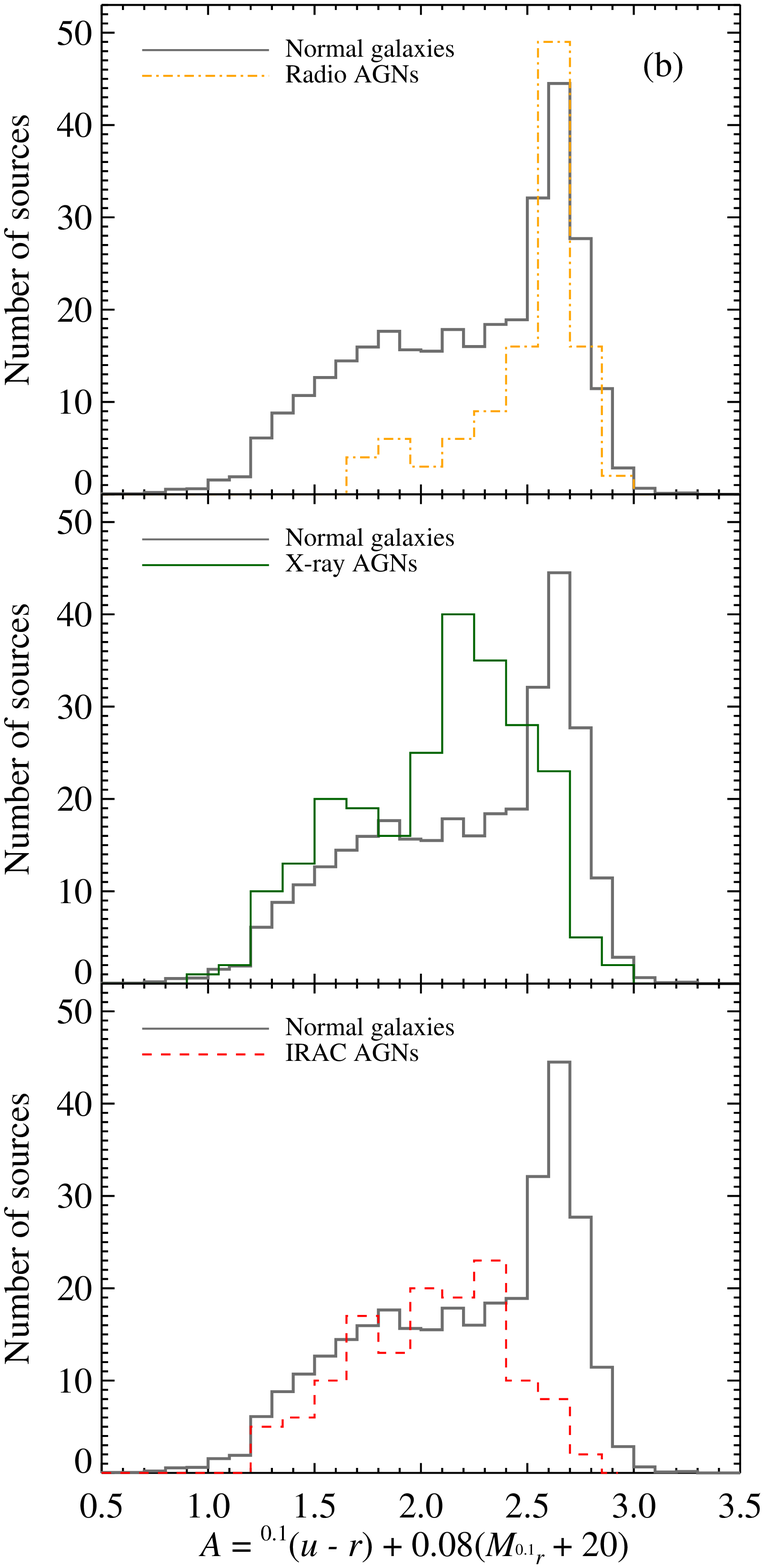}
\caption{({\em a}) Optical colors and absolute magnitudes of AGNs with
extended optical counterparts.  Contours, black points, and
dashed lines are as in Figure~\ref{fcol}.  Orange circles, green stars,
and red squares show radio, X-ray, and IR AGNs, respectively.  The arrows
show the typical correction for nuclear contamination for different
host galaxy colors, which range from 0 to 0.3 mag (Appendix
\ref{appendix}). ({\em b}) Distribution in $A$ for AGNs selected in
the three wavebands, compared to all AGES main sample galaxies at
$0.25<z<0.8$ (scaled by $1/25$, thick gray line).  The radio AGN
color distribution peaks along the red sequence, while X-ray AGNs are
found preferentially in the ``green valley'' between the red sequence
and the blue cloud.  The distribution of IR AGNs is similar to that of
X-ray AGNs, although they are typically found in somewhat less luminous galaxies with very few on
  the red sequence, and show a less pronounced peak in the ``green
valley''.
\label{fagncol}} \vskip0.5cm
\end{figure*}

The color-magnitude diagrams for AGN hosts are shown in
Figure~\ref{fagncol}{\em a}. AGNs of all types have optical colors and
luminosities that lie within, or close to, the parameter space
occupied by normal galaxies, however their distributions in
color-magnitude space are markedly different from the total galaxy
population, and vary significantly for AGNs selected in different
wavebands.

Radio AGNs are concentrated on the luminous end of the red sequence,
indicating that they reside in massive early-type galaxies, with only
a few sources detected in the blue cloud (Figure~\ref{fagncol}, {\em
top panels}).  The mean and dispersion in \urs\ and \mrone\ are
$(\average{C},  \sigma_C)=(2.6, 0.2)$ and $(\average{M},
\sigma_M)=(-22.0, 0.6)$, respectively.  In contrast, X-ray AGNs are
found throughout the galaxy color-magnitude space, with a
disproportionate number found in the ``green valley''
(Figure~\ref{fagncol}, {\em center panels}).  X-ray AGN hosts have
$(\average{C}, \sigma_C)=(2.2, 0.4)$ and $(\average{M},
\sigma_M)=(-21.3,0.7)$.  The fraction of AGES main sample galaxies
containing X-ray AGNs is $4.6\%\pm0.7\%$ in the ``green valley''
(having $A$ within $\pm 0.2$ mag of the division between red and blue
galaxies).  In contrast, the X-ray AGN fraction is $2.1\%\pm0.2\%$ for
galaxies with redder or bluer colors.  As we discuss below, this
enhanced X-ray AGN activity in green galaxies may reflect the
connection between these objects and SMBH accretion.

IR-selected AGNs have a color distribution similar to that of X-ray
AGNs, although they have a less pronounced excess in the ``green
valley'', and have very few hosts on the red sequence
(Figure~\ref{fagncol}, {\em bottom panels}), with $(\average{C},
\sigma_C)=(2.1,0.4)$.  The hosts of IR AGNs have smaller average
optical luminosities than those X-ray AGNs, with $(\average{M},
\sigma_M)=(-21.1,0.6)$.  The sources that are selected as both X-ray
and IRAC AGNs are found in a similar region of color-magnitude space
to the full X-ray and IR selected samples, with $(\average{C},
\sigma_C)=(2.2,0.4)$ and $(\average{M}, \sigma_M)=(-21.2,0.6)$.

We conclude that IR and radio AGNs occupy almost entirely separate
regions of the color-magnitude space, with radio AGNs in luminous red
sequence galaxies, and IR AGNs in the less luminous blue and ``green'' hosts.
X-ray AGNs are found in galaxies that are slightly redder and more
luminous than the hosts of IR-selected AGNs.

\section{Correlation analysis}
\label{corr}

Another important clue to the evolution of AGNs and their host galaxies
comes from the large-scale environments in which AGNs are found.  We
derive the environments of AGNs and galaxies through a spatial
two-point correlation analysis, which probes the clustering of sources
on different comoving scales.  We first calculate the autocorrelation
of AGES main sample galaxies (and the red and blue samples
separately), to determine the absolute linear bias of AGES
galaxies relative to dark matter.  Using the absolute bias, we then
derive the characteristic masses of the dark matter halos which host
galaxies of different types.

We next derive the cross-correlation of AGNs with AGES galaxies, to
determine the {\em relative} bias between AGNs and galaxies, from
which we derive the absolute bias and characteristic dark matter halo
masses for the AGNs themselves.  AGN-galaxy
cross-correlation has been used since the first studies of quasar
clustering \citep[e.g.,][]{bahc69qso} and has two main advantages, as
discussed in \citet{coil07a}.  First, it does not require a full
understanding of the AGN selection function, which may be complicated,
particularly for AGNs selected in multiple wavebands.  Instead, we
need only know the selection function for the galaxies, which is well
determined (Section~\ref{galsample}).  The second advantage is that
AGN-galaxy cross-correlation has much greater statistical power than
an AGN autocorrelation, owing to the large number of galaxies in the
redshift survey (greater than 6000) compared to the number of AGNs (a few
hundred).

The two-point correlation function $\xi(r)$ is defined as the
probability above Poisson of finding a galaxy in a volume element $dV$
at a physical separation $r$ from another randomly chosen galaxy, such
that
\begin{equation}
dP=n[1+\xi(r)]dV,
\end{equation}
where $n$ is the mean space density of the galaxies in the sample.  To
calculate the autocorrelation of AGES galaxies, we derive $\xi(r)$
using the \citet{land93} estimator:
\begin{equation}
\xi(r)=\frac{1}{{\rm DD}}({\rm DD}-2{\rm DR}+{\rm RR}),
\label{eqnxidef1a}
\end{equation}
where DD, DR, and RR are the number of data-data, data-random,
and random-random galaxy pairs, respectively, at a separation $r$.  We
account for the sparse sampling of fainter galaxies by multiplying
each pair by the product of the statistical weights of the observed (or
random) galaxies.  That is, when correlating galaxy sample 1 against
galaxy sample 2, the weighted number of data-data pairs is
\begin{equation}
{\rm DD} = \displaystyle\sum_{i\in {\rm D_1 D_2}} W^{i}_1 W^{i}_2,
\label{eqnweight}
\end{equation}
where $W^{i}_1$ and $W^{i}_2$ for each pair are the statistical
weights of the galaxies from samples 1 and 2, respectively.  By
including these weights, we ensure that brightest galaxies do not
overly dominate the correlation function.  We correspondingly include
the random galaxy weights (Section~\ref{random}) in calculating DR and
RR. The random galaxy weights are assigned so that their total
weight equals that of the AGES galaxies, so that DD, DR, and RR
may be compared directly.

Following \citet{coil07a}, for the cross-correlation between
AGNs and galaxies we use the simple estimator
\begin{equation}
\xi(r)=\frac{\rm D_1 D_2}{\rm D_1 R_2} -1
\label{eqnxidef2}
\end{equation}
where ${\rm D_1 D_2}$ is the number of AGN-galaxy pairs, and ${\rm D_1 R_2}$ is
the number of AGN-random pairs.  Because the selection function of the AGN samples is not as well defined as that for the AGES main sample galaxies, we do not assign statistical weights to the AGNs.  Therefore we use only the galaxy weights in calculating the weighted number of pairs, such that
\begin{equation}
{\rm D_1 D_2} = \displaystyle\sum_{i\in {\rm D_1 D_2}} W^{i}_2,
\label{eqnweight2}
\end{equation}
and correspondingly for ${\rm D_1 R_2}$.

In the range of separations $1\lesssim r \lesssim10$ \hminus\ Mpc,
$\xi(r)$ for galaxies is roughly observed to be a power law,
$\xi(r)=(r/r_0)^{-\gamma}$, although recent work has shown evidence
for separate terms in the correlation owing to dark matter halos that
host single galaxies and those that host pairs of galaxies
\citep[e.g.,][]{zeha04, zhen07hod,coil08galclust,
brow08halo,zhen09redhod}.  For simplicity, and in light of the
uncertainties in the AGN-galaxy correlations, we will restrict our
fits to power-law models.  Throughout the paper $r_0$ is given in
comoving coordinates.

\subsection{The projected correlation function}
\label{projcorr}
With a redshift survey, we cannot directly measure $\xi(r)$ in
physical space, because peculiar motions of galaxies distort the
line-of-sight distances measured from redshifts.  To account for these
redshift-space distortions, we instead measure the correlation
function $\xi(r_p,\pi)$, where $r_p$ and $\pi$ are the projected
comoving separations between galaxies in the directions perpendicular
and parallel, respectively, to the mean line of sight between the two
galaxies.  

Redshift space distortions only affect the correlation function along the line of sight, so we integrate $\xi(r_p,\pi)$ along the $\pi$ direction
 to obtain the statistic,
\begin{equation}
w_p(r_p)=2\int_{0}^{\pi_{\rm max}}\xi(r_p,\pi)d\pi,
\end{equation}
which is independent of redshift space distortions
\citep[following][]{davi83}. This estimator has been used to measure
correlations in a number of surveys, for example SDSS
\citep{zeha05a, li06agnclust}, 2SLAQ \citep{wake08radio}, DEEP2
\citep{coil07a, coil08galclust}, and GOODS \citep{gill07c}.  If
$\pi_{\rm max}=\infty$, then we average over all line-of-sight
peculiar velocities, and $w_p(r_p)$ can be directly related to
$\xi(r)$ (for a power-law parametrization) by
\begin{equation}
w_p(r_p)=r_p\left (\frac{r_0}{r_p}\right)^\gamma \frac{\Gamma(1/2)\Gamma[(\gamma-1)/2]}{\Gamma(\gamma/2)}.
\label{eqnwp}
\end{equation}

In practice, we truncate the integral at a finite $\pi_{\rm max}$, to
reduce the noise from integrating over weakly correlated objects at
large $\pi$.  Here we use $\pi_{\rm max}=25$ \hminus\ Mpc.  Because of
the finite $\pi_{\rm max}$, in order to recover $r_0$ and $\gamma$
exactly we must account for redshift-space distortions due to peculiar
motions along the line of sight.  This requires an accounting for
coherent infall of galaxies on large scales, as well as the random
velocities of galaxies on small scales (``fingers of God'').

We account for these redshift-space distortions using the method of
\citet{coil07a, coil08galclust}.  Coherent infall depends on the
parameter $\beta\approx \Omega_{\rm m}^{0.6}/b$ \citep{kais87}, where
$\Omega_{\rm m}$ evaluated at the mean redshift of the sample, and $b$
is the absolute bias relative to dark matter.  For $z=0.5$ (the mean
redshift for the AGNs), $\Omega_{\rm m}=0.59$.  We take $b=1.23$, the
absolute bias of AGES galaxies at $0.25<z<0.8$ (see Section~\ref{absbias}
and Section~\ref{results}), so that $\beta=0.59$.  For the pairwise random
velocities of galaxies, we take $\sigma_{12}=500$ km s$^{-1}$, which
is characteristic of galaxies with luminosities typical of the AGES
sample \citep[e.g.,][]{hawk03corr, coil08galclust}.  For reasonable
variations in $\beta$ ($\pm0.1$) or $\sigma_{12}$ ($\pm 400$ km
\pers), the resulting $w_p(r_p)$ changes by at most a few percent.

We then derive $\xi(r_p,\pi)$ in redshift space as a function of $r_0$
and $\gamma$, including coherent infall and random velocities as
described in Section~4.1 of \citet{hawk03corr}.  We numerically
integrate $\xi(r_p,\pi)$ to $\pi_{\rm max}=25$ \hminus\ Mpc to
determine a model $w_p(r_p)$ which we fit to the data.  For a given
$r_0$ and $\gamma$, this $w_p(r_p)$ differs from that given in
Eqn.~\ref{eqnwp} by less than 4\% for $r_p<10$ \hminus\ Mpc.  Accordingly,
the best-fit $r_0$ and $\gamma$ differ from those using
Eqn.~\ref{eqnwp} by less than a few percent.

\subsection{Calculating relative bias}
\label{relbias}

For each subset of galaxies and AGNs, we also calculate the relative
bias $b_{\rm rel}$ compared to the autocorrelation of all AGES main
sample galaxies.   The relative bias for the red galaxy sample is given by
\begin{equation}
b_{\rm rel}^2=\frac{w_p(r_p) {\: \rm [red-red]}}{w_p(r_p)\: {\rm [gal-gal]}},
\end{equation}
and accordingly for blue galaxies.  For AGNs, since we calculate the AGN-galaxy cross-correlation, the relative bias is
\begin{equation}
b_{\rm rel}=\frac{w_p(r_p)\: {\rm [AGN-gal]}}{w_p(r_p)\: {\rm [gal-gal]}}.
\end{equation}
In each case we derive $b_{\rm rel}$ as a function of $r_p$, and
calculate the mean biases averaged over scales of 1--10 \hminus\ Mpc
and 0.3--1 \hminus\ Mpc.

The galaxy autocorrelation varies with redshift, owing to the
evolution of large-scale structure, and because the use of a
flux-limited sample means we selected more luminous galaxies at higher
$z$.  This can affect bias measurements, since the redshift
distribution of the AGN tends to peak at higher $z$ than that for the
galaxies.  To account for this, for each AGN sample we assign weights
to the AGES galaxies and the random galaxy sample, so that their
(weighted) redshift distribution is the same as that of the AGNs in
bins of size $\Delta z=0.1$.  Including these redshift weights, we
rederive the autocorrelation of the AGES main AGES sample,
correspondingly weighting the random galaxy sample, and multiplying
each galaxy-galaxy, galaxy-random, or random-random pair by these
additional weights. We then use this redshift-matched autocorrelation
function when calculating the bias of AGNs relative to galaxies.

Furthermore, for AGNs with extended optical counterparts, it is
interesting to compare the AGN-galaxy cross-correlation with that
measured for galaxies with similar color classification, redshift, and
absolute magnitude.  To create this comparison sample, we assign
additional weights to the AGES main sample galaxies, so that their
(weighted) distribution in color, absolute magnitude, and redshift is
the same as that for the AGN hosts (in bins of 0.25 mag, 0.25 mag, and
0.1 respectively).  We then calculate the cross-correlation $w_p(r_p)$
between this comparison sample and all AGES galaxies (again
multiplying each galaxy-galaxy or galaxy-random pair by these
additional weight) and derive the bias (on scales 1--10 and 0.3--1
\hmpc) for the AGNs relative to this cross-correlation function.

\subsection{Absolute bias and dark matter halo mass}
\label{absbias}
As discussed in Section~\ref{intro}, the mass of the parent dark matter
halo may be a key parameter in determining the evolution of galaxies
and their central black holes.  We therefore use our clustering
results to estimate $M_{\rm halo}$ for our different subsets of
galaxies and AGNs.  The masses of dark matter halos are reflected in
their {\em absolute} bias relative to the dark matter distribution.
To determine absolute bias \citep[following][]{coil08galclust} we
first calculate the two-point autocorrelation of dark matter as a
function of redshift, using the code of \citet{smit03dm}, and assuming
our standard cosmology, with $\sigma_8=0.9$ and the slope of the
initial fluctuation power spectrum, $\Gamma=\Omega_m h=0.21$.  We
derive $w_p(r_p)$ as a function of redshift for the dark matter,
integrating to $\pi=25$ \hminus\ Mpc as for the data.  We assume an
uncertainty in the model dark matter correlation function of 5\% at
each scale.

For each subset of AGNs, we next divide the projected autocorrelation
function for galaxies that are matched in redshift (Section~\ref{relbias}) by
the $w_p(r_p)$ for the dark matter, averaged over the redshift
distribution of the sample.  The mean ratio between these two
correlation functions on scales $1<r_p<10$ \hminus\ Mpc gives $b_{\rm
gal}^2$, where $b_{\rm gal}$ is the absolute linear bias of AGES galaxies
with given redshift distribution.  The absolute bias of the AGN sample
is then $b_{\rm abs}=b_{\rm gal}b_{\rm rel}$, where $b_{\rm rel}$ is
the bias relative to AGES galaxies, determined by the
cross-correlation (Section~\ref{relbias}).

Finally, we use $b_{\rm abs}$ to derive the characteristic mass of the
dark matter halos hosting each subset of galaxies or AGNs.  The halo
masses presented in this paper represent the virial mass in the sense
of a top-hat spherical collapse model \citep[see e.g.,][and references
therein]{peeb93book}.  To obtain halo mass, we first convert $b_{\rm
abs}$ to the quantity $\nu=\delta_c/\sigma(M)$, where $\sigma(M)$ is
the linear theory rms mass fluctuation in spheres of radius
$r=(3M/4\pi \overline{\rho})^{1/3}$ ($\overline{\rho}$ is the
background density) and $\delta_c\approx 1.69$ is the critical
overdensity required for collapse.  We convert $b_{\rm abs}$ to $\nu$
using Eqn.~(8) of \citet{shet01halo}, and use $\nu$ to derive $M_{\rm
halo}$ following Appendix A of \citet{vand02macc}.  If we use a
different relation between $b_{\rm abs}$ and $\nu$ \citep{tink05}, we
obtain estimates for $M_{\rm halo}$ that are similar, although
slightly larger by 0.1--0.2 dex.

We  note that to estimate $M_{\rm halo}$, we have averaged the
absolute bias on scales 1--10 \hminus\ Mpc.  On these scales, the dark
matter and galaxy correlation functions can have somewhat different
shapes, so that the relative bias is a function of scale
\citep[e.g.,][]{cres09bias}.  To verify our method for estimating
$M_{\rm halo}$ using the 1--10 \hmpc\ bias, we compared the $b_{\rm
abs}$--$M_{\rm halo}$ relation of \citet{shet01halo} to a relation
derived from the dark matter simulations of \citet{padm08qsored}.
Using the simulation results for $z\approx0.5$, we divided the
$w_p(r_p)$ for halos of different masses (evaluated between 1 and 10
\hmpc) to the $w_p(r_p)$ from the dark matter model of
\citet{smit03dm} [calculated using the same cosmology assumed by
\citet{padm08qsored}].  For a given 1--10 \hmpc\ bias, the
corresponding $M_{\rm halo}$ from the simulations matches the
prediction of \citet{shet01halo} to better than $0.08$ dex.  This
confirms that measuring bias from 1--10 \hmpc\ scales allows
sufficiently accurate estimates of $M_{\rm halo}$.  Finally, we
check that our estimates of $M_{\rm halo}$ are relatively
insensitive to our choice of $\sigma_8$.  Decreasing $\sigma_8$
results in weaker dark matter clustering and so larger $b_{\rm abs}$
for galaxies with a given observed $w_p(r_p)$.  However, decreasing
$\sigma_8$ also reduces the $M_{\rm halo}$ corresponding to a given
$b_{\rm abs}$, so there is little change in our estimates of $M_{\rm
halo}$.  If we change $\sigma_8$ from 0.9 to 0.8, our $M_{\rm halo}$
estimates for AGES galaxies and AGNs vary by less than $\pm0.1$ dex.

In addition to top-hat virial mass, another commonly used mass
estimator is $M_{\rm 200}$, which is equal to the mass contained
within the radius where the overdensity is 200 times the critical
density.  For our chosen cosmology, $M_{\rm 200}$ is slightly smaller
than the top-hat virial mass, by $\approx$20\% \citep{whit01mhalo}.

\subsection{Uncertainties}

\label{uncertainties}

We determine uncertainties in $w_p(r_p)$ directly from the data, in
two ways.  For the galaxy autocorrelations, errors are derived from
jackknife resampling, which generally approximates the variance in
$w_p(r_p)$ derived from simulated galaxy catalogs
\citep[e.g.,][]{zeha05a, coil08galclust}.  We divide the sample into five
separate regions (large enough to sample all the appropriate scales)
and recalculate $w_p(r_p)$ excluding each region in turn.  The
uncertainty in $w_p(r_p)$ is taken to be the variance between the
jackknife samples.  However, when calculating $w_p(r_p)$ for the AGN
samples, jackknife resampling involves only a small number of sources,
so the uncertainties on different scales vary significantly.
Therefore, for the optically extended AGNs we instead create a set of
100 random comparison galaxy samples matched in color, absolute
magnitude, and redshift (as described in Section~\ref{relbias}), but
containing only one galaxy corresponding to each AGN.  We calculate
$w_p(r_p)$ for each of the 100 samples, and take the variance to be
the uncertainty in $w_p(r_p)$.  These uncertainties are typically
$\sim$2 times the corresponding jackknife errors.  We similarly
calculate the variance in the bias (relative to all galaxies, and to
the matched sample) and the best-fit power law parameters, to account
for covariance in different bins of $r_p$.  For optically unresolved
AGNs, we cannot create a matched sample of galaxies, so we simply
repeat the calculation for the optically extended sources of each type
(radio, X-ray, and IR), but vary the sample size to match the
corresponding number of unresolved AGNs and take as the uncertainties
the variance in $w_p$ and bias between the 100 samples.

We perform power-law $\chi^2$ fits to $w_p$ (using the uncertainties
derived from the above resampling) over the range $0.3<r_p<10$
\hminus\ Mpc, from which we obtain the best-fit $r_0$ and $\gamma$.
We estimate the uncertainties in $r_0$ and $\gamma$ by the variance in
the best-fit parameters between the jackknife samples (for the galaxy
autocorrelation) or the 100 comparison samples (for the AGN-galaxy
cross-correlation), to account for covariance between bins of $r_p$.
We note that our jackknife resampling appears to underestimate the
uncertainties in the fit parameters by about 3\% of the values of
$r_0$ and $\gamma$, compared to variance in simulated galaxy catalogs
with comparable source densities and number of objects
\citep{coil08galclust}.  We therefore include a systematic uncertainty
of $\pm3\%$ on the best-fit parameters and on the absolute bias. This
gives total uncertainties for the galaxy autocorrelations that are
similar to those derived from simulations, but has a negligible effect
on the (much larger) errors for the cross-correlation of AGNs and
galaxies.

\section{Correlation results}
\label{results}

\begin{deluxetable*}{lccccccccccc}
\tabletypesize{\scriptsize}
\tablewidth{7in}
\tablecaption{Correlation results\tnm{a} \label{tabcorr}}
\tablehead{
\colhead{} &
\colhead{} &
\colhead{} &
\colhead{} &
\colhead{} &
\colhead{} &
\multicolumn{4}{c}{Relative bias} &
\colhead{} &
\colhead{Characteristic} \\
\colhead{} &
\colhead{} &
\colhead{} &
\colhead{} &
\colhead{} &
\colhead{} &
\multicolumn{2}{c}{vs.\ all galaxies\tnm{c}} &
\multicolumn{2}{c}{vs.\ matched sample\tnm{d}} &
\colhead{Absolute bias} &
\colhead{halo mass\tnm{e}}  \\
\colhead{Subset} &
\colhead{$N_{\rm src}$} &
\colhead{$\average{z}$\tnm{b}} &
\colhead{$r_0$} &
\colhead{$\gamma$} &
\colhead{$\chi^2_\nu$} &
\colhead{$1<r_{\rm p}<10$} &
\colhead{$0.3<r_{\rm p}<1$} &
\colhead{$1<r_{\rm p}<10$} &
\colhead{$0.3<r_{\rm p}<1$} &
\colhead{$1<r_{\rm p}<10$} &
\colhead{($\log{h^{-1} M_{\sun}}$)}
}
\startdata
\multicolumn{12}{c}{\em Galaxies} \\
All & 6262 & 0.39 & $ 4.6\pm 0.2$ & $ 1.8\pm 0.1$ &  1.9 &  1 & 1 &\nodata & \nodata & $1.19\pm0.04$ & $12.6^{+ 0.1}_{- 0.1}$ \\
Red & 3146 & 0.41 & $ 5.3\pm 0.2$ & $ 2.1\pm 0.1$ &  2.2 & $1.13\pm0.05$ & $1.45\pm0.06$ & \nodata & \nodata & $1.37\pm0.08$ & $13.0^{+ 0.1}_{- 0.1}$ \\
Blue & 3116 & 0.38 & $ 3.8\pm 0.2$ & $ 1.6\pm 0.1$ &  1.8 & $0.89\pm0.08$ & $0.71\pm0.08$ & \nodata & \nodata & $1.04\pm0.09$ & $12.2^{+ 0.3}_{- 0.4}$ \\
\multicolumn{12}{c}{\em All AGNs} \\
Radio & 122 & 0.57 & $ 6.3\pm 0.6$ & $ 1.8\pm 0.2$ &  3.5 & $1.51\pm0.14$ & $1.70\pm0.25$ & \nodata & \nodata & $2.03\pm0.20$ & $13.5^{+ 0.1}_{- 0.2}$ \\
X-ray & 362 & 0.51 & $ 4.7\pm 0.3$ & $ 1.6\pm 0.1$ &  0.8 & $1.08\pm0.12$ & $0.66\pm0.15$ & \nodata & \nodata & $1.40\pm0.16$ & $12.9^{+ 0.2}_{- 0.3}$ \\
IRAC & 238 & 0.48 & $ 3.7\pm 0.4$ & $ 1.5\pm 0.1$ &  0.4 & $0.75\pm0.14$ & $0.46\pm0.18$ & \nodata & \nodata & $0.95\pm0.18$ & $11.7^{+ 0.6}_{- 1.5}$ \\
X-ray and IRAC\tnm{f} & 118 & 0.53 & $ 3.9\pm 0.7$ & $ 1.6\pm 0.3$ &  0.7 & $0.77\pm0.21$ & $0.51\pm0.33$ & \nodata & \nodata & $0.99\pm0.27$ & $11.9^{+ 0.7}_{- 5.9}$ \\
\multicolumn{12}{c}{\em AGNs (galaxies)} \\
Radio & 113 & 0.55 & $ 6.1\pm 0.5$ & $ 1.8\pm 0.2$ &  3.4 & $1.46\pm0.15$ & $1.68\pm0.28$ & $1.03\pm0.11$ & $0.91\pm0.15$ & $1.95\pm0.21$ & $13.5^{+ 0.2}_{- 0.2}$ \\
X-ray & 241 & 0.48 & $ 4.6\pm 0.4$ & $ 1.6\pm 0.1$ &  0.3 & $1.05\pm0.12$ & $0.65\pm0.22$ & $1.01\pm0.11$ & $0.66\pm0.23$ & $1.34\pm0.15$ & $12.8^{+ 0.2}_{- 0.3}$ \\
IRAC & 133 & 0.42 & $ 3.1\pm 0.7$ & $ 1.2\pm 0.3$ &  0.3 & $0.73\pm0.18$ & $0.26\pm0.33$ & $0.70\pm0.17$ & $0.34\pm0.36$ & $0.90\pm0.20$ & $11.6^{+ 0.8}_{- 5.6}$ \\
\multicolumn{12}{c}{\em AGNs (unresolved)\tnm{g}} \\
X-ray & 121 & 0.57 & $ 5.0\pm 0.7$ & $ 1.7\pm 0.2$ &  1.3 & $1.14\pm0.17$ & $0.69\pm0.25$ & \nodata & \nodata & $1.52\pm0.23$ & $13.0^{+ 0.3}_{- 0.4}$ \\
IRAC & 105 & 0.55 & $ 4.1\pm 0.7$ & $ 2.0\pm 0.2$ &  2.2 & $0.70\pm0.18$ & $0.81\pm0.27$ & \nodata & \nodata & $0.93\pm0.24$ & $11.6^{+ 0.8}_{- 5.6}$
\enddata
\tablecomments{All units for $r_{\rm p}$ and $r_0$ are in  \hminus\ Mpc (comoving).}

\tnt{a}{Results in the top three rows are for the autocorrelation of
all AGES main sample galaxies, red AGES galaxies, and blue AGES
galaxies, derived using the \citet{land93} estimator
(Eqn.~\ref{eqnxidef1a}).  Values in the other rows are for the {\em
cross}-correlation of the given AGN sample with all AGES main sample
galaxies, derived using the estimator given by
Eqn.~(\ref{eqnxidef2}).}

\tnt{b}{Median redshift for the sources in the sample.} 

\tnt{c}{Bias
relative to  AGES main sample galaxies with matched redshift
distributions (\S~\ref{relbias}), averaged over the scales shown.}

\tnt{d}{Bias relative to AGES main sample galaxies with matched
distributions in color, absolute magnitude, and redshift (\S~\ref{relbias}),
averaged over the scales shown.}

\tnt{e}{Virial mass in the sense of a top-hat collapse model, assuming
a cosmology with $\Omega_m=0.3$, $\Omega_\Lambda=0.7$, and
$\sigma_8=0.9$.}

\tnt{f}{AGNs selected both in the X-rays and IR.}

\tnt{g}{Optically unresolved radio AGNs are not included in the clustering analysis, owing to the small sample size and corresponding large uncertainties.}

\end{deluxetable*}

In this section we discuss the results of the correlation analysis and
the characteristic dark matter halo masses for galaxies and AGNs.

\subsection{Galaxy autocorrelation}
\label{galcorr}

\begin{figure}
\epsscale{1.1}
\plotone{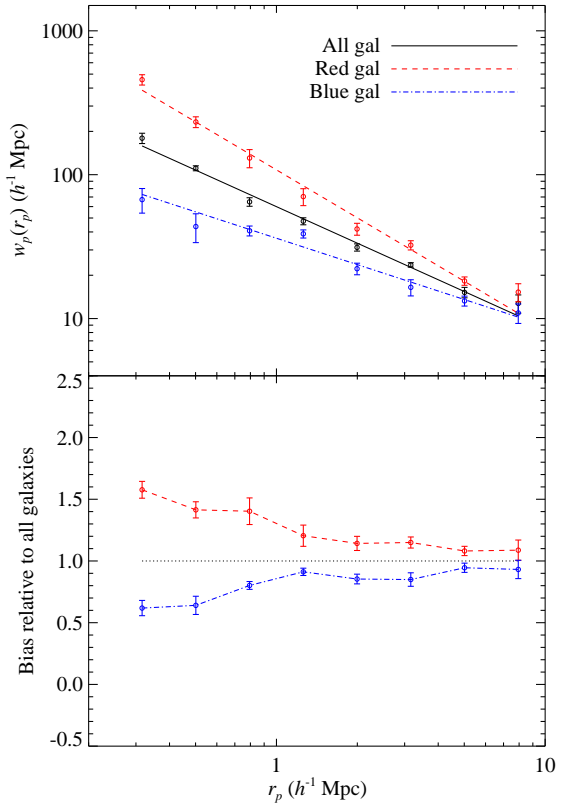}
\caption{Top panel shows galaxy autocorrelations in the AGES
galaxy sample.  The points show the projected autocorrelation function in
the redshift interval $0.25<z<0.8$ for all galaxies in the main AGES
sample ({\em black}), red AGES galaxies ({\em red}), and blue AGES
galaxies ({\em blue}).  Uncertainties are taken from jackknife
resampling, and lines show power-law fits to the data.  The bottom
panel shows bias relative to the main galaxy sample, defined as the
square root of the ratio between $w_p(r_p)$ for the autocorrelation of
red (or blue) galaxies and $w_p(r_p)$ for the autocorrelation of all
AGES galaxies.  Real-space fit parameters and average bias values are
given in Table~\ref{tabcorr}.
\label{fgalredblue}}
\vskip0.5cm
\end{figure}

\begin{figure}
\epsscale{1.1}
\plotone{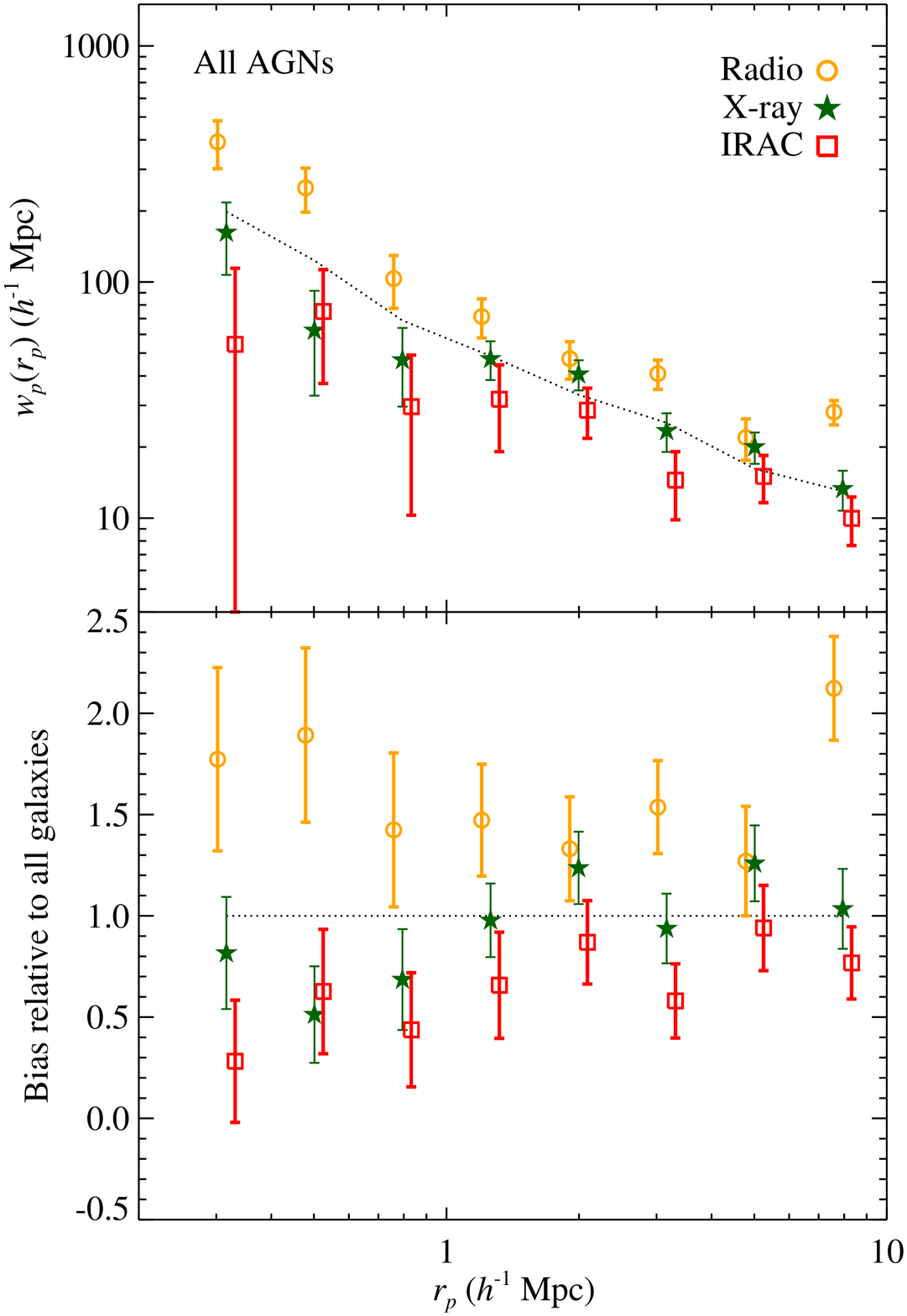}
\caption{Projected cross-correlations between AGNs and all AGES
galaxies. The top panel shows the projected cross-correlation
 between radio AGNs (orange circles), X-ray AGNs (green
stars), and IR AGNs (red squares), and all AGES galaxies in the
redshift interval $0.25<z<0.8$.  For comparison, the dotted lines show
galaxy autocorrelations for the main AGES galaxy samples, weighted to
have similar weighted redshift distributions to the combined sample of
AGNs. The bottom panel shows bias relative to AGES galaxies, defined
as the ratio between $w_p(r_p)$ for the AGN-galaxy cross-correlation
and $w_p(r_p)$ for the galaxy autocorrelation.
\label{frdhixstern}}
\vskip0.5cm
\end{figure}

As a first exercise, we measure the autocorrelation of the main AGES
galaxy sample, as well as for the red and blue galaxy samples
separately.   The resulting $w_p(r_p)$ for the galaxy autocorrelations are shown
in Figure~\ref{fgalredblue}, and the parameters of power-law fits to the
data (including the reduced $\chi^2$) are given in
Table~\ref{tabcorr}.  The range of absolute magnitudes for the red and
blue galaxy samples are $-19\lesssim M_{^{0.1}r} - 5\log{h}\lesssim-23$ and
$-19\lesssim M_{^{0.1}r} - 5\log{h}\lesssim-22.5$, respectively.  The
autocorrelation of the main galaxy sample is best fit by
$r_0=4.6\pm0.2$ \hmpc\ and $\gamma=1.8\pm0.1$. This is broadly
consistent with previous results [see, e.g., Table 4 in
\citet{brow03redcl}, Table 1 of \citet{coil08galclust},
\citet{brow08halo}, and references therein]. The corresponding
absolute bias indicates that typical AGES galaxies reside in dark
matter halos with $M_{\rm halo}\simeq 10^{13}$ \hmsun.

Dividing the galaxy sample into red and blue subsets, we find that red
galaxies show a significantly stronger autocorrelation and a steeper
slope than the blue galaxies, reflecting the well-established
relationship between color and clustering.  The difference in
clustering between red and blue galaxies is not simply a luminosity
effect.  If we select random samples of red and blue galaxies that are
uniformly distributed in bins of 0.25 mag between
$-19>M_{^{0.1}r}>-21.25$, the relative bias between red and blue
galaxies changes by only 2\%.  We note that the $w_p(r_p)$ is not
exactly a power law, especially for the red galaxy sample.  The
apparent upturn of $w_p(r_p)$ on scales less than 1 \hminus\ Mpc is
indicative of an additional ``one-halo'' term in the correlation
function, corresponding to dark matter halos that contain multiple
galaxies
\citep[e.g.,][]{zeha04,zhen07hod,coil08galclust,brow08halo,zhen09redhod}.

As a check, we also perform the galaxy autocorrelation using an
alternative clustering measure, the $\omega$ statistic of
\citet{padm07w}.  This statistic has the advantage of being less
sensitive than $w_p(r_p)$ to large-scale structures and the effects of truncating the
line-of-sight integral at $\pi_{\rm max}$.  In addition, it is
possible to evaluate $\omega(R)$ by converting the integrals to
Riemann sums, and so it does not require binning the data.  Using the
$(2,2)$ filter preferred in \citet{padm07w}, and evaluating their
Eqn.~(29), we derive $\omega(R)$ and calculate jackknife uncertainties
as described above.  We perform a power-law fit to $\omega$ and derive
$r_0$ and $\gamma$ using Eqn.~(24) of \citet{padm07w}.  The best-fit
values are $r_0=4.8$ and $\gamma=2.0$, within 5\% and 7\%,
respectively ($\sim$$1 \sigma$) of the values derived from the
$w_p(r_p)$ fit.  We note, that since $\omega(R)$ involves an integral
over $r_p$ as well as $\pi$, the bins of $\omega(R)$ are not
independent.  We therefore use $w_p(r_p)$ to estimate the AGN-galaxy
cross-correlation and relative bias as a function of scale.

\subsection{AGN-galaxy cross-correlation}
\label{agncorr}

\begin{figure}
\epsscale{1.1}
\plotone{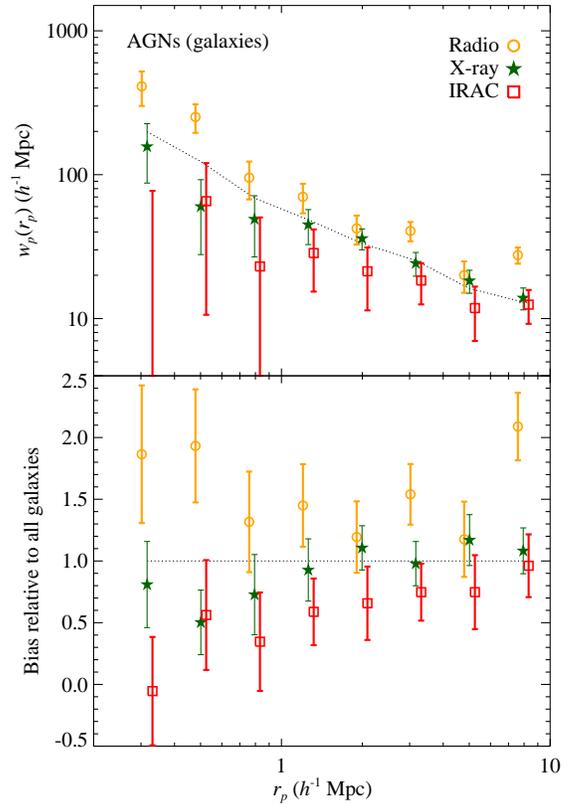}
\caption{Same as Figure~\ref{frdhixstern}, for only those sources with extended optical counterparts (Section~\ref{agnhost}).   Real-space fit parameters and average bias values are given in Table~\ref{tabcorr}.
\label{frdhixsterngal}}
\vskip0.5cm
\end{figure}

\begin{figure}
\epsscale{1.1}
\plotone{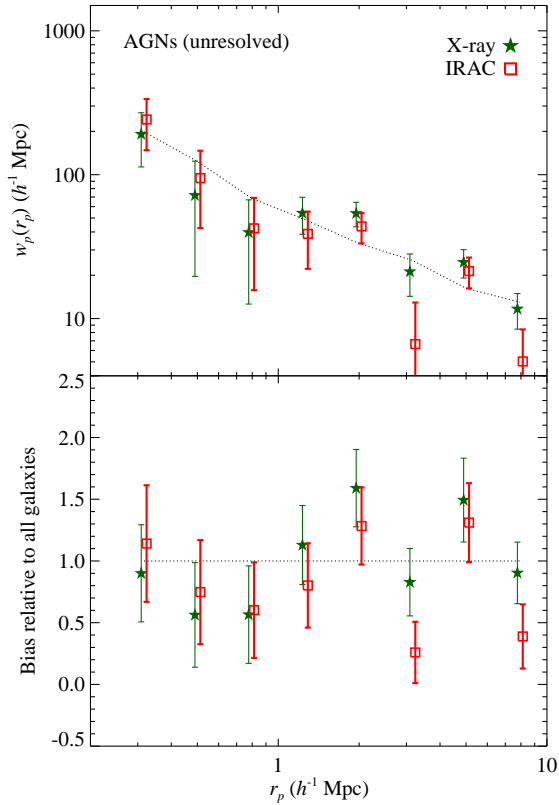}
\caption{Same as Figure~\ref{frdhixstern}, for only those sources with
unresolved optical counterparts (Section~\ref{agnhost}).  There are only
nine radio-selected, optically unresolved AGNs, so the corresponding
$w_p(r_p)$ values have large uncertainties and are not included here.
Real-space fit parameters and average bias values are given in
Table~\ref{tabcorr}.
\label{fxsternqso}}
\vskip0.5cm
\end{figure}

We next measure the cross-correlation of AGNs with AGES galaxies,
which gives us the bias of AGNs relative to galaxies, from which we derive
the absolute bias and $M_{\rm halo}$ for the AGNs.  We first perform
the AGN-galaxy cross-correlations for all AGNs (including those with
extended or unresolved counterparts).  The $w_p(r_p)$ values are shown
in Figure~\ref{frdhixstern}, and the parameters of power-law fits to the
correlation functions are given in Table~\ref{tabcorr}.  We further
divide the AGN samples into those with extended or unresolved optical
counterparts (Figs.~\ref{frdhixsterngal} and \ref{fxsternqso}).  For
optically extended sources, we compare the clustering of AGN hosts to
a control sample of normal galaxies that are appropriately weighted to
have similar colors, absolute magnitudes, and redshifts as the AGN
hosts (Section~\ref{relbias}).  Here we discuss the results for the
different classes of AGN, which show significant differences in
clustering.

\subsubsection{Radio AGNs}
\label{rdcorr}

The radio AGNs are most strongly clustered, with significant bias
relative to normal galaxies indicating that they reside in massive
dark matter halos ($M_{\rm halo}\sim3\times10^{13}$ \hmsun)
characteristic of large galaxy groups or small clusters.  There is an
increase in bias on scales less than 0.5 \hmpc, similar to that seen for red
AGES galaxies.  This suggests that on small scales, the
cross-correlation of radio AGNs with galaxies is dominated by pairs of
objects in the same dark matter halo, which is to be expected if these
sources reside in massive systems
\citep[e.g.,][]{zeha04,zhen07hod,coil08galclust,brow08halo,zhen09redhod}.

Interestingly, compared to a control sample of AGES galaxies matched
in color, \mrone, and $z$, the radio AGNs show no significant
difference in clustering on all scales from 0.3 to 10 \hmpc.  This is
apparently in contrast to recent studies that found that radio AGNs
are more strongly clustered (and have correspondingly larger halo
masses) compared to matched control samples of galaxies.  For example,
\citet{mand09agnclust} selected radio AGNs from SDSS and found that
their dark matter halos are roughly twice as massive as those for
galaxies matched in redshift, stellar mass, and stellar velocity
dispersion.  \citet{wake08radio} found a similar effect for more
luminous radio AGNs at $0.4<z<0.7$ taken from the 2SLAQ Luminous Red
Galaxy survey, with the radio sample inhabiting $\sim2$ times larger
halos than the control galaxies matched in luminosity and color.  

We note that the differences between these studies and our own results
are only marginally significant; if the bias of the \bootes\ radio
AGNs were increased by 1.5$\sigma$, this would correspond to a factor
of 2 larger halo mass relative to the control galaxies, and would
match the results of \citeauthor{mand09agnclust}\ and
\citeauthor{wake08radio} Nonetheless, it may be worth noting that
\citeauthor{mand09agnclust}\ studied a subsample of high stellar-mass
objects that corresponds closely to the range of $P_{1.4\; \rm{GHz}}$
and host galaxy mass probed in \bootes\ \citep[assuming mass-to-light
ratios from][]{kauf03stel}, and found that for these objects, the
difference in clustering between the radio AGNs and the control sample
is small.

These results may hint at an interesting point regarding the fueling
of radio AGNs.  In particular, the importance of environment in
triggering radio activity may depend on the radio luminosity relative
to the galaxy's stellar mass.  Compared to the \bootes\ radio AGNs, the
full \citeauthor{mand09agnclust}\ sample  extends to lower stellar masses,
while the \citeauthor{wake08radio}\ AGNs have similar host galaxy
absolute magnitudes but higher radio luminosities \citep[their median
$P_{1.4\; \rm{GHz}}$ is $\sim$3 times higher than in our
sample;][]{sadl07radio}.  Therefore, both samples probe sources with a
higher typical ratio of radio power to stellar mass than the \bootes\
sample (or in the \citeauthor{mand09agnclust}\ high-mass sample), for
which we see little clustering difference between radio AGNs and
control galaxies.

This suggests that for galaxies with low radio luminosities (relative to their
stellar masses), environment may have little effect on the fueling of
radio activity.  This is certainly true in the limit of very
low luminosity radio AGNs, which are found in virtually every massive
galaxy \citep{sadl89radio}.  However, the fueling of AGNs with higher
radio power relative to galaxy mass, which are less common than the
low-power sources \citep{best05}, may be more strongly dependent on
the mass of the host dark matter halo.  In the future, a larger study
that examines radio AGN clustering over a wide range of radio
luminosities and environments will allow us to better understand how
radio activity is triggered in halos of different masses.

\subsubsection{X-ray AGNs}
\label{xcorr}

For the X-ray AGNs, the clustering is consistent with that of typical
AGES galaxies on scales of 1--10 \hmpc, indicating that they reside in
halos with $M_{\rm halo}\sim 10^{13}$ \hmsun, characteristic of poor
to moderate-sized galaxy groups.  There is little difference in the
clustering between the optically unresolved and and optically extended
X-ray AGN samples.  This suggests that obscured or optically faint AGNs
show similar clustering to their optically bright counterparts.  This
is analogous to the well-established results for optical quasars, for
which clustering is only weakly (if at all) dependent on optical
luminosity \citep[e.g.,][]{adel05qsoclust, porc06qsoclust,
myer07clust1, daan08clust}, although unlike in optical quasar samples,
some of our X-ray AGNs are likely optically faint because of obscuration,
rather than low intrinsic luminosity. Relative to a matched sample of
normal galaxies, the optically extended X-ray AGNs show no significant
bias.

However, on scales less than 1 \hmpc, the clustering of X-ray AGNs differs
significantly from that of normal galaxies.  The slope of
the AGN-galaxy cross-correlation function for X-ray AGNs is
significantly flatter than the autocorrelation function for normal
AGES galaxies, such that the X-ray AGNs are significantly antibiased
on small scales (0.3--1 \hmpc).  This antibias on small scales is
found for all X-ray AGNs relative to all AGES galaxies, and also for
optically extended X-ray AGNs, relative to the control sample
of normal galaxies with matched optical properties.

A similar result was recently obtained by \citet{li06agnclust} for
optically selected narrow-line AGN in SDSS.  Specifically,
\citet{li06agnclust} found that on scales greater than 1 \hmpc, the
cross-correlation between optical AGNs and galaxies is similar to a
control sample of normal galaxies, but that the AGN-galaxy clustering
is weaker on scales less than 1 \hmpc.  Comparing to mock galaxy catalogs
taken from cosmological simulations, \citet{li06agnclust} showed that
the small-scale antibias can  naturally be explained if the AGNs are
preferentially located in central galaxies within their dark matter
halos, which would serve to suppress the ``one-halo'' term
corresponding to strong clustering on small scales.  While statistical
uncertainties in the clustering analysis prevent us from performing a
similar sophisticated comparison here, the consistent antibias on
small scales for the X-ray AGNs suggests that they may also
preferentially reside in central galaxies.

\subsubsection{IR AGNs}
\label{ircorr}

For the IR-selected AGNs, the cross-correlation with AGES galaxies is
weaker and has a flatter slope than for radio or X-ray AGNs. The
estimated dark matter halo masses are $M_{\rm halo}\lesssim 10^{12}$
\hmsun.  These halo mass estimates are uncertain because at these
redshifts and halo masses, $b_{\rm abs}$ is a weak function of $M_{\rm
halo}$.  However, we can conclude that IR AGNs typically reside in
halos that are significantly smaller than those that host radio or
X-ray AGNs.  We also note that the reduced $\chi^2$ for the fit to
$w_p(r_p)$ is quite low (0.4) for the IR AGNs, possibly because of
covariance between the $w_p(r_p)$ on different scales that is not
reflected in the errors on the individual points.  The uncertainties
in the fit parameters and the bias are derived directly from
resampling, and so account for any covariance between scales.

In contrast to the radio or X-ray AGNs, optically extended IR AGNs are
significantly antibiased relative to normal galaxies with similar
optical properties.  This suggests that the process that triggers
IR-bright AGN activity depends not only on host galaxy properties but
also on local environment, with AGNs preferring underdense regions.
The optically extended IR AGNs are less biased (more strongly
antibiased) on small scales (0.3--1 \hmpc) compared to larger scales
(1--10 \hmpc), reflected in the flat slope of the cross-correlation
function.  Keeping in mind that the relative bias on the two scales
differ by only $1.5\sigma$, this may suggest that IR AGNs preferentially
reside in the central galaxies of their (relatively small) dark matter
halos.

\subsubsection{X-ray and IR AGNs}
\label{xircorr}

Finally, we examine the AGN-galaxy cross-correlation for objects that
are selected as both X-ray and IR AGNs.  Owing to the limited sample
size, we do not divide these sources into optically extended and
unresolved subsets.  The clustering parameters, bias, and
characteristic $M_{\rm halo}$ for these sources are given in
Table~\ref{tabcorr} (to avoid confusion, the cross-correlation
results for this sample are not shown in Figure~\ref{frdhixstern}). AGNs that
are selected in both the IR and X-rays show essentially identical
clustering to the full sample of IR-selected AGNs, with significant
antibias relative to AGES galaxies.  This suggests that the
IR-selected AGNs are weakly clustered, regardless of whether they are
selected in X-rays, and are found in smaller dark matter halos than
typical X-ray AGNs (most of which are not selected with IRAC).  We
discuss the implications of the clustering results for all the AGN
samples in Section~\ref{discussion}.

\section{X-ray spectra and Eddington ratios}
\label{xedd}

In this section, we probe the properties of the accretion process in
the different classes of AGNs, by measuring average X-ray spectra and
Eddington ratios.  There is growing evidence that black hole accretion
can occur in different states, with a key parameter being the
Eddington ratio \citep[e.g.,][]{chur05smbh, merl08agnsynth,
kauf09modes}, and AGNs in these different modes may play distinct
roles in galaxy evolution.  In addition, many AGNs are obscured by gas
and dust; according to ``unified models'', this obscuration originates
in a roughly toroidal structure that is intrinsic to the accretion
flow \citep[e.g.,][]{anto93, urry95}.  We first compute average X-ray
spectra, which can be used to detect the absorption by gas or can
possibly indicate different accretion modes.  We then derive
bolometric AGN luminosities ($L_{\rm bol}$) for the different classes
of AGNs, which we use along with the estimates of black hole masses to
derive Eddington ratios.

\subsection{X-ray stacking analysis}
\label{stack}

\begin{deluxetable*}{lccccc}
\tabletypesize{\scriptsize}
\tablewidth{4.1in}
\tablecaption{X-ray hardness ratios and spectral shapes \label{tabxray}}
\tablehead{
\colhead{} &
\colhead{} &
\multicolumn{2}{c}{All} &
\multicolumn{2}{c}{$<20$ counts} \\
\colhead{Subset} &
\colhead{$N_{\rm src}$\tnm{a}} &
\colhead{HR} &
\colhead{$\Gamma$\tnm{b}} &
\colhead{HR} &
\colhead{$\Gamma$\tnm{b}}}
\startdata
\multicolumn{6}{c}{\em All AGNs} \\
Radio &   103 & $-0.44\pm 0.04$ & $ 1.6\pm 0.1$ & $-0.42\pm 0.10$ & $ 1.5\pm 0.2$ \\
X-ray &   318 & $-0.34\pm 0.01$ & $ 1.4\pm 0.1$ & $-0.24\pm 0.02$ & $ 1.1\pm 0.1$ \\
IR &   198 & $-0.34\pm 0.02$ & $ 1.4\pm 0.1$ & $-0.11\pm 0.04$ & $ 0.9\pm 0.1$ \\
\multicolumn{6}{c}{\em AGNs (galaxies)} \\
Radio &    95 & $-0.37\pm 0.06$ & $ 1.4\pm 0.1$ & $-0.36\pm 0.11$ & $ 1.4\pm 0.2$ \\
X-ray &   212 & $-0.21\pm 0.02$ & $ 1.1\pm 0.1$ & $-0.15\pm 0.03$ & $ 1.0\pm 0.1$ \\
IR &   114 & $-0.10\pm 0.04$ & $ 0.8\pm 0.1$ & $ 0.11\pm 0.06$ & $ 0.4\pm 0.1$ \\
\multicolumn{6}{c}{\em AGNs (unresolved)} \\
Radio &     8 & $-0.51\pm 0.06$ & $ 1.7\pm 0.1$ & $-0.71\pm 0.27$ & $ 2.2\pm 0.6$ \\
X-ray &   106 & $-0.46\pm 0.02$ & $ 1.6\pm 0.1$ & $-0.41\pm 0.03$ & $ 1.5\pm 0.1$ \\
IR &    84 & $-0.44\pm 0.02$ & $ 1.6\pm 0.1$ & $-0.31\pm 0.05$ & $ 1.3\pm 0.1$
\enddata

\tnt{a}{Number of sources in stacked sample; excludes sources in \chandra\ exposures with high background.}
\tnt{b}{Effective power law photon index, derived from HR using Eqn.~\ref{eqngamma}.}

\end{deluxetable*}

To estimate the average X-ray spectra for AGNs, we use an X-ray
stacking analysis. Unabsorbed, Type 1 AGNs typically have power-law
X-ray spectra with $\Gamma\simeq1.8$ \citep{tozz06}, but intervening
neutral gas preferentially absorbs soft X-rays and so hardens the
X-ray spectrum.  Alternatively, a harder X-ray spectrum may indicate a
different accretion mode than is typical for optically bright Seyferts
or quasars.  Galactic black hole binary systems show multiple X-ray
spectral states with varying spectral hardness that are caused by
changes in the accretion mode \citep[for reviews see][]{remi06araa,
nara08adaf}.  The ``low-hard'' state found in black hole binaries is
often interpreted as a radiatively inefficient mode
\citep{esin97nova}, such as an ADAF \citep{nara95}.  Unusually hard
X-ray spectra observed in an AGN (particular for a low accretion rate)
may indicate a similar mode of accretion
\citep[e.g.,][]{hopk09lowlum}, as has been suggested for very
low Eddington AGNs in red galaxies in AGES \citep{brand05}.

Even for the X-ray-detected sources in our AGN sample, we cannot
accurately determine spectral shapes for individual sources since most
have $\leq$10 counts.  However, we can use X-ray stacking to
calculate average spectral properties for subsets of sources.  Specifically, we calculate the hardness ratio, defined as
\begin{equation}
{\rm HR}=\frac{H-S}{H+S},
\end{equation}
where $H$ and $S$ are the observed counts in the hard (2--7 keV) and
soft (0.5--2 keV) bands, respectively.  

The stacking procedure is described in Section~5.1 of \citet{hick07abs}.
Stacked counts are defined as the number of (background-subtracted)
photons detected within $r_{90}$ from the source position, where
$r_{90}$ is an approximation of the 90\% point-spread function (PSF)
energy encircled radius at 1.5 keV, and varies as\footnotemark:
\begin{equation}
r_{90}=1\arcsec+10\arcsec(\theta/10\arcmin)^2.
\label{eqnrad}
\end{equation}
 \footnotetext{\chandra\ Proposer's Observatory Guide
http://cxc.harvard.edu/proposer/POG/} We subtract a small background
of 3.0 count s$^{-1}$ deg$^{-2}$ source$^{-1}$ in the 0.5--2 keV band
and 5.0 count s$^{-1}$ deg$^{-2}$ source$^{-1}$ in the 2--7 keV band
\citep[note that we do not include sources found in 11 pointings with
high flaring background; see Section~5.1 of][]{hick07abs}.  In order to
maximize the number of source counts, we do not limit the stacking to
the central 6\arcmin\ around the pointing center for each observation
\citep[unlike in][]{hick07abs}.  
 Count uncertainties are calculated using the approximation
$\sigma_{X}=\sqrt{X+0.75}+1$, where $X$ is the number of counts in a
given band \citep{gehr86}.  Uncertainties in HR are derived by
propagating these count rate errors.

To estimate an effective power-law photon index $\Gamma$, we use PIMMS
to calculate the HR that would be observed for different values of
$\Gamma$, given the ACIS Cycle 4 on-axis response function.  For the small
Galactic $N_{\rm H}$ toward this field ($\sim$$10^{20}$ \cdens), the
relationship between HR and $\Gamma$ can be approximated by
\begin{equation}
{\rm HR}=0.29-0.46\Gamma.
\label{eqngamma}
\end{equation}
It is possible that a few very bright sources can dominate our
estimates of the average HR, so we also calculate HR for only those
sources with less than 20 counts.  The average HR values and the corresponding
$\Gamma$ estimates are given in Table~\ref{tabxray}.  As a check, we
perform the same analysis for the X-ray-detected sources using the
observed counts in the \citet{kent05} catalog [rather than the full
stacking procedure of \citet{hick07abs}], and obtain essentially
identical results.

As shown in Table~\ref{tabxray}, X-ray and IR AGNs with unresolved
optical counterparts have relatively soft average spectra, with
$\Gamma\simeq1.6$.  The bright optical emission in these sources
suggests that the nucleus is relatively unobscured by dust, and their
X-ray spectra are typical of X-ray AGNs that have little intervening
absorbing gas.  Optical obscuration and X-ray absorption are strongly
correlated for most AGNs \citep{tozz06}, so these sources are
consistent with being a population of unobscured Seyfert galaxies.  In
contrast, the X-ray and IR AGNs with extended optical counterparts
have significantly harder average X-ray spectra, with
$0.8<\Gamma<1.4$.  These hard X-ray spectra may be caused by
absorption by gas, with $N_{\rm H}\sim10^{22}$ \cdens\
\citep{hick07abs}, or alternatively, these sources might have
intrinsically hard X-ray spectra owing to radiatively inefficient
accretion.  In either case, it would be expected that the sources
would have relatively weak optical emission, as observed.  We conclude
that the average X-ray spectra of the X-ray and IR AGNs are consistent
with the properties of their optical counterparts.  

Among the radio-selected AGNs, the six sources also detected in X-rays
have relatively soft spectra with ${\rm HR}\simeq-0.4$, corresponding to a
typical AGN spectrum with little X-ray absorption.  However, the radio
AGNs that are not individually detected in X-rays have a softer
stacked spectrum; the average number of background-subtracted counts
per source are $0.41\pm 0.08$ in the 0.5--2 keV band and $0.10\pm
0.04$ in the 2--7 keV band.  If all this flux comes from the AGN, it
corresponds to a soft spectrum with $\Gamma\simeq1.9$.  The lack of
hard X-ray flux in these sources provides an upper limit to the
contribution from an AGN, assuming $\Gamma=1.8$.  The average AGN flux
for these radio galaxies must be $S_{\rm 0.5-7\; keV}<
2.1\times10^{-15}$ \flux\ (3 $\sigma$ upper limit).  For the range in
redshifts observed for the radio AGNs, this corresponds to an upper
limit on $L_X$ of $(0.5$--$5)\times10^{42}$ \ergs.

\subsection{Eddington ratios}

\label{edd}

We further use the observed luminosities to estimate Eddington ratios,
which can provide insights as to the accretion states of the AGNs.
For a central object with mass $M_{\rm BH}$, the Eddington luminosity
is given by \citep{shap83}
\begin{equation}
L_{\rm Edd}=1.3\times10^{38} (M_{\rm BH}/M_{\sun})\; {\rm erg\; s^{-1}}.
\end{equation}
The Eddington ratio is defined as $\lambda=L_{\rm bol}/L_{\rm Edd}$,
where $L_{\rm bol}$ is the bolometric accretion luminosity of the
system.  Luminous quasars tend to have relatively high $\lambda\gtrsim0.1$ \citep[e.g,][]{mclu04mbh, koll06}, but some Seyfert galaxies,
LINERs, and low-luminosity X-ray and radio AGNs can have
$\lambda\lesssim10^{-2}$, and even as low as $\lambda\sim10^{-8}$ \citep[e.g.,][]{ho02mbh,sori06agnx,siko07radio}.

We calculate a rough estimate of $\lambda$ for the different classes
of AGNs with extended optical counterparts (for which we can estimate
the galaxy stellar mass, and thus $M_{\rm BH}$).  To estimate \lbol\
for X-ray AGNs, we scale from the observed \lx.  For radio AGNs that
are not detected in X-rays, we use the X-ray stacking results
(Section~\ref{stack}) to derive approximate upper limits on the AGN
luminosity.  We take the $3 \sigma$ upper limit on the average nuclear
0.5--7 keV flux, and derive the corresponding upper limit on the
luminosity for each source.

\begin{figure}[t]
\epsscale{1.1}
\plotone{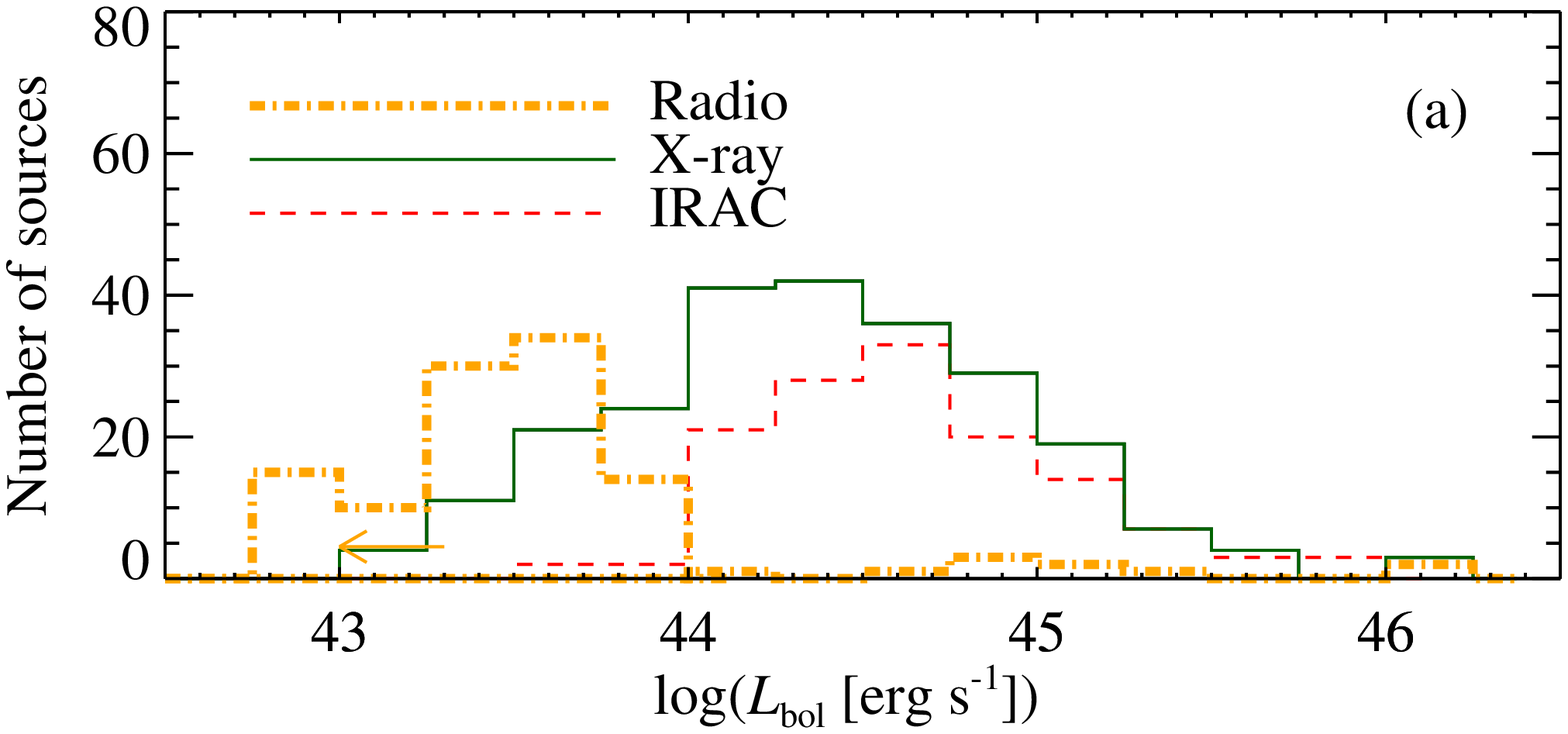}
\plotone{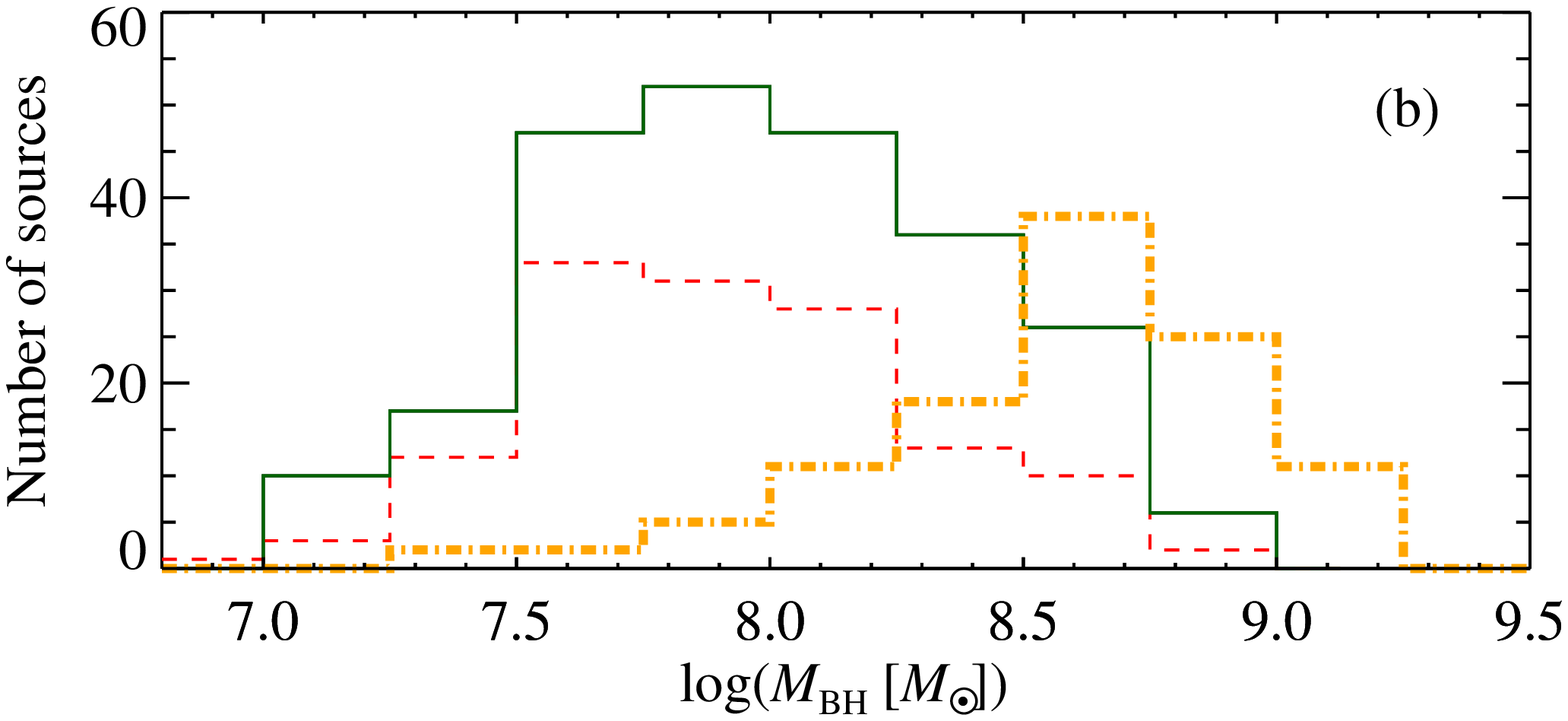}
\plotone{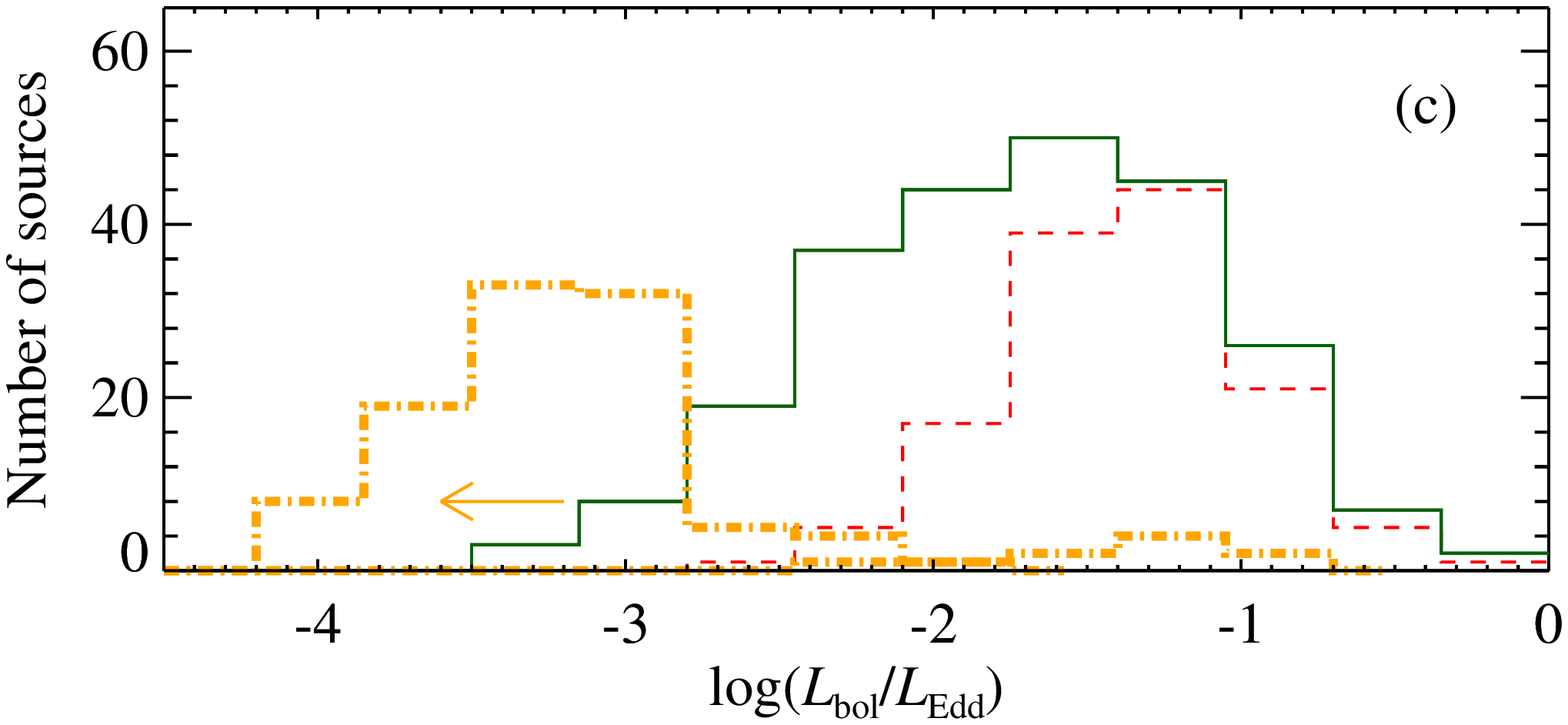}
\caption{Distributions in (a) \lbol, (b) $M_{\rm BH}$, and (c) Eddington ratios for AGNs with extended optical counterparts, for radio, X-ray, and IR AGNs.  The \lbol\ and $L_{\rm bol}/L_{\rm Edd}$ estimates for the X-ray undetected radio AGNs show only upper limits derived from X-ray stacking (Section~\ref{stack}).   See text in Section~\ref{edd} for details.   Radio, X-ray, and IR AGNs have progressively smaller typical black hole masses and higher Eddington ratios. \label{fagng_hist} } 
\vskip0.5cm
\end{figure}

We next obtain \lbol\ by scaling from \lx.  We first convert \lx\ from
the 0.5--7 band to the 0.5--8 keV band by multiplying by 1.1, which is
roughly valid for power-law spectra with photon index
$0.8<\Gamma<1.8$.  We then convert to \lbol\ using the
luminosity-dependent bolometric corrections of \citet{hopk07qlf},
which are in the range ${\rm BC}_{\rm 0.5-8\; keV}\simeq10$--20.  This
correction assumes that the intrinsic broadband SED of these AGNs is
similar to that of unobscured quasars. If instead the source has an
SED that is more dominated by X-ray emission, this correction may
overestimate \lbol\ \citep[e.g.,][]{vasu07bolc}.  The distributions in \lbol\
for the X-ray sources (and limits for the radio AGNs not detected in
X-rays) are shown in Figure~\ref{fagng_hist}{\em a}.

\begin{figure}[t]
\epsscale{1.1}
\plotone{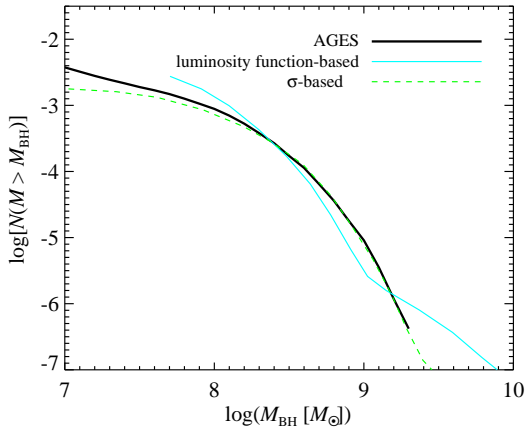}
\caption{Black hole mass function for $0.25<z<0.8$, derived from $M_{\rm BH}$ measurements described in Section~\ref{edd}, compared to local mass function estimates in the literature.  The thick black line shows $\phi(M_{\rm BH})$ derived from the AGES data, including rough completeness corrections. The solid cyan and dashed green lines show local estimates of $\phi(M_{\rm BH})$ compiled in Figure~11 of \citet{laue07bhmass}, derived from luminosity functions and distributions in bulge stellar velocity dispersions, respectively.  Our rough estimates of $M_{\rm BH}$ give a BH mass function that is similar to those derived from other methods. \label{fbh_massfunc}} 
\vskip0.5cm
\end{figure}

For IR AGNs that are not detected in X-rays, \lbol\ is derived from
the rest-frame 4.5 \micron\ luminosity.  Using the \citet{hopk07qlf}
scaling somewhat overestimates \lbol, compared to the values derived
from the X-rays (for those sources that are both X-ray and IR
AGNs). This may be owing to the fact that the host galaxy can
contribute some of the observed IR flux.  We therefore decrease the IR
bolometric corrections by a factor of 2 so that \lbol\ derived from
the X-rays and IR are consistent, and use these corrections for all
the IR AGNs.  The distributions in \lbol\ for the IR AGNs are also
shown in Figure~\ref{fagng_hist}{\em a}.

For an estimate of $M_{\rm BH}$, we first determine the bulge
luminosity in the $B$ band ($L_{B, \rm bul}$) and then scale, using
the relation of \citet{marc03}, between $L_{B, \rm bul}$ and
$M_{\rm BH}$.  We derive the absolute $B$ band magnitude of the
galaxy, $M_B$, by adding the rest-frame $^{0.1}(g-r)$ color to
the evolution-corrected absolute magnitude in the $^{0.1}r$ band
(Section~\ref{kcorr}), and then subtracting 0.1 mag for the small conversion
between the $^{0.1}g$ and $B$ bands.  We then calculate $L_{B}$ (in
$L_{\sun}$), taking the absolute magnitude of the Sun in the $B$ band
to be 5.48.

We further require an estimate of the ratio of bulge to total
luminosity in the $B$ band, $(B/T)_B$, which can vary from less than
0.1 for Scd galaxies to 1 for ellipticals.  Here we use the
morphological analysis of the Millennium Galaxy Catalog (MGC) by
\citet{alle06morph}, who calculated structural parameters (including
bulge and disk luminosities) for a large sample of galaxies.  Using a
subset of 8243 galaxies from the MGC with $-24<M_B<-18$ (roughly the
magnitude distribution of our AGN host galaxies), we calculate the
average $(B/T)_B$ in bins of $u-r$ color.  $(B/T)_B$ varies from
$\simeq$0.1 at $u-r=1$ (the bottom of the blue cloud) to $\sim0.5$ at
$u-r=3$ (the top of the red sequence), with variance of
$\Delta(B/T)_B\sim0.2$ in each color bin.  Using this relation and the
rest-frame $^{0.1}(u-r)$ colors (making a slight correction of 0.1 mag
to convert to $u-r$), we derive $(B/T)_B$, and thus $L_{B, \rm bul}$,
for the AGN hosts.

Finally, we derive $M_{\rm BH}$ from $L_{B, \rm bul}$ using the
relation of \citet{marc03} for their ``Group 1'' sample of galaxies.
The corresponding distributions in $M_{\rm BH}$ and $\lambda$ are
shown in Figs.~\ref{fagng_hist}{\em b} and \ref{fagng_hist}{\em c}.
Owing to the scatter in the above relations, these $M_{\rm BH}$
estimates may be uncertain by as much as 0.5 dex for individual
galaxies, but we expect the average distribution to be accurate to
within $\sim0.2$ dex.  The method for measuring $M_{\rm BH}$ likely
introduces a further systematic factor of $\sim0.3$--0.5 dex, but this
does not affect the {\em relative} distributions in $M_{\rm BH}$ and
$\lambda$ for the different samples, which are primarily of interest
here.

As a check on these $M_{\rm BH}$ estimates, we derive a rough SMBH
mass function from the AGES galaxies and compare with those derived by
previous works.  For galaxies with $M_{\rm BH} \lesssim 10^9 \msun$
the optical luminosity is low enough that they are not detected out to
$z=0.8$.  To account for this, we use the observed distribution of
black hole masses with redshift to estimate the typical volume probed
for sources as a function of $M_{\rm BH}$.  Including the galaxy
sampling weights (Section~\ref{weights}), we derive the mass function
$\phi(M_{\rm BH})$.  We directly compare this to local black hole mass
functions compiled by \citet{laue07bhmass}, and find that they closely
agree (Figure~\ref{fbh_massfunc}), suggesting that on average our BH
mass estimates are reasonably accurate.  We note that we can directly
compare our BH mass function for objects at $0.25<z<0.8$ to
$\phi(M_{\rm BH})$ derived locally, because the high-mass end of the
BH mass function has evolved only slightly since the epochs probed by
the typical AGES galaxies.  The observed AGN ``downsizing''
\citep[e.g.,][]{hasi05,barg05} implies that BH growth at $z<0.5$ is
dominated by relatively small black holes, so that the change in the
BH mass function from $z\sim0.5$ to $z=0$ is only 0.2 dex at $M_{\rm
BH}\sim10^7$ $\msun$, and negligible for $M_{\rm BH}>10^8$ $\msun$
\citep{merl08agnsynth}.

Figure~\ref{fagng_hist}({\em c}) shows that the different classes of
AGNs have significantly different characteristic Eddington ratios.
X-ray AGNs have a wide range of $10^{-3}\lesssim \lambda \lesssim 1$,
while the IR AGNs have higher Eddington ratios, almost all having
$\lambda \gtrsim 10^{-2}$. The X-ray undetected radio AGNs have very
low upper limits on $\lambda$, with most sources having $\lambda <
10^{-3}$.  Although these estimates of $\lambda$ are highly uncertain,
particularly for individual galaxies, we can conclude that the
population of X-ray AGNs extends to lower Eddington ratios than that
of the IR AGNs, while the radio AGNs have very low Eddington ratios.

Finally, we note that the sources that are selected as both X-ray and
IR AGNs have a distribution in $\lambda$ that is almost identical to
that of the full IR AGN population.  The median values of $\lambda$
for the IR AGNs that are detected and undetected in X-rays differ by
only 0.05 dex.  We note as well that the clustering of the X-ray
and IR AGNs is essentially identical to that for the full IR AGN
sample (Section~\ref{xircorr}), and that the stacked X-ray flux from the
X-ray undetected IR AGNs is consistent with having low-level X-ray
emission (as shown in Appendix \ref{appendix_ir}).  Therefore, the
IR-selected sources may represent a single coherent population of
AGNs, distinct from typical X-ray AGNs, most of which are not selected
with IRAC.

\section{Discussion}
\label{discussion}

In this paper we have used observations across a broad range of wavelengths
together with spectroscopic redshifts to provide a general picture of
the populations of accreting SMBHs, their host galaxies, their
environments, and their accretion modes at moderate redshifts
($0.25<z<0.8$).  As discussed in Section~\ref{intro}, there is evidence for
a  connection between accretion onto SMBHs and the evolution of
their host galaxies.  Therefore the properties of AGN and their hosts
that we observe at $z\sim0.5$ provide a ``snapshot'' in the
co-evolution of these objects over cosmological time.

\subsection{A simple model of AGN and galaxy evolution}

In this section we present a simple ``cartoon'' model of AGN and host galaxy
evolution that explains, in broad terms, the observations described
in this paper.  We base this model on four ideas that have received
much discussion in the literature on AGN and galaxy evolution, and are
supported by considerable observational and theoretical evidence:

\begin{enumerate}

\im In the early universe, galaxies initially form as systems that are
rich in gold gas and have rotation-dominated dynamics (although they
can be clumpier and more turbulent than present-day disk galaxies).
This scenario is supported by observations of high-redshift galaxies
\citep[e.g.,][]{ravi06morph, elme07disk} as well as theoretical
arguments for the dissipational collapse of cold clouds
\citep[e.g.,][]{whit78condens, fall80disk}, and detailed numerical
models \citep[e.g.,][]{robe04disk, robe06merge, hopk09disk}.

\im Optical quasars are found in dark matter halos with a
characteristic mass of $M_{\rm halo}\sim3\times10^{12}$ $\msun$ that
remains effectively constant with redshift.  In large surveys, the
bias of quasars is observed to increase with redshift, consistent with
this characteristic parent dark matter halo mass
\citep{porc04clust,croo05,porc06qsoclust, coil07a,
myer07clust1,shen07clust, daan08clust, padm08qsored}.  Quasars
represent phases of accretion at high Eddington ratios \citep[$\lambda
> 0.1$;][]{mclu04mbh,koll06}, in which massive black holes accrete the
bulk of their final mass
\citep[e.g.,][]{yu02smbh,shan09agnbh,yu08bhacc}.

\im The processes that fuel quasars are also responsible for the
creation of stellar bulges.  There is evidence that the quasar phase
is often preceded by dust-obscured phases of very rapid star
formation, which would be manifest observationally as ultraluminous
infrared galaxies (ULIRGs) or submillimeter galaxies (SMGs)
\citep{sand96, smai97smg}.  The process that disturbs the galaxy and
fuels the star formation and accretion can also likely create a
dynamically hot stellar bulge.  Possible physical processes include
major mergers of gas-rich galaxies \citep[e.g.,][]{kauf00merge,
spri05, hopk06merge} and disk instabilities \citep[e.g.,][]{mo98disk,
bowe06gal, genz08ifsz2}

\im After the creation of the stellar bulge, the star formation in a
galaxy is quenched on a relatively short timescale.  The quenching can
occur because the available gas gets used up in starbursts, or because
the dark matter halo has increased in mass so that the gas in the
system is heated to the high virial temperature of the halo and cannot
cool efficiently \citep[e.g.,][]{rees77gas, birn03virial}.  Feedback
from an AGN may also help to blow away gas and quench star formation
\citep[e.g.,][]{hopk06merge}.  Further mechanical feedback (likely
also from an AGN) prevents subsequent gas cooling and keeps the system
``red and dead'' \citep[e.g.,][]{chur05smbh,bowe06gal,
crot06,khal08feedback}.

\end{enumerate}

\begin{figure*}[t]
\plotone{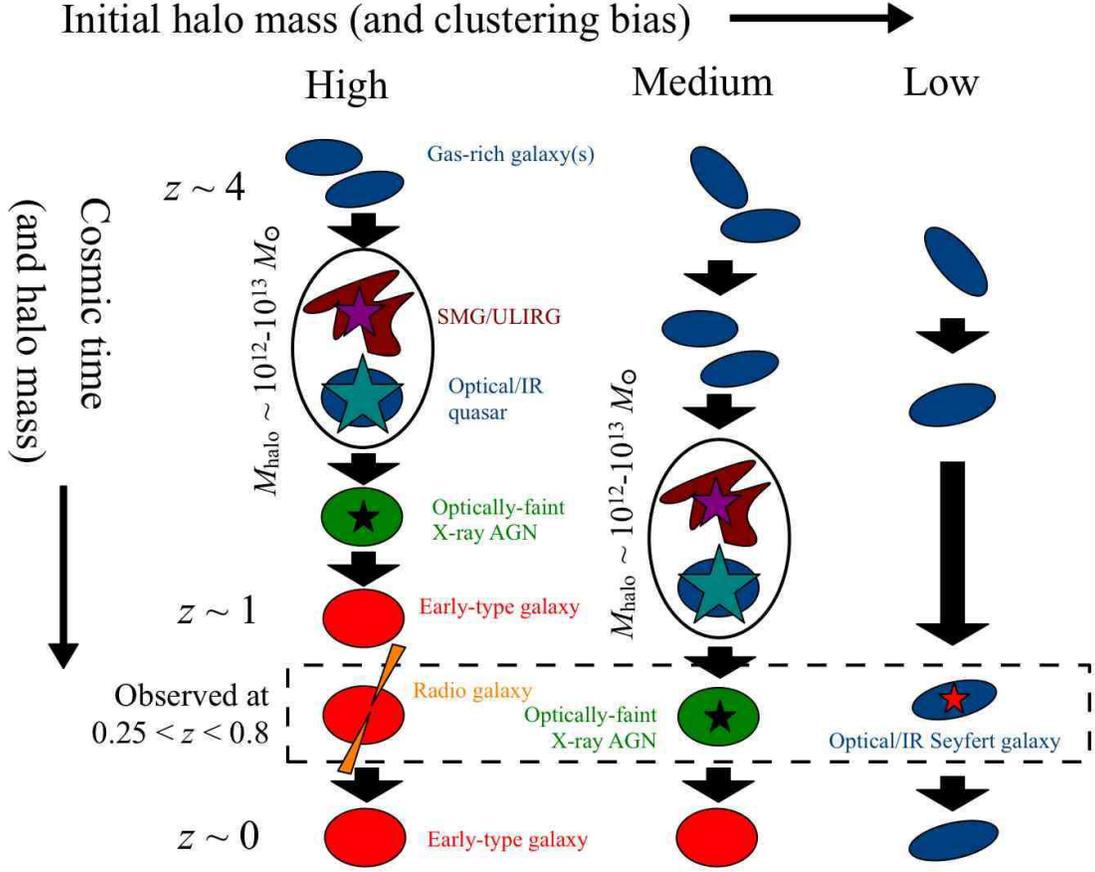}
\vskip0.6cm
\caption{\footnotesize Schematic for a simple picture of AGN and host galaxy
evolution.  The picture consists of an evolutionary sequence that
occurs at different redshifts for halo masses of different sizes.  In
this scenario, luminous AGN accretion occurs preferentially (through a
merger or some secular process) when a host dark matter halo reaches a
critical $M_{\rm halo}$ between $10^{12}$ and $10^{13}$ \hmsun\ (this
phase is indicated by the solid ovals).  Once a large halo reaches
this critical mass, it becomes visible as a ULIRG or SMG (owing to a
burst of dusty star formation) or (perhaps subsequently) as a
luminous, unobscured quasar.  
The ULIRG/quasar phase is associated
with rapid growth of the SMBH and formation of a stellar spheroid, and
is followed by the rapid quenching of star formation in the galaxy.
Subsequently, the young stellar population in the galaxy ages
(producing ``green'' host galaxy colors), and the galaxy experiences
declining nuclear accretion that may be associated with an X-ray AGN.
Eventually the aging of the young stars leaves a ``red and dead''
early-type galaxy, which experiences intermittent ``radio-mode'' AGN
outbursts that heat the surrounding medium.  For ``medium'' initial
dark matter halos, the quasar phase and formation of the spheroid
occur later than for the system with high halo mass, so that at
$z\sim0.5$ we may observe the ``green'' X-ray AGN phase.  Even smaller
halos never reach the threshold mass for quasar triggering; these
still contain star-forming disk galaxies at $z\lesssim 0.8$, and we
observe some of them as optical or IR-selected Seyfert galaxies.  The
dashed box indicates the AGN types (in their characteristic dark
matter halos) that would be observable in the redshift range
$0.25<z<0.8$.
\label{fschem}} 
\end{figure*}

These ideas imply a relatively straightforward evolutionary sequence
for massive galaxies and their central black holes.  Most aspects of
this picture have been addressed in much more detail elsewhere
\citep[e.g.,][]{kauf00merge,spri05,crot06,bowe06gal, mona07model,
hopk08frame2,hopk08frame1,some08bhev}; here we present only a simple overview and
show that it is generally consistent with our results.  Comparisons of
the observations with more detailed model predictions will be left for
future work.

\subsection{Evolutionary sequence}
\label{sequence}

In our simple picture, the sequence of AGN and galaxy evolution begins
in the early niverse ($z\gtrsim 6$), where galaxies are formed as
gas-rich, rotation-dominated systems and contain small ``seed''
central black holes.  As their parent dark matter halos grow up to
masses $\sim$$10^{12}$ $\msun$, these galaxies retain their general
morphological and dynamical properties, including flattened morphologies,
small stellar bulges, and small central black holes.  This growth of
these galaxies occurs through the accretion of further cold gas or
through minor mergers (with mass ratios $>$3:1) with other gas rich
systems \citep{robe06merge,hopk09disk}.  As the galaxies grow, their
dark matter halos also increase in mass through the hierarchical
growth of structure.  The largest halos grow fastest, so that the most
massive overdensities at high redshift remain the most massive at
lower redshift, and so on for smaller masses \citep[e.g.,][]{pres74,
lace93merge}.

When a galaxy's parent dark matter halo reaches a critical $M_{\rm
halo}$ between $10^{12}$ and $10^{13}$ $\msun$, a dramatic event
occurs that triggers luminous quasar activity and thus rapid growth of
the central black hole, as well as producing a dynamically hot stellar
system (the classical bulge).  One candidate for such an event is the
major merger of two gas rich galaxies; the characteristic quasar halo
mass of $M_{\rm halo}\sim 3\times 10^{12}$ $\msun$ corresponds to the small group environment where major mergers of galaxies with
masses $\sim M_{*}$ are most common \citep{hopk08frame1}.  Models of
mergers appear to naturally produce many observed properties of
quasars and the resulting spheroidal remnants, as well as their
abundance and redshift evolution \citep{kauf00merge, spri05,
hopk08frame1}, and there is evidence for merger remnants hosting
individual quasars \citep[e.g.,][]{riec08wetqso, arav08qsosb}.

However, in addition to mergers, other types of events such as disk
instabilities \citep[e.g.,][]{bowe06gal} may also possibly trigger
efficient accretion onto the black hole and grow the stellar bulge.
We also note that other processes, such as the accretion of recycled gas
from evolved stars \citep{ciot07flare} may trigger a quasar after the formation
of the bulge.  In the context of our observations at $z\sim0.5$, the
details of the triggering mechanism are relatively unimportant.  It is
sufficient only that the primary mechanism for quasar accretion and
the growth of the stellar bulge occur together {\em at a
characteristic dark matter halo mass}, at high redshift for the
largest overdensities, and at progressively later times for smaller and
smaller overdensities.

After the event that builds the bulge and grows the central black
hole, star formation in the galaxy must be quenched in order to
produce the observed population of passively evolving, bulge-dominated
galaxies with old stellar populations that are found on the red
sequence \citep[e.g.,][]{nela05red,thom05epoch}, which is detected out
to $z\sim2$ \citep{zirm08red,krie08red}.  The cessation of star
formation must necessarily occur after or simultaneous to the
formation of the bulge, since essentially all passively evolving
galaxies contain stellar bulges \citep{bell08quench}\fnm.  It is
possible that this quenching is a direct result of the event that
fuels the quasar.  Processes that drive gas onto the black hole can
also power intense star bursts \citep[as are often associated with
luminous AGNs, e.g.][]{alex05,brand06b} that can use up the galaxy's
supply of cold gas and thus halt star formation.  In addition, high virial temperature of the (growing) dark matter halo
can limit further accretion of cold gas and thus star formation
\citep[e.g.,][]{rees77gas, catt06crit, fabe07lfunc}.  Simulations suggest that the
transition to this ``hot halo'' regime occurs at $M_{\rm
halo}\sim10^{12}$ \citep[e.g.,][]{birn03virial, deke06shock}, close to
the typical $M_{\rm halo}$ for quasar activity.  For our simple
picture, the details of this quenching are not important; it is
sufficient that soon after the formation of a massive stellar bulge,
star formation in the galaxy rapidly ceases.

\fnt{We note that the \citet{bell08quench} result is only for
{\em central} galaxies in groups or clusters; see Section~\ref{caveats} for a discussion.}

If the timescale for the quenching of star formation is short, the
galaxy's optical spectrum will be a composite of emission from the old
stellar population plus a fading contribution from the younger stars,
producing ``green'' colors that become redder as the young stellar
population evolves.  The timescale for the galaxy to evolve red colors
typical of early-type galaxies is $\sim$1--2 Gyr
\citep{newb90sb,barg96sfclust,bowe98colmag}.  This phase may be
accompanied by further accretion onto the central black hole, which
can evolve on a similar timescale \citep{hopk05life1}, and may
contribute to the galaxy's transition to the red sequence.
Observations of local ($z<0.1$) early-type galaxies suggest that
optically-detected AGN activity coincides with the decline of star
formation and the transition from blue to red colors
\citep{scha07feed}, as well as the destruction of molecular clouds
\citep{scha09agnmol}.  The decline of the average black hole accretion
rate from its peak in the quasar phase (when $\dot{M} \sim
\dot{M}_{\rm Edd}$) can occur (with significant variability) over
timescales of $\sim$1 Gyr \citep{hopk05life1}, finally reaching
accretion rates of $\dot{M}\lesssim 10^{-2} \dot{M}_{\rm Edd}$ or
lower.  At these low accretion rates (and corresponding low
luminosities), the SED of an AGN is increasingly dominated by X-ray
emission \citep{stef06alphaox,vasu07bolc}, and may even enter a new,
radiatively inefficient mode of accretion \citep{chur05smbh}.
Therefore, these fading AGNs in ``green'' galaxies may be most
efficiently found as X-ray AGNs.

AGN activity may serve to prevent further bursts of star
formation.  As mentioned above, at this stage the massive stellar
bulge of the galaxy will be surrounded by a halo of hot, virialized
gas.  While most of this gas will have very long cooling times, near
the center (where the gas density is highest and cooling most
efficient), gas cooling could fuel further star formation.  Such
cooling must be suppressed to explain the old stellar populations
observed in red-sequence galaxies and the observed dearth of massive,
blue star-forming galaxies at $z<1$ \citep{fabe07lfunc}.  Once cooling
is suppressed, the red-sequence galaxies can further grow through
``dry'' mergers  that
would not induce further bursts of star formation.

One promising mechanism for suppressing cooling in red sequence
galaxies is feedback from AGNs, which can inject mechanical energy
that reheats the cooling gas \citep{tabe93cool, binn95cool,
chur02cool}.  While there is indirect evidence for such feedback from
optically or X-ray-selected AGNs
\citep[e.g.,][]{scha07feed,bund08quench}, the process of AGN feedback
is observed most directly in radio AGNs. In many elliptical galaxies
and galaxy clusters, relativistic jets from AGNs (detected in the
radio) inflate buoyant bubbles and drive shocks in the hot surrounding
medium \citep[e.g.,][]{fabi03,mcna05nature, form07m87,mcna07araa}.
The heating rate from this process can be sufficient to balance the
cooling from the hot halo, and might therefore quench star formation
\citep{birz04, raff06feedback, cros08radiox}.  The Eddington ratios
for accretion in these ``radio-mode'' outbursts are generally small
($\lambda \lesssim10^{-3}$), and the observed radiative power is far
smaller than the inferred mechanical power of the jets
\citep{chur02cool, birz04}.  It has been proposed that these
radiatively inefficient, jet-dominated outbursts may be fueled by
accretion directly from the hot gas halo (in contrast to the accretion
of cold gas that likely powers high-Eddington, optically luminous
AGNs), and so is only possible in massive galaxies with large dark
matter halos \citep[e.g.,][]{tass08thesis}.

In summary, the simple picture described above suggests that massive
galaxies undergo the following evolutionary sequence: gas-rich,
star-forming, rotation-dominated system $\rightarrow$ SMG/ULIRG/quasar
$\rightarrow$ ``green'' spheroid with declining AGN activity
$\rightarrow$ red spheroid with intermittent radio AGN
activity. Figure~1 of \citet{hopk08frame1} shows a schematic of these
various phases, for the case in which the quasar is triggered by major
mergers.  The SMG/ULIRG/quasar phase evolves on a short timescale
($\lesssim$$10^8$ yr) and occurs in dark matter halos with a
characteristic $M_{\rm halo}$ between $10^{12}$ and $10^{13}$ $\msun$.
The subsequent ``green'' spheroid phase evolves over somewhat longer
times ($\sim$1 Gyr), while the phase of intermittent radio AGN
activity can last for the subsequent lifetime of the system.  A
schematic of this sequence is given in Figure~\ref{fschem}.  The columns
represent systems for different initial halo masses, for which the
evolutionary sequence occurs at different redshifts: large dark matter
halos reach the critical mass $10^{12}$--$10^{13}$ $\msun$ (and thus
trigger quasars) at relatively high redshifts, while for smaller
systems the sequence occurs later.  Therefore, at $z\sim0.5$, {\em
objects in different stages of this sequence should be in dark matter
halos of different sizes}, which can be probed through their
clustering.

A key element of this simple model is that quasar activity, bulge
formation, and the quenching of star formation occur when dark matter
halos reach the critical $M_{\rm halo}$.  Therefore, galaxies residing
in halos that have not reached this mass will remain as gas-rich
disks, and will not experience a luminous quasar phase and the
associated rapid black hole growth.  A schematic of the evolution of
low halo mass systems is given in the rightmost column of
Figure~\ref{fschem}.  Although these objects have never hosted luminous
quasars, they can be seen as low-luminosity AGNs; various processes
such as the stochastic accretion of cold gas clouds \citep{hopk06low}
can fuel low-level (and likely intermittent) AGN activity, while
galaxy-scale processes, such as the formation of galactic bars,
can enhance the gas density in the nuclear regions and
preferentially help drive accretion \citep[e.g.,][]{ohta07bar}. 

Accretion in low halo mass systems can proceed in a wide range of
accretion rates; however, owing to the small black hole masses
($M_{\rm BH}\lesssim10^{7}$ $\msun$) only the sources with relatively
high accretion rates will be detected at $z\sim0.5$ in a wide-field
survey such as \bootes.  The observational evidence for AGN
``downsizing'', in which small black holes dominate the accretion
density of the universe at low redshift
\citep[e.g.,][]{hasi05,barg05}, suggests that AGNs in less-luminous,
disk-dominated galaxies should dominate the population of
high-Eddington-rate AGNs at $z<0.8$.  These high-Eddington AGNs would
likely be identifiable by their mid-IR colors.

\subsection{Comparison to observations}

We next compare the evolutionary scenario described above to our
observational results for the host galaxies, clustering, and Eddington
ratios of the different classes of AGNs.

\begin{enumerate}
\im {\em Radio AGNs.} Most radio AGNs in our survey at $0.25<z<0.8$
are found in luminous red sequence galaxies and are strongly
clustered.  In the evolutionary picture described above, radio AGNs
are a key part of the late stages of massive galaxy evolution, after
the bulge has formed and star formation has ceased.  Therefore at
$z\sim0.5$, radio AGNs would be expected to be found primarily in
massive red sequence galaxies in large dark matter halos.
Accordingly, we find that radio AGNs are located in relatively large
dark matter halos with $M_{\rm halo}\sim 3\times10^{13}$ \hmsun,
corresponding to the large galaxy groups or small clusters.  The radio
AGNs are clustered similarly to a control sample of galaxies with
matched luminosities and colors, suggesting that intermittent radio
activity may be common, at least for the host galaxies and radio
luminosities probed by our sample.

We note that a small fraction ($\sim20\%$) of the radio AGNs with
extended optical counterparts are found in blue galaxies, which is
contrary to this simple picture.  However, some of these are are
likely radio-loud Seyfert galaxies and quasars, which make up
$\sim$10\% of those populations, and which may represent a mode of
accretion more similar to radiatively-efficient Seyferts than
mechanically-dominated radio galaxies \citep[e.g.,][]{merl08agnsynth, dono09radio}.

\im {\em X-ray AGNs.} X-ray AGNs in our survey at $0.25<z<0.8$ are
found in host galaxies throughout the color-magnitude diagram, but the
distribution peaks in the ``green valley'' along with a tail of bluer
galaxies.  The large-scale (1--10 \hmpc) clustering indicates that the
X-ray AGNs inhabit dark matter halos of mass $\sim10^{13}$ \hmsun,
similar to the clustering for normal AGES galaxies.  The X-ray AGNs
are unbiased relative to a control sample with matched galaxy
properties, although there is a (marginally-significant) antibias
relative to AGES galaxies on scales $<1$ \hmpc, suggesting that the
X-ray AGNs may preferentially reside in central galaxies.  In our
simple evolutionary picture, central galaxies in halos with $M_{\rm halo}\sim
10^{13}$ \hmsun\ are likely to have recently experienced the buildup
of the stellar bulge and quenching of star formation, followed by a
decline in the accretion rate onto the black hole.  The properties of
X-ray AGNs, with hosts typically showing ``green'' colors and with a
wide range of Eddington ratios ($10^{-3}\lesssim \lambda \lesssim 1$)
are generally consistent with this picture.

\im {\em IR AGNs.} IR-selected AGNs in our observations are found in
relatively low-luminosity galaxies and are weakly clustered, with
characteristic $M_{\rm halo}\lesssim 10^{12}$ \hmsun.  There is a
significant antibias relative to a matched control sample of normal
galaxies.  The Eddington ratios of these sources are relatively high
($\lambda\gtrsim10^{-2}$), and the clustering and Eddington ratios do
not depend strongly on whether the AGNs are also detected in X-rays.
In our simple picture of AGN evolution, at $z<0.8$ most high-Eddington
AGNs will be found in objects with small black holes, in environments
that have not yet reached the critical halo mass for the quenching of
star formation.  In our observations we identify IR-selected AGNs with
weakly clustered star-forming galaxies, in general agreement with this
picture.

\end{enumerate}

\subsection{Caveats}
\label{caveats}

The simple evolutionary scenario described above thus corresponds
generally to our observational results of radio, X-ray, and
IR-selected AGNs.  However, by its nature our simple model of AGN and
galaxy evolution ignores many details and is subject to several
important caveats.  In this section we discuss several of these issues
and how they may be addressed with future observations.

\begin{enumerate}
\im {\em Central versus satellite galaxies.} The model described above
may be only strictly valid for {\em central} galaxies, defined as the
dominant galaxy in the parent dark matter halo.  It is possible that
the quenching of star formation can happen in different ways for
satellite galaxies that fall into larger dark matter halos; ram
pressure stripping, strangulation, and starvation can cause a disk
galaxy to lose its cold gas and cease star formation \citep[see][and
references therein]{vand08quench}.  Further, there is evidence that
for satellite galaxies, star formation and AGN activity depend more
strongly on the stellar mass of the galaxy than the mass of the halo
\citep{pasq09agnhalo}.  Successful models of AGN and galaxy evolution
will likely need to explain the evolution of central and satellite
galaxies separately.  We note, however, that at least for X-ray and
IR-selected AGN in our sample, the observed clustering suggests that
they may preferentially reside in central galaxies (Section~\ref{agncorr}).

\im {\em Critical halo mass.} Our simple evolutionary model includes
``critical'' dark matter halo mass $M_{\rm crit}$ between $10^{12}$
and $10^{13}$ $\msun$ at which a galaxy undergoes luminous quasar
activity and the growth of the stellar bulge.  This idea is motivated
by observations of quasar clustering and models of hot gaseous halos,
but the details of this process are certainly more complicated than
can be described by a single critical halo mass.  Indeed, galaxy
evolution models in which quenching occurs at a halo mass threshold
may fail to reproduce the density of passively-evolving galaxies at
high redshift, as well as relative numbers of red and blue galaxies
\citep{hopk08frame2}.  Further, evidence from the evolution of
luminosity function on the red and blue sequence suggests that
quenching may occur with different mechanisms, and in different mass
halos \citep{fabe07lfunc}.  Quasar triggering and bulge formation
probably occur by some more complex process that {\em preferentially}
occurs at a given halo mass.  The details of this process cannot be
constrained with the observations presented here, but they could be
tested in the future by more detailed work.

\im {\em Extinction and galaxy colors.} The above discussion makes
note of the ``green'' galaxy host colors of the X-ray AGNs as evidence
that these sources have undergone recent  star formation.
However, in calculation of galaxy colors, we have not taken into
account the effects of dust extinction intrinsic to the galaxy.  It is
possible that some galaxies in the ``green valley'' are in fact
actively star-forming galaxies that are reddened by dust.  In a
multiwavelength study of galaxies at $0.05<z<1.5$ in the GOODS-North
field, \citet{cowi08int} found a significant population of galaxies
(mainly {\it Spitzer} 24 \micron\ sources) that have observed colors
in the ``green valley'', but lie in the blue cloud after correcting
for extinction.  However, in our sample the X-ray AGNs in ``green''
galaxies show clustering consistent with that of typical quiescent
``green'' galaxies with similar colors (Section~\ref{agncorr}), and
stronger clustering than typical blue cloud galaxies.  Thus it is
unlikely that the majority of the ``green'' hosts of X-ray AGNs are
dust-reddened blue cloud galaxies. In the future, a more detailed
study of the optical-IR SEDs and average optical spectra of these
objects would allow us to put more strict limits on the population of
dust-reddened star-forming galaxies.
\end{enumerate}

\section{Summary}
\label{summary}

In this paper we explored the links between SMBH accretion, host
galaxy evolution, and large-scale environment, by studying the host
galaxies of AGNs detected at redshifts $0.25<z<0.8$ in the AGES
survey.  The AGNs were selected using observations in the radio,
X-ray, and IR wavebands.  Key results from this paper are as follows:

\begin{enumerate}
\im  AGNs in the AGES sample at $0.25<z<0.8$ that are selected in
different wavebands comprise generally distinct populations of
sources.  Radio AGNs are generally not selected in the other bands,
while there is 30\%--50\% overlap between the X-ray and IR-selected AGNs.

\im The host galaxies of optically faint AGNs are significantly different between the three AGN samples.  Radio AGNs are found in luminous red-sequence galaxies.  X-ray AGNs are found in galaxies of all colors, with a peak in the ``green valley''. IR AGN hosts are relatively bluer and less luminous than those of the X-ray or radio AGNs.

\im The two-point cross-correlation between AGNs and galaxies is
significantly different for the three classes of AGNs.  We estimate that
radio, X-ray, and IR AGNs are found in dark matter halos with
characteristic masses $\sim$$3\times10^{13}$ \hmsun, $\sim$$10^{13}$ \hmsun, and
$\lesssim$$10^{12}$ \hmsun, respectively.

\im X-ray and radio AGNs with extended galaxy counterparts are
clustered similarly to samples matched in color, absolute magnitude,
and redshift, indicating that they inhabit environments typical of
their parent host galaxy populations.  Conversely, IR AGNs are weakly
clustered relative to a matched galaxy sample, suggesting that the
process that triggers IR-bright AGN accretion depends on local
environment as well as host galaxy type, and is more likely to occur
in regions of lower galaxy density.

\im The average X-ray spectra of X-ray and IR AGNs are consistent with
the properties of their optical counterparts. Optically bright AGNs
have soft X-ray spectra with photon index $\Gamma\simeq1.6$, typical of unabsorbed AGNs.
Optically faint AGNs have harder average spectra, which likely
indicates either absorption by intervening gas, or a radiatively
inefficient mode of accretion.  

\im Most radio AGNs in our sample have massive black holes
($M_{\rm BH}>10^8$ $\msun$) and very small Eddington ratios ($\lambda < 10^{-3}$). X-ray
AGNs have smaller typical $M_{\rm BH}$ values that extend down to
$\sim$$10^7$ $\msun$, and $\lambda$ between $10^{-3}$ and 1.  IR AGNs
have relatively small black holes ($3\times10^7\lesssim M_{\rm BH}
\lesssim 3\times10^{8}$ $\msun$) and high Eddington ratios
($\lambda>10^{-2}$).

\im Sources that are selected as both X-ray and IR AGNs have host
galaxy properties, clustering, and Eddington ratios that are similar
to those for the full IR AGN population.  This suggests that the
IR-bright AGN may represent a single population of AGNs, distinct from
typical X-ray AGNs (most of which are not selected with IRAC).

\im The host galaxies, clustering, and Eddington ratios of the three
classes of AGNs are generally consistent with a simple picture of AGN
and galaxy evolution in which SMBH growth and the quenching of star
formation occur preferentially at a characteristic dark matter halo
mass $M_{\rm halo}\sim10^{12}$--$10^{13}$ $\msun$, and are followed by a
decline in the AGN accretion rate and a change to a
radiatively inefficient, mechanically dominated accretion mode.

\end{enumerate}

\begin{acknowledgements}
We thank our colleagues on the NDWFS, AGES, IRAC Shallow Survey and
XBo\"otes teams.  We thank Matthew Ashby, Anca Constantin, Alexey
Vikhlinin, and Steven Willner for fruitful discussions, and thank the
referee, Philip Hopkins, for helpful suggestions.  We also thank
Martin White for sharing dark matter simulation results.  The NOAO
Deep Wide-Field Survey, and the research of A.D. and B.T.J. are
supported by NOAO, which is operated by the Association of
Universities for Research in Astronomy (AURA), Inc.\ under a
cooperative agreement with the National Science Foundation.  This
paper would not have been possible without the efforts of the
\chandra, \spitzer, KPNO, and MMT support staff.  Optical spectroscopy
discussed in this paper was obtained at the MMT Observatory, a joint
facility of the Smithsonian Institution and the University of Arizona.
R.C.H. was supported by an SAO Postdoctoral Fellowship, NASA GSRP
Fellowship, and Harvard Merit Fellowship, and by {\em Chandra} grants
GO5-6130A and GO5-6121A.
\end{acknowledgements}

\appendix

\section{Estimating nuclear contamination for AGN host galaxies}
\label{appendix}

Here we describe in detail our method to estimate the contribution of
active nuclei to the optical colors of X-ray and radio-selected
AGNs.  For unobscured, optically bright Seyfert galaxies and quasars,
the optical emission from the nucleus is generally significantly bluer
than the host galaxy, so that the combined AGN plus galaxy is bluer
than the galaxy alone.  For our study of the colors and luminosities
of AGN host galaxies (Figure~\ref{fagncol}) we have chosen sources that
are optically extended (and so not dominated by a central point
source), but the nucleus may make some contribution to their
integrated optical flux.  

To test for nuclear contamination, we use the optical imaging to
calculate the optical flux and color profiles of the different types
of sources.  Here we use aperture photometry derived using the
technique of \citet{brow07red, brow08halo}, with the $B_W$,
$R$, and $I$-band photometry smoothed to a common PSF with FWHM
1\farcs35 (to avoid PSF differences between bands), and
uncertainties estimated using a Monte Carlo technique.

\begin{figure}
\epsscale{0.5}
\plotone{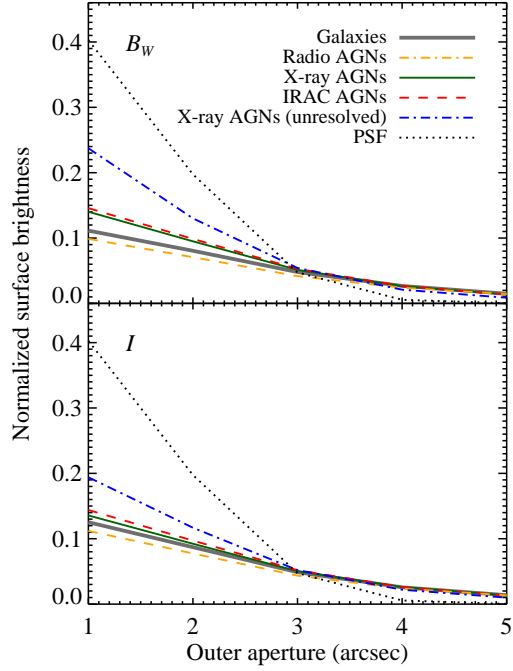}
\caption{Average flux profiles for subsets of galaxies and AGNs from
the AGES survey, in the $B_W$ (top) and $I$ (bottom) bands.  Fluxes
are calculated in annular apertures of width 1\arcsec\ (the $x$-axis
gives the outer diameter of the annulus). Shown are the profiles for
normal AGES galaxies (black solid line), AGNs with extended optical
counterparts (radio, orange dotted line; X-ray, green solid line; IR,
red dashed line), X-ray AGNs with unresolved optical counterparts
(blue dot-dashed line), and a 1\farcs35 (FWHM) Gaussian PSF (black
dotted line).  The AGNs in extended sources have relatively flat flux
profiles similar to normal galaxies, suggesting that they have little
nuclear contamination.
\label{fprofile}}
\end{figure}

Evidence for some nuclear contamination is given in
Figure~\ref{fprofile}, which shows the average $B_W$ and $I$-band flux
profiles for normal galaxies, and radio, X-ray, and IR AGNs with
extended optical counterparts.  For comparison, we also show profiles
for X-ray sources with unresolved optical morphologies (see
Section~\ref{counterparts}), as well as the expected profile for a Gaussian
PSF with FWHM 1\farcs35, typical of the NDWFS observations.  The
profiles are calculated in annuli of 1\arcsec\ width.  Normal galaxies
have broad, extended profiles, while optically unresolved X-ray AGNs
have more strongly peaked profiles indicating that they have a
significant contribution from a nuclear point source.  The radio AGNs
have a broad average profile similar to that of quiescent galaxies,
indicating that on average, they have little or no nuclear
contamination (this may be expected, given that most radio AGNs also
show little nuclear activity in the X-rays or IR).  In contrast, X-ray
and IR-selected AGNs show a slight excess in the $B_W$ band over the
quiescent galaxies inside a 3\arcsec\ diameter, suggesting a moderate
contribution from a nuclear point source.  There is only a weak excess
for the X-ray and IR AGNs in the $I$ band, however, indicating that
the nuclear contribution is bluer than the host galaxy.  Therefore
nuclear emission may affect the integrated colors of the source.

\begin{figure}[t]
\epsscale{0.35}
\plotone{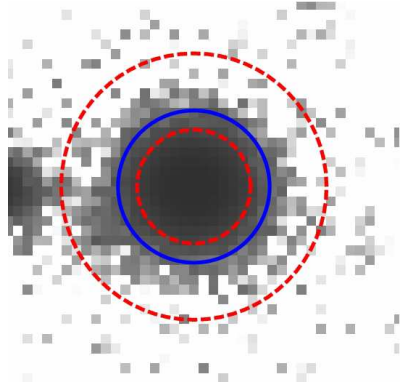}
\caption{NDWFS $I$-band image of a randomly selected X-ray AGN with an extended optical counterpart at $z=0.55$.  The circles show the apertures used for determining the
color contamination from the nucleus; the solid blue line shows the
inner 4\arcsec\ aperture, while the red dashed lines show the annulus
with inner and outer radii of 3\arcsec\ and 7\arcsec, respectively.
\label{faperim}}
\end{figure}

For the radio, X-ray, and IR AGNs, we wish to estimate the extent of
this optical color contamination by the active nucleus, and thus
produce corrected rest-frame \urs\ colors for the host galaxies alone.
To this end, we compare the optical colors observed in the center of
AGN-hosting galaxies to that at the outskirts, and compare these to
quiescent galaxies.  First, for all AGES galaxies we derive the
quantity $C_{3-7}$, which is the $B_W-I$ color in an annular aperture
of inner and outer diameters 3\arcsec\ and 7\arcsec, respectively
(Figure~\ref{faperim}).  Since the observations have a PSF of width
1\farcs35 (FWHM), excluding the central 3\arcsec\ removes most of
the nuclear contribution while including enough flux from the
outskirts of the galaxy to derive sufficiently accurate colors.  We
then calculate $\Delta C$, the difference between $C_{3-7}$ and $C_4$
(the $B_W-I$ 4\arcsec\ aperture colors).  We would expect nuclear
contamination to make $C_4$ bluer, but to have little effect on
$C_{\rm 3-7}$, so that contamination should increase $\Delta C$.

\begin{figure}[t]
\epsscale{1.1} 
\plotone{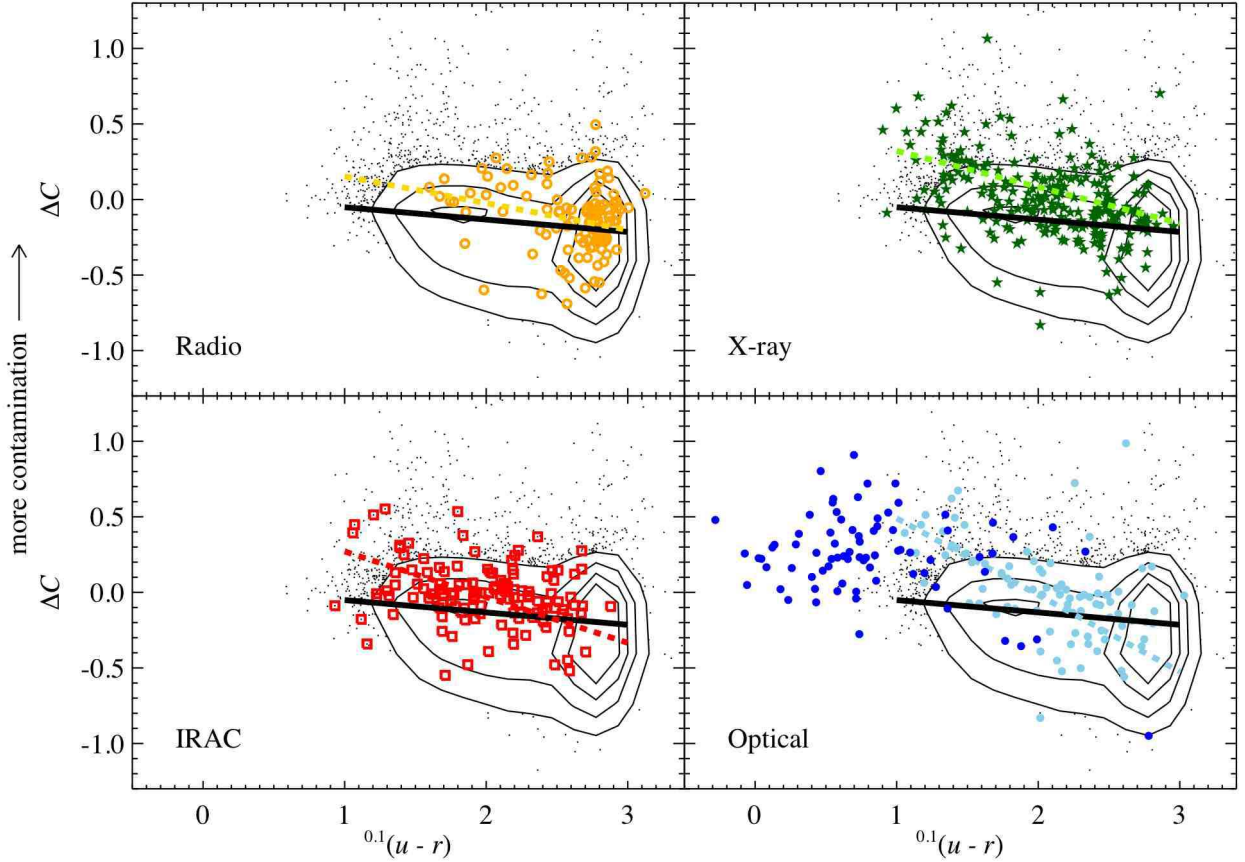}
\caption{Color profiles of galaxies and AGNs in the AGES sample.
Shown is $\Delta C$ (the difference in $B_W-I$ color between the
3\arcsec--7\arcsec\ annulus and the central 4\arcsec\ aperture) vs.\
rest-frame color \urs, as derived in Section~\ref{kcorr}.  The contours and
points represent normal galaxies, while the open circles, stars, and
squares represent radio, X-ray, and IR AGNs with extended optical
counterparts. The bottom right panel shows the same quantities for
AGNs with broad lines in their optical spectra.  Optically unresolved
sources are shown in dark blue, and optically extended sources are
shown in light blue. In all panels, the thick lines show linear fits to $\Delta C$
vs.\ \urs\ for the three sets of sources (see the text for details).
\label{faper}}
\end{figure}

We note that the color profiles of galaxies vary with galaxy type;
early-type galaxies are redder in the centers (where they are
dominated by old, higher metallicity stars) compared to late-type
galaxies.  Figure~\ref{faper} shows $\Delta C$ versus the rest-frame
$^{0.1}(u-r)$; contours and points show this distribution for
normal galaxies, while stars and circles represent X-ray and radio
AGNs, respectively.  For normal galaxies, $\Delta C$ increases
slightly with bluer \urs; the reddest galaxies have $\Delta
C\simeq -0.1$ while the bluest have $\Delta C\simeq 0$.  In contrast,
the radio, X-ray, and IRAC AGNs show stronger increases in $\Delta C$
with bluer colors, indicating that some X-ray AGNs have significant
nuclear emission.  The solid black and dashed green, orange, and red
lines show linear average fits to the trend of $\Delta C$ with \urs.
The trends are roughly linear; there is no significant difference if
we compute the average $\Delta C$ in bins of \urs, so for simplicity
we use the linear fit.

Using these fits, we can derive the average nuclear contamination of
the observed color as a function of the integrated \urs\ obtained for
X-ray and radio AGN.  For a given \urs, the $B_W-I$ 4\arcsec\ color of
the host galaxy ($C_{\rm gal}$) is related to the observed 4\arcsec\
color ($C_4$) by

\begin{equation} 
C_{\rm gal} = C_4 + (\Delta C)_{\rm corr},
\end{equation}
where $(\Delta C)_{\rm corr}$ is a function of \urs, and is simply the
difference between the dashed green line and the solid black line in
Figure~\ref{faper}.  Including all radio, X-ray, and IR AGNs, the best
linear fit gives:

\begin{equation}
(\Delta C)_{\rm corr}=0.58+0.20 [^{0.1}(u-r)].
\label{eqncorr}
\end{equation}
We note that, as is clear in Figure~\ref{faper}, \dccorr\ versus \urs\
is slightly different for the three classes of AGNs, but for simplicity we use
the fit that includes for all AGNs.

\begin{figure}
\epsscale{0.7}
\plotone{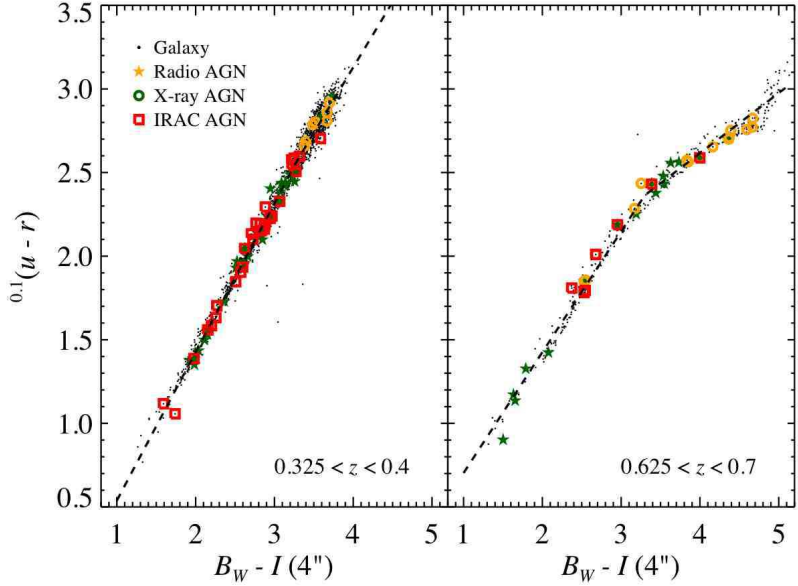}
\caption{$K$-corrected rest-frame color \urs\ vs.\ observed $B_W-I$ 4\arcsec\ aperture color, for AGES main sample galaxies (points), and AGNs with extended optical counterparts: radio (circles), X-ray (stars), and IRAC (squares).  Sources and broken power-law fits are shown for two redshift bins. 
\label{fcolrest}}
\end{figure}

For normal galaxies and  AGNs, the average $\Delta C$ increases
slightly with redshift; if we divide the sample into sources with
$0.25<z<0.5$ and $0.5<z<0.8$, the average $\Delta C$ differs by up to
$\simeq$$0.2$ (depending on \urs).  However, the change with redshift
is the same for both normal galaxies and AGNs, so \dccorr\ [as a
function of \urs] changes by very little (less than 0.04).  For simplicity,
we therefore use the single parameterization for \dccorr\ versus \urs\
(Eqn.~\ref{eqncorr}) for all $0.25<z<0.8$.   

For comparison to the radio, X-ray, and IR AGNs with extended
counterparts, we also derive $\Delta C$ for X-ray sources at $0.25<z<0.8$
with broad emission lines in their AGES optical spectra, indicating
that they are optically unobscured AGNs\footnotemark.  These objects
are shown in the lower right panel of Figure~\ref{faper}.  Sources with
unresolved morphologies (broad-line, unresolved AGNs) are shown in
dark blue, and those with extended morphologies are shown in light
blue.  The broad-line, optically unresolved AGNs typically have
$\Delta C \simeq\ 0.5$ and have $^{0.1}(u-r)<1$, and they do not
show a trend in $\Delta C$ with \urs, while the optically extended
sources show a similar trend to the radio, X-ray, and IR AGNs in the
other four panels, although the broad-line AGNs have slightly stronger
average contamination.  Since most AGNs with extended galaxy counterparts
have smaller $\Delta C$ and redder \urs\ than is typical of optically
unresolved AGNs with broad lines, we conclude that these sources are
not as dominated by a nucleus (as also indicated in
Figure~\ref{fprofile}, and that their colors may be robustly corrected
for nuclear contamination.

\footnotetext{As mentioned in Section~\ref{intro}, in this paper we
generally do not make use of optical spectroscopic data for selecting
AGNs, because of significant selection effects.  In estimating nuclear
color contamination, however, these sources are useful because the
spectroscopy provides evidence for some optical contribution from the
AGN.}

To correct the colors for nuclear contamination, we use
Eqn.~\ref{eqncorr} to derive \dccorr\ for each source, and so
determine a ``corrected'' $C_4$, then convert this to a corrected {\em
rest-frame} color \urs.  To determine how \urs\ varies with $C_4$, we
divide AGES main sample galaxies into seven bins of redshift from 0.25 to
0.8, and fit the dependence of \urs\ on $C_4$ in each bin.  A broken
power-law model with a break at $C_4=3.3$ gives a good fit for all
redshifts; examples of two fits are given in Figure~\ref{fcolrest}.  The
relation between \urs\ and $C_4$ becomes shallower with increasing
redshift, with slopes given by:

\begin{equation}
\frac{\Delta ^{0.1}(u-r)}{\Delta C_4} = 
\begin{cases}
0.73 + 1.27(z-0.70)^2 & (C_4 < 3.3) \\
0.33 + 3.17(z-0.72)^2 & (C_4 \geq 3.3)
\end{cases}
\label{eqnslope}
\end{equation}

We thus calculate the correction to \urs\ by multiplying \dccorr\
(from Eqn.~\ref{eqncorr}) by ${\Delta ^{0.1}(u-r)}/{\Delta
C_4}$ (from Eqn.~\ref{eqnslope}).  The typical corrections are small,
varying from $\simeq$0.3 for the bluest AGNs, to less than 0.05 for the
reddest AGNs.  The effects of these nuclear contamination corrections
on the host-galaxy color distribution of the AGN sample are shown in
Figure~\ref{fagncol}.

\section{Limits on star burst contamination of IR-selected AGN sample}
\label{appendix_ir}

As discussed in Section~\ref{irac}, it is possible that the IRAC-selected
AGN sample may contain some sources whose IR emission is powered by
star formation instead of nuclear accretion.  For the IRAC AGNs that
are also detected in X-rays (roughly half of the total sample,
and$\sim$$1/3$ of those with extended optical counterparts), the X-ray
emission is an unambiguous indicator of nuclear accretion.  In
addition, $\sim$40\% of IRAC-selected AGNs show clear evidence of
nuclear emission on the basis of their unresolved optical
morphologies.  However, for  83 (or 30\%) of the IR-selected AGNs
that are optically extended and not detected in X-rays, it is useful
to estimate the possible fraction of star burst contaminants.

For this we also use X-ray emission, but are limited to X-ray
stacking because the sources are not individually detected.  As
discussed in Section~\ref{stack}, the X-ray undetected IRAC AGNs have a
high average luminosity ($\average{L_X}\sim10^{42}$ \ergs) and hard
X-ray spectrum with ${\rm HR}=0.21\pm0.21$ (or $\Gamma\simeq 0.2$),
characteristic of  absorbed AGNs.  The stacked signal is
significantly harder and more luminous than the typical X-ray spectra
of star burst galaxies, which have $L_X\lesssim3\times10^{41}$ \ergs\
and $\Gamma\gtrsim 1.4$ \citep[e.g.,][]{rana03}.  This indicates that
most of the stacked X-ray flux from these sources is from AGNs, but
there remains a possibility than an X-ray-faint population of
star bursts contributes significantly to the stacked flux.

To test this possibility, we extrapolate the X-ray flux distribution
for the IRAC AGNs with extended counterparts that are X-ray detected,
and check that the stacked signal is consistent with this
extrapolation.  As a constraint, we use the {\em distribution} of
source counts at each IRAC AGN position.  This analysis constrains not
only the average X-ray flux of a sample, but also the shape of its
underlying flux distribution \citep[for a detailed discussion of this
technique, see][]{hick07c}.

\begin{figure}
\epsscale{1.1} 
\plottwo{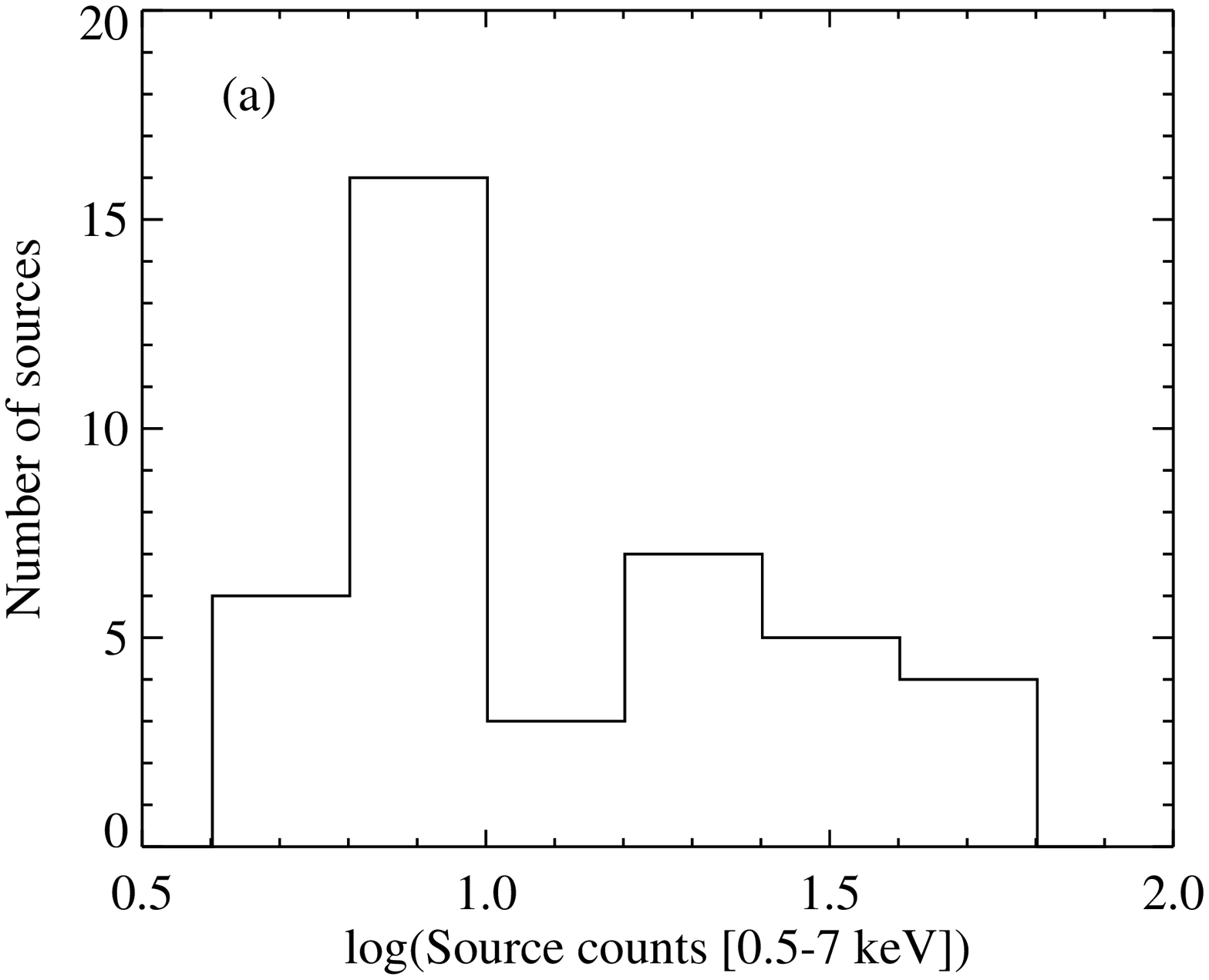}{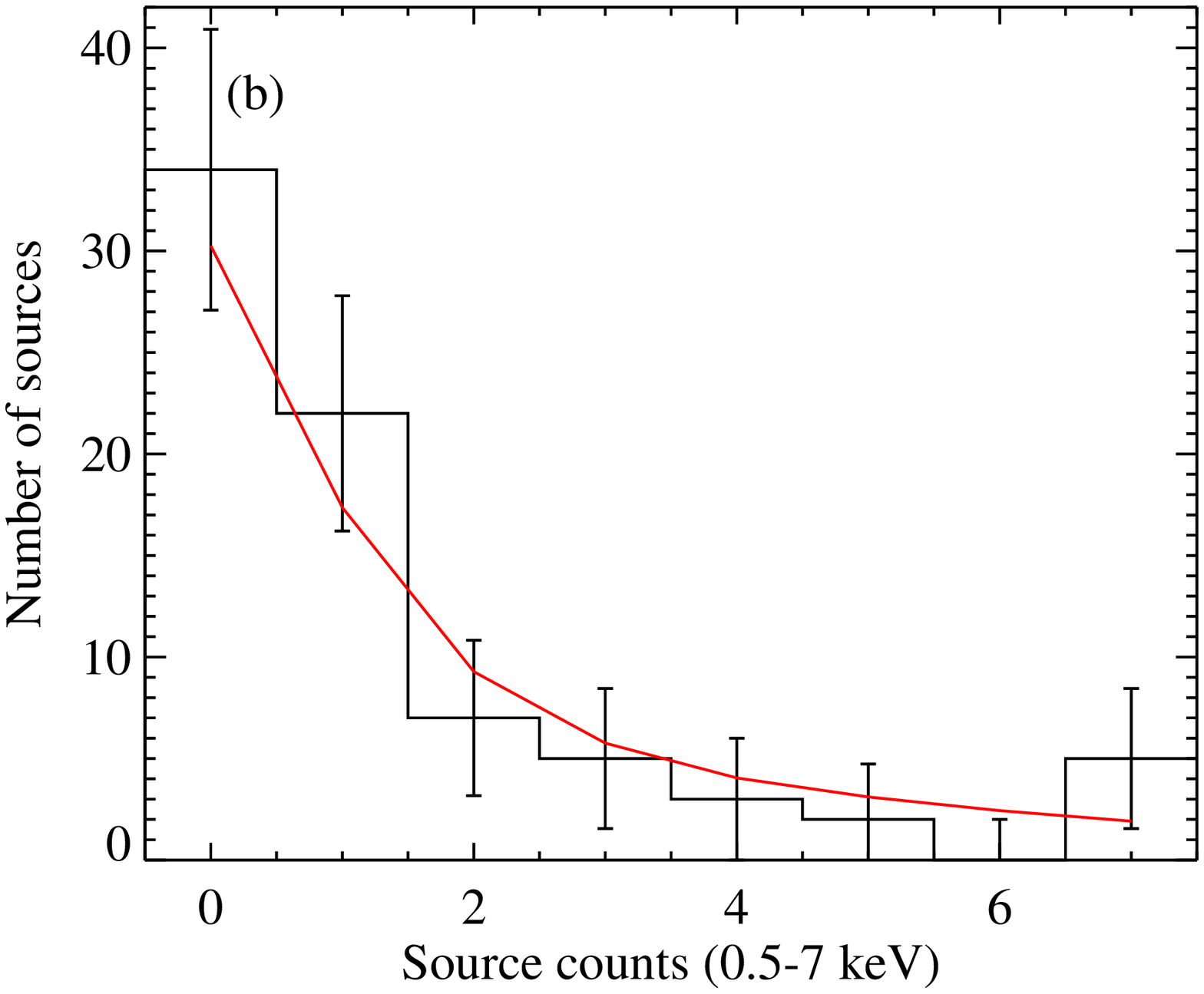}
\caption{({\em a}) Histogram of source counts for the 50 IR AGNs with
extended optical counterparts that are detected in X-rays.  We
extrapolate the roughly log-uniform distribution to lower fluxes, to
model the observed counts for the X-ray undetected sources. ({\em b})
Histogram of total observed counts (source plus background) within
$r_{90}$ for the 82 IR AGNs with extended optical counterparts that
are inside the \xbootes\ coverage area and are not detected in X-rays.
The red line shows the predicted counts distribution from a simple
extrapolation of a log-uniform flux distribution observed for the
detected sources.
\label{firstack}}
\end{figure}

The X-ray flux distribution for the 50 IRAC-selected AGNs with
extended counterparts that are detected in X-rays is roughly
log-uniform (Figure~\ref{firstack}{\em a}).  The IR luminosity and
redshift distributions of the X-ray- detected sources are similar to
those of their X-ray undetected counterparts, so it is reasonable to
expect that the shape of the X-ray flux distribution of the IRAC AGNs
may extend to fainter fluxes.  We therefore create a log-uniform model
flux distribution for the 82 X-ray undetected IRAC AGNs that have
extended optical counterparts and lie within the X-ray coverage area.
The model sources have fluxes from 0.02 to 4 count source$^{-1}$ (29
sources dex$^{-1}$ in flux).  We randomly assign the model fluxes to each
source in the sample.

We next calculate an estimated background X-ray flux from each
undetected IRAC AGN, given the average background surface brightness
in the 0.5--7 keV band (Section~\ref{stack}), and the area of the
extraction aperture (given by the 90\% energy encircled radius).  The
background varies from 0.01 to 0.1 count source$^{-1}$.  Adding the
background to the flux from the model flux distribution, we calculate
a predicted distribution of source counts at the IRAC AGN positions.
We repeat this calculation 10 times, with different permutations of
the source and background fluxes, and average the results to produce a
model counts distribution.

This model distribution shows remarkably good agreement with the total
number of observed source counts and their distribution, as shown in
Figure~\ref{firstack}{\em b} (note that the model is a direct
extrapolation of the observed flux distribution, and has not been
scaled fit the observed counts).  The X-ray emission from the
undetected IRAC AGNs is therefore fully consistent with an
extrapolation of the observed flux distribution for the X-ray-detected
sources.

Of course, there remains the possibility of a population of
star-forming galaxies at the faint end of the sample.  To put limits
on such a population, we repeat the above calculation but replace
the faintest model AGNs with star burst galaxies, with X-ray fluxes of
0.05 count source$^{-1}$ in 5 ks (corresponding to $L_X\sim10^{41}$ \ergs).
We calculate the variation in the $C$-statistic [\citet{cash79}, see
also \citet{hick07c}] as a function of the fraction of sources
replaced by star bursts.  The best match to the data is achieved for
the case of no star burst contamination; to 95\% confidence, the
non-AGN fraction must be less than 60\%.  This implies that for the full
sample of IRAC AGNs with extended optical counterparts, the
contamination is less than 40\%.  

In addition to the X-ray constraints on contamination, we also note
that (1) these sources would have to be highly extinguished star bursts
in order to have the AGN-like IRAC colors, and (2) the
optically extended IR AGNs show small but significant nuclear
contamination in the optical.  On the whole, these facts suggest that
the star burst contamination of the IRAC AGNs with extended
counterparts is most likely quite small ($\lesssim$$20\%$). This implies
that among the full IRAC AGN sample (including optically unresolved
sources), star burst contamination is at most $20\%$ and more likely
$\lesssim$$10\%$.

\clearpage

\end{document}